\documentclass[a4paper,11pt]{article}

\usepackage{lmodern}

\usepackage[utf8]{inputenc}
\usepackage[english]{babel}
\usepackage[myheadings]{fullpage}
\usepackage[T1]{fontenc}
\usepackage{subfig}
\usepackage{fancyhdr}
\usepackage{graphicx}
\usepackage{sectsty}
\usepackage{url}
\usepackage[usenames, dvipsnames, table, xcdraw]{xcolor}
\usepackage{amsmath}
\usepackage{amssymb}
\usepackage{calrsfs}
\usepackage{bbold}
\usepackage{dsfont}
\usepackage{esint}
\usepackage{indentfirst}
\usepackage{xfrac}
\usepackage{bigints}
\usepackage{slashed}
\usepackage{breqn}
\usepackage{autobreak}
\usepackage{setspace}
\usepackage[colorlinks]{hyperref}
\usepackage{microtype}
\usepackage{quoting}
\usepackage{bigints}
\usepackage{float}
\usepackage{array}
\usepackage[export]{adjustbox}
\usepackage[section]{placeins}
\usepackage[capitalise]{cleveref}
\usepackage{enumitem}
\usepackage{scalefnt}
\usepackage{booktabs}
\usepackage{caption}
\usepackage{makecell}
\usepackage{authblk}

\usepackage[normalem]{ulem}

\usepackage{cite}

\usepackage{xspace}

\usepackage{colortbl}
\definecolor{myblue}{rgb}{0.8,0.85,1}
\definecolor{light-gray}{gray}{0.95}

\hypersetup{colorlinks,
linkcolor={red!50!black},
citecolor={blue!50!black},
urlcolor={blue!80!black}}

\long \def \blockcomment #1\endcomment{}

\def\a{\alpha}
\def\b{\beta}
\def\c{\chi}
\def\d{\delta}

\def\g{\gamma}
\def\h{\eta}

\def\l{\lambda}
\def\m{\mu}
\def\n{\nu}

\def\p{\pi}

\def\r{\rho}

\def\t{\tau}

\def\D{\Delta}

\def\G{\Gamma}

\def\P{\Pi}

\makeatletter
\newcommand{\thickhline}{\noalign {\ifnum 0=`}\fi \hrule height 1pt
\futurelet \reserved@a \@xhline}
\newcolumntype{"}{@{\hskip\tabcolsep\vrule width 1pt\hskip\tabcolsep}}
\makeatother

\allowdisplaybreaks

\def\beq {\begin{equation}}
\def\eeq {\end{equation}}
\def\bea {\begin{eqnarray}}
\def\eea {\end{eqnarray}}

\def\nn {\nonumber}

\def\lbl{\label}

\def\beq{\begin{equation}}
\def\eeq{\end{equation}}
\def\bqry{\begin{eqnarray}}
\def\eqry{\end{eqnarray}}

\def\seEq#1{Eq.~(\ref{#1})}

\def\seneq#1{~(\ref{#1})}

\def\rcite#1{Ref.~\cite{#1}}
\def\rcites#1{Refs.~\cite{#1}}

\AtBeginDocument{}

\title{\LARGE {\bf \boldmath
Quark-hadron duality and the determination of $\alpha_s$ from
hadronic $\tau$ decay: facts ${\mathit vs.}$ myths}\\ }

\author[a]{Diogo Boito}
\author[b]{Maarten Golterman}
\author[c,d]{Kim Maltman}
\author[e]{Santiago Peris}\vspace{0.5cm}

\affil[a]{\it Instituto de F\'isica de S\~ao Carlos, Universidade de
S\~ao Paulo, CP 369, 13560-970, S\~ao Carlos, SP, Brazil\vspace{0.2cm}}

\affil[b]{\it Department of Physics and Astronomy, San Francisco State
University\\
San Francisco, CA 94132, USA\vspace{0.2cm}}

\affil[c]{\it Department of Mathematics and Statistics,
York University\\ Toronto, ON Canada M3J~1P3\vspace{0.2cm}}

\affil[d]{\it CSSM, University of Adelaide, Adelaide, SA~5005
Australia\vspace{0.2cm}}

\affil[e]{\it Department of Physics and IFAE-BIST, Universitat Aut\`onoma de
Barcelona\\
E-08193 Bellaterra, Barcelona, Spain
\vspace{0.3cm}}

\date{}

\begin{document}

\begin{flushright}
{\small \today}
\end{flushright}

\vspace*{-0.7cm}
\begingroup
\let\newpage\relax
\maketitle
\endgroup

\vspace*{-1.0cm}
\begin{abstract}
\noindent

Non-perturbative effects have a small but non-trivial
impact on the determination of the strong coupling from hadronic $\tau$
decay data. Several approaches have been proposed to take these into account,
the two most important of which are the ``truncated OPE'' approach and
``DV-model'' approach. Recently, Pich and Rodr\'iguez-S\'anchez have
raised a number of criticisms of the latter approach, including, most notably,
claims of the existence of (i) a supposed instability with respect to
variations of the model for incorporating quark-hadron duality violations,
and (ii) an alleged redundancy in the fitting strategy employed in the
DV-model approach. In this paper, we investigate these criticisms
and show they fail to survive more detailed scrutiny of the mathematical
or numerical arguments that underpin them. We also show that, while the
redundancy claim does not apply to the DV-model approach, it does,
in fact, apply to the truncated OPE approach. In particular, the $\alpha_s$ value
determined in the latter turns out to derive purely from perturbation
theory, with no role played by the non-perturbative
condensates determined in the rest of the analysis. This leads to the
conclusion that a revision of the conventional understanding of what
is learned from truncated OPE analyses is necessary and that only very
limited self-consistency checks are possible within this framework.
These observations raise new, non-trivial issues for the truncated OPE
approach.

\end{abstract}

\thispagestyle{empty}

\clearpage
\vspace*{0.0cm}

\setcounter{page}{1}
\section{Introduction}

The determination of the strong coupling, $\alpha_s$, from inclusive
hadronic $\tau$ decays is among the most precise determinations
relying on experimental data~\cite{ParticleDataGroup:2022pth}. The general
strategy for this determination was developed in
Refs.~\cite{Tsai:1971vv,Shankar:1977ap,Floratos:1978jb,Nachtmann:1978zf,Narison:1979pf,Krasnikov:1982ea,Schilcher:1983ae,Bertlmann:1984ih,Braaten:1988hc,Braaten:1988ea,Narison:1988ni,Braaten:1991qm}, and involves the use of finite energy
sum rules (FESRs), in which weighted integrals over hadronic spectral
functions obtained from experiment are equated to QCD-based
integrals evaluated along a closed contour in the complex plane
of the hadronic invariant-mass-squared variable $s=q^2 \equiv -Q^2$.
Through the use of this general strategy it was possible to
show, already in the 90s, that the QCD prediction for the
inclusive hadronic decay width is dominated by perturbative
contributions, in spite of the relatively low scale of the process
set by the $\tau$ mass, $m_\tau\simeq 1.8$ GeV. Since the coupling
at this scale is relatively large, $\alpha_s(m_\tau^2)\sim 0.3$, the
perturbative series is very sensitive to the value of $\alpha_s$,
allowing a rather precise determination, one that becomes even
more precise after evolving it to the $Z$ mass scale.
The comparison with other determinations at the $Z$ mass
provides strong quantitative evidence for asymptotic freedom.

Of course, the fact that $\alpha_s(m_\tau^2)$ is relatively large also
means the perturbative series may converge relatively slowly at the
$\tau$ mass scale, making it likely that numerically non-negligible
non-perturbative (NP) effects are also present. It is thus necessary
to take these into account in a precise determination of $\alpha_s$.
The decay width, for example, is in principle sensitive to the values
of dimension six and eight operator-product-expansion (OPE) condensates.
In the first analyses, which produced results for $\alpha_s(m_\tau^2)$
with a limited precision of order 10\%, order-of-magnitude estimates for
these contributions were sufficient. Naturally, as the precision improved
over the years, other, previously neglected, effects had to be considered.
Beyond non-perturbative OPE contributions, one should also expect
quark-hadron duality violations (DVs) to show
up~\cite{Poggio:1975af,Bigi:1998kc,Cata:2005zj,Cata:2008ye},
since for $s$ near $m_\tau^2$, one clearly sees residual oscillations,
reflecting the presence of resonances, in the experimental spectral
functions, and such oscillations are not captured by the OPE. Nevertheless,
DV effects were assumed negligible in the first $\tau$-based determinations.
This represented a reasonable first approximation since the weight function
associated with the decay width---the kinematic weight---has a
double zero at the timelike point on the OPE contour and hence suppresses
contributions from the vicinity of the Minkowski axis, where DVs are
expected to be largest~\cite{Poggio:1975af}. However, given the
current goal in precision, the absence
in a modern analysis of a quantitative study of the strength of this
suppression leaves open the possibility of a potentially relevant, but
unquantified, DV-related systematic uncertainty in the value
obtained for $\alpha_s$.

In the late 90s, the LEP experiments ALEPH~\cite{ALEPH:1997fek,ALEPH:1998rgl}
and OPAL~\cite{OPAL:1998rrm} measured the exclusive-mode spectra
for all dominant hadronic $\tau$-decay modes and obtained the
inclusive vector ($V$) and axial-vector ($A$) hadronic spectral functions
(supplementing the spectra for measured exclusive-modes with
Monte Carlo simulated data for subleading modes). This made it
possible to include in the analysis additional pseudo-observables
beyond the decay width~\cite{LeDiberder:1992zhd}, since FESRs can
be constructed for spectral integrals involving any analytical
weight function, $w(s)$, and integration ranges in $s$ extending
from threshold up to any $s_0\leq m_\tau^2$.
In practice, the weights are taken to be polynomials in the variable
$x=s/s_0$. The FESRs in question then have the form
\begin{equation}
\int_{\rm th}^{{s_0}} {\frac{ds}{s_0}}\, w(s/s_0)\, \rho (s)\, =\,
{\frac{-1}{2\pi i}}\oint_{\vert z\vert =s_0}
{\frac{dz}{s_0}}\, w(z/s_0)\Pi (z)\, ,
\label{basicfesr}\end{equation}
where $\Pi (q^2)$ is the relevant kinematic-singularity-free ($V$, $A$ or
$V+A$) polarization of the corresponding $V$, $A$ or $V+A$ current-current
two-point function and $\rho (s)$ is the corresponding spectral function.
The FESR relation is a consequence of Cauchy's theorem and holds when
exact forms are used for $\Pi$ and $\rho$. In the applications discussed
below, $\rho (s)$ is taken from experiment and the right-hand (theory)
side evaluated using the OPE approximation for $\Pi(z)$, possibly
supplemented by a model for residual DV contributions.

Working with weights which are polynomials in $x=s/s_0$ has the
advantage that integrated OPE contributions of different dimensions,
$D$, scale differently with $s_0$ (for the form above, as $1/s_0^{D/2}$).
In addition, neglecting logarithmic contributions suppressed by additional
powers of $\alpha_s$, the FESR based on a polynomial weight of degree $N$
receives non-perturbative OPE contributions only up to dimension $D=2N+2$,
with terms proportional to $x^m$ in $w(x)$ producing unsuppressed OPE
contributions proportional to $C_{2m+2}/s_0^{m+1}$, with
$C_D$ the relevant dimension-$D$ OPE condensate. It has become standard
to use several such polynomial-weighted moments of the spectral functions
in a single fit, with the goal of determining, from the data, both
the perturbative parameter $\alpha_s$ and non-perturbative parameters
$C_D$ simultaneously.

The first concrete implementation of this general idea was introduced
in Refs.~\cite{LeDiberder:1992zhd,ALEPH:1998rgl}, and employed by both
ALEPH and OPAL. In those analyses, it was necessary to assume that
certain in-principle-present higher-dimension OPE condensate contributions
could be safely neglected in order to ensure the number of OPE parameters
to be fit was less than the number of spectral integrals employed. For
this reason, we refer to this approach as the truncated-OPE strategy
(tOPE). Various versions of this approach remain in use today by some
groups~\cite{Pich:2016bdg,Ayala:2022cxo}.

The idea behind the tOPE strategy is to restrict the analysis to
moments with weights $w(x)=w(s/s_0)$ having at least a double zero
on the Minkowki axis (dubbed ``doubly pinched'') and work at the
highest possible $s_0$, $s_0\lesssim m_\tau^2$, in order to avoid,
as much as possible, contributions from DVs, which are
subsequently neglected. These weights typically include the kinematic
weight, which has degree $3$ and produces perturbative as
well as dimension $D=6$ and $8$ OPE contributions. The width alone,
which is determined by the kinematically weighted integral
with $s_0=m_\tau^2$, provides insufficient input to allow all of
$\alpha_s$ and the $D=6$ and $8$ condensates $C_6$ and $C_8$ to be
fitted. Additional moments,
involving a linearly independent set of polynomial weights, thus need
to be included. These necessarily include weights of degree $N>3$,
and hence higher-$D$ OPE contributions, unsuppressed by additional
powers of $\alpha_s$, proportional to condensates $C_D$ with $D$ up to
$2N+2>8$ on the theory side of the corresponding FESRs. Since little
or nothing is known about such $D>8$ condensates, adding a
new higher-degree weight thus also increases the number of OPE
parameters to be fit. The number of these OPE parameters, therefore,
always exceeds the number of independent doubly pinched weighted
$s_0=m_\tau^2$ spectral integrals, and tOPE analyses are forced to
truncate the OPE series, assuming the highest $D$ OPE contributions
can be neglected, on the grounds that, for polynomials in the variable
$x=s/s_0$, such contributions scale with higher inverse powers of
$s_0=m_\tau^2$. An example is provided by the classic $km$
spectral weight analyses carried out by ALEPH and OPAL, which employed 5
such doubly-pinched weights with degree up to 7. The theory sides of the
corresponding FESRs thus, in principle, depend on 8 OPE parameters:
$\alpha_s$ and the 7 non-perturbative condensates, $C_4$, $C_6$, $C_8$,
$C_{10}$, $C_{12}$, $C_{14}$ and $C_{16}$ which determine the
unsuppressed $D=4,\cdots ,\, 16$ OPE contributions. With only five
$s_0=m_\tau^2$ spectral integrals available, the OPE has to be
truncated, in this case at $D=8$, to keep the number of fit
parameters less than the number of spectral integrals used to fit them.
OPE contributions involving a sufficiently large number of higher
dimension condensates which are in principle present (in the literature
those of dimension 10, 12, 14 and 16), are thus set to zero. The
obvious question is how reliable this truncation is.

Another strategy for the extraction of $\alpha_s$ was introduced
in Ref.~\cite{Boito:2011qt}, and employed subsequently in
\rcites{Boito:2012cr,Boito:2014sta,Boito:2020xli}. The
rationale behind it is different. Since the OPE is at best
an asymptotic expansion, with high-dimension coefficients
expected to grow roughly factorially, it will become unreliable
at sufficiently high dimension. Uncontrolled truncation and/or
use of the series at high dimension, which might be reasonable
for a convergent series, is thus potentially dangerous for
the OPE and introduces some model dependence.
Although higher-order pinching is useful for suppressing DVs,
this comes at the cost of using higher-order polynomials,
which, in turn, are sensitive to the contribution from
higher-dimension condensates. Further suppressing DVs
by increasing the level of pinching thus eventually becomes
counter-productive, at some point bringing into play condensates
of dimension high enough that use of the OPE is no longer valid.

In view of the above discussion, it is important to avoid
moments which produce high-dimension OPE contributions. The choice
of moments to be used in a given analysis thus involves a trade-off:
while increasing the degree of pinching increases the suppression of
residual DV contributions and makes it safer to neglect them, doing
so eventually brings one into the regime where use of the OPE is no
longer valid, thus limiting how far one can go in implementing this
suppression. Since we do not know, a priori, the dimension
below which use of the OPE remains valid, the question of whether
DVs are sufficiently suppressed is one that requires quantitative
investigation. In the alternate strategy, which we refer to as
the DV strategy, this is done by using a model for the DV
contributions to the spectral function. This provides quantitative
estimates for the residual DV contributions present in the various
FESRs of the analysis. This model has been shown to follow, for
sufficiently large $s$, from general properties widely believed to
hold in QCD at large but finite number of colors~\cite{Boito:2017cnp}:
\begin{equation}\label{MnlargeNc}
M^2_n \sim n \quad, \quad \frac{\Gamma_n}{M_n}\sim \frac{1}{N_c} \ ,
\end{equation}
where $M_n$ is the hadron mass, $\Gamma_n$ the width, and $n$ the
excitation number very high in the spectrum. These properties are exact
in large-$N_c$ QCD in two dimensions, agree with the stringy picture of
hadrons and have a rather good phenomenological support.
Reference~\cite{Boito:2017cnp} also considered subleading corrections
in the spectrum that allow for a matching to perturbation theory and the OPE.

The parameters of the model are extracted self-consistently, together
with $\alpha_s$ and the relevant OPE condensates, using a range of
low-degree moments of the spectral-function data, together with a range
of $s_0$ values, rather than just $s_0= m_\tau^2$ (or a different single
value close to $m_\tau^2$). An advantage of this strategy is that it fully
exploits the $s$ dependence of the data, and the resulting $s_0$-dependence
of the weighted spectral integrals, and is therefore, at least in principle,
able to disentangle OPE contributions of different dimension, something
which, unless further assumptions are made, is not possible in an
analysis using only a single fixed $s_0$. The problem for single-$s_0$
analyses involving the at-least-doubly-pinched weights used in tOPE analyses
in the literature is that any such weight contains at least two positive
powers of $s$ and the associated FESR, therefore, receives contributions from
$\alpha_s$ and at least two NP condensates. A single such FESR, employed at a
single $s_0$, thus provides only one equation, involving a minimum of three
unknowns. Since each additional independent pinch-weighted FESR introduced to
provide additional equations necessarily also introduces a minimum of one new
NP condensate, the number of parameters to be fit ($\alpha_s$ and the relevant
NP condensates) always exceeds the number of single-$s_0$, pinch-weighted
spectral integrals available to fit them. The tOPE approach deals with this
problem by throwing away enough highest-dimension theory-side contributions
that the number of remaining fit parameters becomes less than the number
of single-$s_0$ FESRs employed. Details of the implementation of this by-hand
truncation are provided in Secs.~\ref{kmspecwtsub}, \ref{whatkmwtsub} and
\ref{optwtsub} for three different five-weight tOPE analysis cases.

The main potential issue with the DV approach is the question of the
extent to which the form of the DV model employed is (i) sufficiently
well-motivated theoretically, and (ii) sufficiently robust against
(reasonable and theoretically well-motivated) modifications of the
parametrization used. We address these questions further in this work.

It is important to note that the difference between the tOPE and DV
strategies is not a purely academic issue. If the same experimental
data are used, it is now well known that analyses based on the DV strategy
lead to results for $\alpha_s$ which lie systematically below
those based on the tOPE approach. In this respect, it is to be noted that
the apparent agreement on $\alpha_s$ shown in Fig 9.2 of the PDG section
on QCD~\cite{ParticleDataGroup:2022pth} between the tOPE approach
of ``BP2008-16 FO'' and the updated DV strategy analysis in
``Boito2021 FO'' is fortuitous, as the former is based solely on the
ALEPH spectral data whereas the latter is based on an improved vector
spectral data obtained by combining, via CVC, recent results from
$e^+e^-\to$ hadrons with $\tau$ decay $2\pi$ and $4\pi$ results
from ALEPH and OPAL (see section \ref{theoryreview} and below
for more details).

For many years, the tOPE strategy was not subjected to close scrutiny
and its issues were not investigated in detail until the works
of \rcites{Boito:2014sta,Boito:2016oam}. In those references,
arguments were presented that additional systematic uncertainties not
quantified by existing tOPE analyses, were present in then-current
tOPE results for $\alpha_s$. Several of these arguments remain,
at present, unaddressed.

Beginning with Ref.~\cite{Pich:2016yfh} in 2016, and
continuing in subsequent proceedings and review
volumes~\cite{Pich:2018jiy,Pich:2016yfh,Pich:2020gzz},
the authors of Ref.~\cite{Pich:2016yfh} have presented a number
of claims purporting to support use of the tOPE assumptions and
cast doubt on the reliability of the DV-model-strategy approach.
A recent invited contribution to the anniversary edition of
JHEP~\cite{Pich:2022tca} has, in addition to restating the results
of the earlier 2016 paper and repeating these previous claims
(including several already refuted in the literature), presented
new arguments against the DV-strategy approach. In this paper,
we address, directly and quantitatively, the claims made in
Ref.~\cite{Pich:2022tca}, identifying errors and/or
mis-statements in the arguments which underlie them. In particular, we
present new results which follow from these investigations,
which are shown to expose previously unrecognized issues
with the tOPE approach and to necessitate a major
reinterpretation of the conventional understanding of what
is learned from analyses employing that approach.

We would like to also comment here on recent developments with regard
to the treatment of the perturbative series, as we believe
the recent history of investigations of this issue in the
context of the hadronic decays of the $\tau$ lepton is of some
interest in its own right.

In 1992 it was suggested in Refs.~\cite{Pivovarov:1991rh,LeDiberder:1992jjr}
that the convergence of the perturbative expansion could be improved by
resumming the logarithms associated with the running of $\alpha_s$
on the contour appearing in the FESR, Eq.~(\ref{basicfesr}). This is
achieved by setting the renormalization scale $\mu^2$ equal to the
complex-valued momentum $z=s_0 e^{i \phi}$ appearing on the right-hand
side of Eq.~(\ref{basicfesr}). This prescription was named Contour
Improved Perturbation Theory (CIPT). The standard $\overline{\mathrm{MS}}$
practice of truncating the perturbative series at a fixed order in
$\alpha_s(s_0)$ was then re-named Fixed Order Perturbation Theory (FOPT).
For years CIPT was advocated as the way to deal with the non-trivial growth
of the FOPT $\log (z/s_0)$ factors when moving from the spacelike
toward the timelike point along the contour~\cite{Pich:2013lsa}.
After the works of Refs.~\cite{Beneke:2008ad,Beneke:2012vb},
which provided evidence in favor of FOPT, the focus changed and
it became common practice to \emph{average} the CIPT and FOPT
values of $\alpha_s$ (see, {\it e.g.}, Ref.~\cite{Pich:2022tca}),
as if the difference between these two values measured a systematic
effect in the determination of $\alpha_s$ rather than originating
in a potential systematic issue with the fundamental properties
of one or both of the expansions. This averaging led to a large source
of error in the value of $\alpha_s$ from $\tau$ decay. Fortunately,
recent developments have drastically changed the situation and it has
become clear that the conventional implementation of CIPT is
incompatible with the canonical form of the
OPE~\cite{Hoang:2020mkw,Hoang:2021nlz,Benitez-Rathgeb:2022yqb,Benitez-Rathgeb:2022hfj,Gracia:2023qdy,Golterman:2023oml,Beneke:2023wkq}
and, hence, that results obtained using this scale setting must not be
averaged with determinations based on $\overline{\mathrm{MS}}$-FOPT,
which is fully compatible with the usual form of the OPE.\footnote{We
note that CIPT-based results are no longer considered in the latest
issue of the PDG review on QCD \cite{Huston:2023ofk}.} Therefore,
in this work, we will not consider CIPT any further.

This paper is organized as follows. In Sec.~\ref{theoryreview},
we review the basic theoretical background and set notation for
use in the discussions which follow. In
Sec.~\ref{sec:supposedtautology} we discuss, and prove to be
incorrect, a key claim from Ref.~\cite{Pich:2022tca} regarding
the supposed redundancy of the multi-weight, multi-$s_0$
DV-strategy fits of Refs.~\cite{Boito:2014sta,Boito:2020xli}
and show that such fits, in fact, provide highly non-trivial
constraints on the underlying theoretical representation. As
part of this discussion, we identify the mathematical error
responsible for the erroneous claim in Ref.~\cite{Pich:2022tca}.
In Sec.~\ref{tOPEredundancy}, we investigate in more detail the
issue of ``redundancy,'' showing that, rather than being an issue
for the multi-weight, multi-$s_0$ DV-strategy fits of
Refs.~\cite{Boito:2014sta,Boito:2020xli}, it is, in fact,
a problematic issue for single-$s_0$ tOPE fits. We conclude in
Sec.~\ref{conclusions} with a summary and discussion of our
results. Additional material is presented in six Appendices.
Some technical details
relevant to addressing (and refuting) claims made in
Ref.~\cite{Pich:2022tca} concerning (i) the potential sensitivity
of the results for $\alpha_s$ to details of the form used to model
DV contributions and (ii) the potential impact on the results for
$\alpha_s$ of $\alpha_s$-suppressed logarithmic corrections to
the leading-order behavior of higher-dimension non-perturbative
OPE contributions to the $V$ and $A$ current-current polarization
functions, $\Pi (Q^2)$, are relegated to
Appendices~\ref{sec:OPE-logs},~\ref{app:dim-6 condensates},
and~\ref{sec:DV-parm}. Additional details not covered in the main text are
dealt with in three additional appendices. In particular, more examples
of the problems the tOPE approach has with the redundancy issue are
presented in Appendix~\ref{app:improvedVchannel}.

\section{Theoretical framework}\label{theoryreview}

For the precise extraction of $\alpha_s$ from
$\tau\rightarrow ({\rm hadrons})+\nu_\tau$ it is
convenient to work with decays without net strangeness in the final
state, restricting the analysis to light-quark (isovector) current
contributions. Mass corrections in this case are tiny and can be
safely neglected, as checked numerically in Ref.~\cite{Boito:2011qt}.
Furthermore, the main results from the most recent DV-strategy analysis
are obtained from the $V$ spectral function, for more than
one reason. First, the dominant resonance in the $V$ channel, the
$\rho(770)$, is lighter than the $a_1$ which makes it safer to work
with a description of duality violations that is valid away from narrow
resonance peaks. A second advantage of the $V$ channel is that it
is now possible to improve on the ALEPH and OPAL spectral functions
by using the conserved vector current (CVC) relation, in combination with
recent results from $e^+e^-\to {\rm hadrons}$, to obtain data-driven
estimates for the spectral contributions from subleading $\tau$ decay modes.
As detailed in Ref.~\cite{Boito:2020xli}, this leads to an improved version
of the $V$ spectral function, one that is, moreover,
based entirely on experimental data, without the need for Monte Carlo
simulated subleading-mode input. We employ this improved $V$ channel
spectral function in a number of the discussions below.

The FESR analysis is based on the current-current correlation
functions of the vector $J^{(V)}_\mu(x)=\bar u \gamma_\mu d(x)$
and axial-vector $J^{(A)}_\mu(x)=\bar u \gamma_\mu\gamma_5 d(x)$
currents, defined as
\bqry
\lbl{correl}
\P_{\m\n}^{V/A}(q)&=&i\int d^4x\,e^{iq\cdot x}
\langle 0|T\left\{J^{(V/A)}_\m(x)J^{(V/A)\dagger}_\n(0)\right\}|0\rangle\\
&=&\left(q_\m q_\n-q^2 g_{\m\n}\right)\P^{(1)}(q^2)+q_\m q_\n\P^{(0)}(q^2)
\nonumber\\
&=&\left(q_\m q_\n-q^2 g_{\m\n}\right)\P^{(1+0)}(q^2)+
q^2 g_{\m\n}\P^{(0)}(q^2)\ ,\nonumber
\eqry
where the superscripts (0) and (1) label the spins and we will
often omit the labels $V/A$. The decomposition in the third line is
introduced because $\Pi^{(1+0)}(q^2)$ and $q^2\Pi^{(0)}(q^2)$ are
free of kinematic singularities. The associated spectral functions
are defined as
\beq
\rho^{(J)}(s) = \frac{1}{\pi}\,{\rm Im} \Pi^{(J)}(s)\ .
\eeq
In the non-strange $V$ channel, $\rho_V^{(0)}(s)$ is
doubly chirally suppressed (proportional to $(m_d-m_u)^2$)
and hence safely negligible. A similar, $(m_d+m_u)^2$
suppression is present for all contributions to $\rho_A^{(0)}(s)$
other than that of the pion pole, which must be explicitly
kept. Since $\Pi^{(1+0)}(s)$ is analytic in the complex plane
except for the unitarity cut along the real axis, Cauchy's theorem
implies that the following ``finite-energy sum rules'' (FESRs),
\beq
\lbl{cauchy}
I^{w}_{V/A}(s_0)
\equiv\frac{1}{s_0}\int_0^{s_0}ds\,w(s)\,\r^{(1+0)}_{V/A}(s)
\, =\, -\frac{1}{2\p i\, s_0}\oint_{|z|=s_0}
dz\,w(z)\,\P^{(1+0)}_{V/A}(z)\ ,
\eeq
are valid for any $s_0>0$ and any weight function
$w(z)$ analytic inside and on the contour of integration. Here we
always take $w(z)$ to be a polynomial in $x=z/s_0$.

The spectral functions can be determined experimentally from the
differential versions of the ratios
\beq
\lbl{R}
R_{V/A;ud}=
{\frac{\G [\t\rightarrow ({\rm hadrons})_{V/A;ud}\n_\t (\g ) ]}
{\G [\t\rightarrow e\bar{\n}_e \nu_\tau (\g ) ]}}
\eeq
of the hadronic decay widths to the electron-mode decay width.
Explicitly,
\beq
\lbl{taukinspectral}
{\frac{dR_{V/A;ud}(s)}{ds}}= 12\pi^2\vert V_{ud}\vert^2 S_{\rm EW}\,
{\frac{1}{m_\tau^2}} \left[ w_T(s;m_\tau^2) \rho_{V/A;ud}^{(1+0)}(s)
- w_L(s;m_\tau^2) \rho_{V/A;ud}^{(0)}(s) \right]\ ,
\eeq
where $V_{ud}$ is the relevant CKM matrix element and $S_{\rm EW}$ is
a short-distance electroweak correction. The functions $w_{T,L}(s;m_\tau^2)$
are the $s_0=m_\tau^2$ versions of
\bqry
\lbl{kinweights}
w_T(s;s_0)&=&\left(1-\frac{s}{s_0}\right)^2\left(1+2\,\frac{s}{s_0}\right)\ ,\\
w_L(s;s_0)&=&2\,\frac{s}{s_0}\left(1-\frac{s}{s_0}\right)^2\ .\nonumber
\eqry
As explained above, the pion pole term entirely dominates the
contribution proportional to $w_L(s;m_\tau^2)$ in Eq.~(\ref{taukinspectral}).
In what follows we refer to the weight $w_T(s;s_0)$, which governs the
(non-pion-pole) continuum decay distribution, as the $\tau$ kinematic weight.

The integrals over the spectral functions in the FESRs Eq.~(\ref{cauchy})
are obtained from experimental data, and will be denoted
$I^{w}_{V/A}(s_0)$ in what follows, with the $V/A$ subscripts dropped
where no confusion arises. The most recent and up-to-date $V$ spectral
function was obtained in Ref.~\cite{Boito:2020xli}. It combines
information on the dominant $2\pi$ and $4\pi$ modes measured
by ALEPH~\cite{Davier:2013sfa} and OPAL~\cite{OPAL:1998rrm} with recent
experimental results for many subleading modes, obtained from exclusive
$e^+e^-\to {\rm hadrons}$ cross-section
measurements~\cite{BaBar:2007ceh,BaBar:2017zmc,Achasov:2016zvn,SND:2014rfi,Achasov:2017kqm,BaBar:2018erh,BaBar:2018rkc,Gribanov:2019qgw,BaBar:2006vzy,CMD-3:2013nph,Achasov:2019nws,CMD-3:2017tgb,Achasov:2019duv,BaBar:2007qju,Achasov:2016eyg},
using CVC, and from BaBar data for
$\tau\to K^-K^0\nu_\tau$~\cite{BaBar:2018qry}, leading to a purely
data-based result, with no need of Monte-Carlo input for
any of the decay mode contributions. The exclusive modes entering this
construction represent $99.95\%$ of the total vector branching fraction.
A key advantage of the use of electroproduction cross-section input
(which is possible only for the $V$ channel) is the fact that $s$ is
kinematically unrestricted in $e^+ e^-\rightarrow {\rm hadrons}$. This
is in contrast to the situation for $\tau$ data, where the restricted
phase space near the kinematic endpoint ensures small data samples and
large statistical errors for $s$ near $m_\tau^2$. This is precisely
the region where multiparticle subleading mode contributions become
important. Compared to results obtained using $\tau$ data alone,
the new vector-isovector spectral function thus has significantly
reduced errors in the upper part of the $\tau$ kinematic range, and
we take advantage of this improvement by focusing some of the key
explorations below on the $V$ channel case.

The right-hand side of the FESR relation Eq.~(\ref{cauchy}) is the theory
side, denoted $I^{w}_{V/A;{\rm th}}(s_0)$ in what follows.
For $|s|=s_0$ sufficiently large and far away from the positive real axis,
the polarization function $\Pi^{(1+0)}(s)$ can be approximated by
its OPE. The dominant, dimension zero, term is the
perturbative contribution. This is usually cast in terms of $D_0(z)$,
the perturbative contribution to the scale-invariant Adler function,
which has the form
\beq
\lbl{pertth}
D_0(z)\equiv -z\,\frac{dC_0(z)}{dz}=\frac{1}{4\p^2}
\sum_{n=0}^\infty\left(\frac{\a_s(\m^2 )}{\p}\right)^n\sum_{m=1}^{n+1}
m\,c_{nm}\left(\log\frac{-z}{\m^2}\right)^{m-1}\ .
\eeq

Because of the formal scale-invariance of the Adler function, only
the coefficients $c_{n1}$ are truly independent. They are known
exactly up to $\mathcal{O}(\alpha_s^4)$~\cite{Baikov:2008jh,Herzog:2017dtz},
with the following values for three quark flavors and in the
$\overline{\rm MS}$ scheme: $c_{01}=c_{11}=1$, $c_{21}=1.63982$,
$c_{31}=6.37101$ and $c_{41}=49.07570$. We also include an
$\mathcal{O}(\alpha_s^5)$ contribution, using as input for the
fifth-order coefficient the value $c_{51}=283\pm 142$, which
covers the range of estimates found in dedicated
studies~\cite{Beneke:2008ad,Boito:2018rwt,Caprini:2019kwp,Jamin:2021qxb}.
In Eq.~(\ref{pertth}) the $\alpha_s(\mu^2)$ running can be computed
with the four- or five-loop QCD $\beta$
function~\cite{Chetyrkin:2017lif,Herzog:2017ohr} with only tiny numerical
differences.

When the perturbative expansion of the Adler function is inserted on
the theory side of the FESR~\cite{Beneke:2008ad}, a prescription for the
renormalization-scale setting is required. It is well known that the most
widely employed prescriptions, FOPT (a strict fixed-order expansion)
and CIPT (which resums the running of the coupling along the contour of
integration), produce discrepant results for $\alpha_s$. Thanks to
significant progress on this problem in the last three
years~\cite{Hoang:2020mkw,Hoang:2021nlz,Benitez-Rathgeb:2022yqb,Benitez-Rathgeb:2022hfj,Gracia:2023qdy,Golterman:2023oml,Beneke:2023wkq},
this issue is now very well understood and it is clear that it is FOPT that
must be used in conjunction with the usual OPE unless the Adler
function is properly modified by using a renormalon-free scheme for
the gluon condensate, which strongly suppresses the inconsistency of CIPT,
following Refs.~\cite{Benitez-Rathgeb:2022yqb, Benitez-Rathgeb:2022hfj,Beneke:2023wkq}.

Higher-dimension OPE corrections start at dimension $D=4$,
with terms proportional to the gluon and chiral
condensates,\footnote{The former dominates the sum for typical values
of the gluon condensate found in the literature.} and can be
generically written as
\beq
\lbl{OPE}
\P^{(1+0)}_{\rm OPE}(z)=\sum_{k=0}^\infty \frac{C_{2k}(z)}{(-z)^{k}}\ .
\eeq
The coefficients $C_{2k}(z)$ for $k\geq 2$ contain both the relevant
dimension $D=2k$ condensates and their Wilson coefficients, the
logarithmic $z$ dependence of which can, in principle, be obtained
perturbatively. We discuss these $\alpha_s$-suppressed logarithmic
corrections --- and their potential impact on extractions of $\alpha_s$
from $\tau\rightarrow ({\rm hadrons})+\nu_\tau$ data --- in
Appendix~\ref{sec:OPE-logs}. Neglecting these corrections, the
coefficients $C_{2k}$ are $z$ independent. In this approximation, the
condensate $C_{D}$ contributes to the theory side of Eq.~(\ref{cauchy})
only if the weight $w(z)$ contains a term proportional to $z^n$, with
$D=2(n+1)$. It is customary, since the work of Ref.~\cite{Beneke:2012vb},
to avoid weight functions that contain the monomial $z$, since those are
maximally sensitive to the gluon condensate and, thus, to its
infra-red renormalon, which leads to instabilities in the perturbative
series~\cite{Boito:2020hvu,Benitez-Rathgeb:2022yqb,Benitez-Rathgeb:2022hfj}.

As already mentioned, the OPE representation for $\Pi^{(1+0)}(z)$
is expected to break down in the vicinity of the positive real $z=s=q^2$
axis~\cite{Poggio:1975af}. This breakdown is related to the analytic
continuation from the Euclidean~\cite{Shifman:2000jv,Boito:2017cnp},
where it is well defined, to the Minkowski axis. In the process,
additional oscillatory terms expected to be damped as
$s\rightarrow\infty$, are generated. We refer to these as the duality
violation (DV) contributions, and account for this contribution in
the DV-strategy approach by defining the duality-violation contribution
to $\Pi^{(1+0)}(z)$ as
\beq\label{DVdifference}
\Delta(z) \equiv \Pi^{(1+0)}(z)- \Pi_{\rm OPE}^{(1+0)}(z)\ .
\eeq
Focusing on the $V$ channel to be specific, commonly accepted properties
of QCD, together with general observations about analytic continuation,
lead one to expect $\mathrm{Im}\,\Delta_V(s)$ to
have an oscillatory form, modulated by an exponential falloff in the
amplitude as $s\to \infty$. For polynomial weights, this expected
exponential suppression allows the resulting DV contribution to the
right-hand side of the $V$ version of Eq.~(\ref{cauchy})
to be rewritten as
\beq
\lbl{FESRDV}
\frac{-1}{s_0}\int_{s_0}^{\infty} ds \, w(s)\ {\frac{1}{\pi}}\,
\mathrm{Im}\Delta_V(s)\, \equiv\, \frac{-1}{s_0}\int_{s_0}^{\infty}
ds \, w(s)\ \rho_V^{\rm DV}(s)\ .
\eeq
This shows explicitly the effect of pinching and why the use of
a polynomial weight, $w(s)$, with a zero at $s=s_0$ \emph{tends} to
suppress DV contributions.\footnote{It also clarifies how
another, unexpected feature of pinching can occur: the structure of the
DV oscillation may cause the DV contribution to a less pinched integral
to be more suppressed than that to a more pinched integral. See
Ref.~\cite{Peris:2016jah} for an example.}
These observations lead one to consider the expression~\cite{Cata:2005zj}
\beq
\lbl{DV-parametrization}
\r_{V}^{\rm DV}(s)=\frac{1}{\p}\,\mbox{Im}\,
\D_{V}(s)=e^{-\d_{V}-\g_{V} s}\sin(\a_{V}+\b_{V} s)\ ,
\eeq
as a parametrization of DVs for sufficiently large $s$.The four parameters
$\delta_V$, $\gamma_V$, $\alpha_V$, and $\beta_V$ are to be obtained
self-consistently from the data, together with $\alpha_s$ and the
condensates contributing to the FESR(s) under consideration. Such a
determination is possible only for fits involving a range of $s_0$ values.
An essential assumption, therefore, is that the asymptotic form
(\ref{DV-parametrization}) remains valid for $s$ down to the lowest $s_0$,
$s_0^{\rm min}$, in the chosen fit window.{\footnote{This assumption can be
tested using a range of multi-weight, multi-$s_0$ fits. The results show
that use of the asymptotic form is consistent with data for $s_0^{\rm min}$
down to $\sim 1.55\ {\rm GeV^2}$~\cite{Boito:2014sta,Boito:2020xli}.}}
This assumption, of course, has to be checked.

Note that, after using the result Eq.~(\ref{FESRDV}), the basic
FESR equation, Eq.~(\ref{cauchy}), takes the form
\beq
\lbl{finalFESR}
\int_0^{s_0}\!\!\frac{ds}{s_0}\,w(s)\,\left[\rho^{(1)}_{V}(s)\right]_{\rm EXP}
\!\!\!=\, -\frac{1}{2\pi i}\oint_{|z|=s_0} \frac{dz}{s_0}\,
w(z)\,\left[\Pi^{(1)}_{V}(z)\right]_{OPE}-\int_{s_0}^{\infty}
\frac{ds}{s_0} \, w(s)\ \rho_V^{\rm DV}(s)\ .
\eeq
This shows that DV contributions to the FESRs\footnote{The DV term in Eq. (\ref{finalFESR}) is the result of deforming, for the contribution of the DV difference in Eq.~(\ref{DVdifference}), the $|z|=s_0$ contour used in Eq.~(\ref{basicfesr}) to a contour of infinite radius (where the DVs vanish exponentially) avoiding the physical cut.} of a given analysis involve
the model DV form $\rho_V^{\rm DV}(s)$ not only for $s$ in the range between
the lowest and highest $s_0$ in the analysis fit window, but also in the
region above the highest such $s_0$, including the region above $s=m_\tau^2$,
where no data is even available, and hence no data is influencing the fit.
Although $\rho_V^{\rm DV}(s)$ is exponentially damped at large $s$, the
value of the parameter governing that damping, $\gamma_V$, obtained from the V
channel fits of Ref.~\cite{Boito:2020xli}, $\simeq 0.6$ GeV$^{-2}$, is not
large enough to make DV contributions to the variously weighted DV integrals
coming from the region above the maximum $s_0$ in the analysis numerically
negligible. The value of $\alpha_s$ obtained from the fits thus depends on
the DV ansatz for $s$ not only in the region of the spectral integrals
entering the analysis, but also the values above that, as do the results of
self-consistency tests resulting from the use of multi-weight, multi-$s_0$
fits.

The same general form as Eq.~(\ref{DV-parametrization}), with different,
channel-specific DV parameters, is, of course, expected for the $A$ channel.
The generic oscillatory, exponentially suppressed DV form has been employed
not only in determinations of $\alpha_s$ but also of chiral low-energy
constants~\cite{Cata:2008ye,Cata:2008ru,Boito:2011qt,Boito:2012cr,Boito:2012nt,Boito:2014sta,Boito:2015fra,Boito:2020xli},
as well as by the authors of Ref.~\cite{Pich:2022tca}, albeit in
a simplified form~\cite{Gonzalez-Alonso:2010kpl,Gonzalez-Alonso:2016ndl}.
We emphasize that {\it all} strategies resort to a model for duality
violations --- ignoring duality violations, as done in the tOPE approach,
corresponds to using the model $\rho_{\rm DV}(s)\equiv 0$ (for $s>s_0$), and
makes an assumption on their impact, rather than investigating that
impact.

While the functional form in Eq.~(\ref{DV-parametrization}) is
based on the expected general properties of DVs, it is relevant to
investigate how this expression may result from QCD and, very importantly,
what form possible corrections may take. Although no first-principles
version of such an investigation exists, it is reassuring that
one \emph{does} obtain this expression from the expected Regge
behavior of QCD at large (but finite) $N_c$ (see Eq.~(\ref{MnlargeNc})).
The experimentally extracted values of the DV parameters in
(\ref{DV-parametrization}) are, moreover, in good agreement with
expectations based on the values of the parameters governing the masses
in the radial Regge trajectories~\cite{Masjuan:2012gc,Boito:2017cnp}
(see also Ref.~\cite{Peris:2021jap}), providing a rather nontrivial
test of this connection.

The analysis of Ref.~\cite{Boito:2017cnp} obtained as possible
corrections to (\ref{DV-parametrization}) the expression
\beq
\lbl{DV-parametrization-gen1}
\r_{V}^{\rm DV}(s)=\left[1+\mathcal{O}\left(\frac{1}{N_c},
\frac{1}{s},\frac{1}{\log s} \right) \right]
e^{-\d_{V}-\g_{V} s}\sin\left(\a_{V}+\b_{V} s +\mathcal{O}(\log s)\right)\ ,
\eeq
as $s\rightarrow \infty$.\footnote{We emphasize that the correction
written schematically as $\mathcal{O}(1/\log (s))$ in
Eq.~(\ref{DV-parametrization-gen1}) ($\mathcal{O}(1/\log (\beta_V s/2\pi ))$
when physical units are restored) is valid {\it{only}} in the limit
$s\to \infty$. The underlying form, valid also at non-asymptotic $s$, is
straightforwardly obtainable, starting from the expression for the term
proportional to $a_{\rm log}$ in the DV contribution to the polarization,
$\Pi_{\rm DV}(s)$, provided in Appendix C of Ref.~\cite{Boito:2017cnp}. As the
divergence as $s\rightarrow 2\pi /\beta_V$ of the asymptotic form makes clear,
this asymptotic form should not be used in any quantitative investigation
of the size of this correction at finite $s$.}
In Appendix~\ref{sec:DV-parm} we discuss how sensitive the extracted value
of $\alpha_s$ is to such corrections, refuting claims about this sensitivity
made in Ref.~\cite{Pich:2022tca}.

\section{On ``tautological tests and fitting the spectral function''}
\label{sec:supposedtautology}

In this section we investigate, and show to be incorrect, two
key claims made in Ref.~\cite{Pich:2022tca} regarding the
DV-strategy analyses in Ref.~\cite{Boito:2014sta} and, more
recently, in Ref.~\cite{Boito:2020xli}. The first of these claims
is that the good agreement observed in these analyses between
results obtained using FESRs with a range of different weight
functions and variable $s_0$ fit windows, is a consequence, not of
the analyses fulfilling non-trivial self-consistency tests on the
underlying theory representations, but rather of the supposed redundancy
of multi-weight versions of such variable-$s_0$ fits. The source of
this erroneous statement is a basic mathematical error in the argument
of Appendix A of Ref.~\cite{Pich:2022tca}, involving the (as it turns
out, incorrect) assumption that a result valid for fits of one type
remains valid for fits of the fundamentaly different type employed
in the analyses of the expanded sets of spectral integrals considered
in Refs.~\cite{Boito:2014sta,Boito:2020xli}. As we will show, the
investigation of this claim, in addition, turns out to expose previously
unrecognized issues with the tOPE approach. As we will see, these issues
necessitate a significant revision of the conventional understanding of
what is learned from standard tOPE analyses in the literature. The second
claim made in Ref.~\cite{Pich:2022tca}, namely that the $\alpha_s$
determinations produced by the DV-strategy analyses of
Refs.~\cite{Boito:2014sta,Boito:2020xli} extract $\alpha_s$
essentially only from the $w_0=1$-weighted spectral integral
at $s_0=1.55$~GeV$^2$, with the rest of the experimental
information (the spectral integrals at the higher $s_0$ in the
$s_0$ fit window) serving only to fix the DV parameters, is also
shown to be incorrect below.

In a given physical channel, a FESR is characterized by the weight function
$w(s)$, and an upper integration limit, $s_0$, appearing in the spectral
integrals, $I^{w}(s_0)$, present on the experimental side of that
FESR. A set of $I^{w}(s_0)$ with fixed weight $w$ but variable $s_0$
is necessarily linearly independent and has a well-defined, invertible
covariance matrix. The same is true of a set of $I^{w}(s_0)$ with fixed
$s_0$ but multiple, linearly independent polynomial weights $w(s)$.
As noted in Appendix A of Ref.~\cite{Pich:2022tca}, however,
a set of $I^{w}(s_0)$ with both multiple weights $w(s)$ and multiple
$s_0$ is generally linearly dependent, producing a singular covariance
matrix and preventing a standard $\chi^2$ fit from being performed
(see also Refs.~\cite{Boito:2014sta,Boito:2020xli}). A non-singular
modification of the covariance matrix must then be
employed\footnote{In general, although a $\chi^2$ fit, using the
full covariance matrix, is the most common one, it is by no means
the only possibility. Other choices exist where a different matrix
is used, as long as this matrix is invertible and the final fit
parameter errors and correlations are obtained by propagating the
full set of underlying correlations. In our case, a standard
$\chi^2$ fit is not possible, so we must resort to another
choice, as explained in the main text.} if one wants
to take advantage of the non-trivial theory representation constraints
which are accessible only in FESR analyses of such variable-weight,
variable-$s_0$ sets. Errors and correlations are then fully taken
into account in the error propagation, using the procedure of
Appendix A in Ref.~\cite{Boito:2011qt}. We will refer to fits
employing such non-standard-$\chi^2$ minimizers as ``$Q^2$ fits''
in what follows ({\it c.f.} Ref.~\cite{Boito:2011qt}). An example of
this type is provided by the fits of the DV-strategy analyses of
Refs.~\cite{Boito:2014sta,Boito:2020xli} which consider ranges of
both $w(s)$ and $s_0$, and work with the minimizing function $Q^2$
constructed using the block-diagonal form obtained by omitting
correlations between spectral integrals involving different weights.

The redundancy claim made in Ref.~\cite{Pich:2022tca} concerns the
multi-weight, multi-$s_0$ block-diagonal-$Q^2$ fits of
Refs.~\cite{Boito:2014sta,Boito:2020xli}. Explicitly, the claim is
that the results for $\alpha_s$ and the DV parameters in those fits
are obtained from the fit to the $w_0=1$ FESR, and that adding further
multi-$s_0$ input represented by the $w_2(x)=1-x^2$ and/or
$w_3(x)=(1-x)^2(1+2x)$ and/or $w_4(x)=(1-x^2)^2$ spectral moments to
the $w_0$ input is redundant and leaves the previously determined
$\alpha_s$ and DV parameter values unchanged with the new spectral
integrals serving only to fix the new OPE parameters $C_6$, $C_8$ and
$C_{10}$. As we will show in what follows this claim is incorrect.

Consider the general fit situation in which one has a set of
experimental data points $d_i\ (i=1,...,n)$ (not all necessarily
linearly independent), and corresponding theory representations, $t_i$,
depending of some parameters $\eta_\alpha$. We have in mind here the
case that the $d_k$ are weighted spectral integrals, with the label
$k$ specifying both the weight $w$ and $s_0$ value, the $t_k(\eta_\alpha )$
are the corresponding theory integrals, with the theory parameters
$\eta_\alpha$ including $\alpha_s$, the relevant OPE condensates, and,
if a model for DV contributions is included, the DV parameters. The
theory parameters are obtained by minimizing a function
\beq
\lbl{fit}
Q^2(\eta_\alpha)=\left(\vec{d}-\vec{t}(\eta_\alpha)\right)^T \widetilde{C}^{-1}
\left(\vec{d}-\vec{t}(\eta_\alpha)\right)\ ,
\eeq
where, if the data set is linearly independent, the matrix $\widetilde{C}$
can be chosen to be the data covariance matrix, $C$, and $Q^2$ reduces
to the standard $\chi^2$ function. For the multi-weight, multi-$s_0$ fits
employed in Refs.~\cite{Boito:2014sta,Boito:2020xli}, where the covariance
matrix $C$ of the full set of spectral integrals is singular,
$\widetilde{C}$ was taken to have the block-diagonal form described
above. Other invertible and positive-definite matrices replacing the
full singular covariance matrix could also have been chosen for the
matrix $\widetilde{C}$.

As is well known, the results obtained from such fits ($\chi^2$ or
$Q^2$) will be the same if one changes basis and shifts from the vectors
$\vec{d},\vec{t}$ to a new set, $\vec{d}\, ',\vec{t}\, '$, related to
the original set by a non-singular matrix $O$, with
$\left(\vec{d}-\vec{t}(\eta_\alpha)\right)= O \left(\vec{d}-\vec{t}
(\eta_\alpha)\right)^'$, provided one uses also the new matrix,
$\widetilde{C}^'$, with $(\widetilde{C}^{-1})'=O^T (\widetilde{C}^{-1}) O$.
We now use this freedom to illustrate how using multi-weight,
multi-$s_0$ fits of the type employed in
Refs.~\cite{Boito:2014sta,Boito:2020xli} provides additional,
non-trivial theory representation constraints not accessible for fits
involving only more limited sets of spectral
integrals for which standard $\chi^2$ fits are possible.

Let us consider, for presentational simplicity, the case where the
first weight, $w(s)$ in Eq.~(\ref{cauchy}), is the monomial $s^n$ and
we perform the fit in a window of $s_0$, $s_{N_0}\leq s_0\leq s_{N_0+N}$,
which includes all $s_0\geq s_{N_0}$ possible for the given experimental
binning.\footnote{In our case a range of $s_{N_0}=s_0^{\rm min}$ is
considered and $s_{N_0+N}\approx m_\tau^2$. We check for the existence
of a region of stability of the results with respect to $s_0^{\rm min}$
and use that region to obtain our final results. The result of the
$V$-channel analysis of Ref.~\cite{Boito:2020xli}, for example, are
obtained from a weighted average over results in a seven-$s_0^{\rm min}$
stability region.} With binned data, the FESR integrals in
Eq.~(\ref{cauchy}) become discrete sums. For simplicity, we assume
a binned data set with constant bin width $\Delta$. One then has the
set of data points given~by
\beq
\lbl{data}
\{I_n(s_{N_0}),\ \dots,\ I_n(s_{N_0+N})\}\ ,
\eeq
for the spectral moments
\beq
\lbl{moms}
I_n(s_k)=\sum_{i=1}^k \r(s_i)s_i^n\D\ ,\qquad s_k\in\{s_{N_0},\dots\ ,
s_{N_0+N}\}\ ,
\eeq
with
\beq
\lbl{delta}
s_{k+1}=s_k+\D\ .
\eeq

Choosing a form for the theory representation, one evaluates the
corresponding (parameter-dependent) theory integrals, $t_n(s_k)$, and
forms the vector
\bqry
\lbl{dataminusth}
&&\left(\begin{array}{c}
I_n(s_{N_0})-t_n(s_{N_0})\\
I_n(s_{N_0}+\D)-t_n(s_{N_0}+\D)\\
I_n(s_{N_0}+2\D)-t_n(s_{N_0}+2\D)\\
\vdots\\
I_n(s_{N_0+N})-t_n(s_{N_0+N})
\end{array}\right)\\&&=\left(\begin{array}{c}
I_n(s_{N_0})-t_n(s_{N_0})\\
I_n(s_{N_0})+(s_{N_0}+\D)^n\r(s_{N_0}+\D)\D-t_n(s_{N_0}+\D)\\
I_n(s_{N_0})+(s_{N_0}+\D)^n\r(s_{N_0}+\D)\D
+(s_{N_0}+2\D)^n\r(s_{N_0}+2\D)\D-t_n(s_{N_0}+2\D)\\
\vdots\\
I_n(s_{N_0})+\dots+s_{N_0+N}^n\r(s_{N_0+N})\D-t_n(s_{N_0+N})
\end{array}\right)\ ,\nonumber
\eqry
with
\beq
\lbl{NDelta}
s_{N_0+N}=s_{N_0}+N\D\ .
\eeq
The vector (\ref{dataminusth}) can be written, alternatively, as
\beq
\lbl{basistransf}
\left(\begin{array}{ccccc}
1 & 0 & 0 & 0 & \dots\\
1 &(s_{N_0}+\D)^n\D & 0 & 0 & \dots \\
1 &(s_{N_0}+\D)^n\D &(s_{N_0}+2\D)^n\D & 0 & \dots \\
\vdots &\vdots & \vdots& &\vdots \\
1 &(s_{N_0}+\D)^n\D &\dots & & s_{N_0+N}^n\D\\
\end{array}\right)\left(\begin{array}{c}
I_n(s_{N_0})-t_n(s_{N_0})\\
\r(s_{N_0}+\D)-\frac{t_n(s_{N_0}+\D)-t_n(s_{N_0})}{(s_{N_0}+\D)^n\D}\\
\r(s_{N_0}+2\D)-\frac{t_n(s_{N_0}+2\D)-t_n(s_{N_0}+\D)}{(s_{N_0}+2\D)^n\D}\\
\vdots\\
\r(s_{N_0+N})-\frac{t_n(s_{N_0+N})-t_n(s_{N_0+N}-\D)}{s_{N_0+N}^n\D}
\end{array}\right)\ ,
\eeq
in terms of an $(N+1)\times (N+1)$ matrix, $M$, times a column
vector which, apart from the contribution from the theory representation
at the $s$ values employed in the fit, $t(s_k)$, contains the spectral
integral at the lowest $s_0$, $I_n(s_{N_0})$, and the values of the
spectral function, $\rho (s)$, at all other $s=s_k$ in the fit window.

To understand how, contrary to what is claimed in
Ref.~\cite{Pich:2022tca}, using multiple weights {\it and} multiple $s_0$
generates additional constraints on the theory representations,
consider now fitting a second such data set, $I_m$, involving the
new weight $s^m$, with $n\ne m$, and again carry out the basis
transformation\seneq{basistransf}. The sets
$(\r(s_{N_0}+\D),\dots,\r(s_{N_0+N}))$ in the two vectors are of
course 100\% correlated, and a standard $\chi^2$ fit to the
full two-weight spectral integral set is thus not possible.
However, it is perfectly possible to carry out a combined
block-diagonal fit in which the matrix $\tilde{C}$ is constructed
from the full covariance matrix $C$ by retaining all correlations
between spectral integrals involving the same weight, but removing
those where the two weights are different. This will only produce
a good fit if ($k=1,\ \dots N$)
\beq
\lbl{cond}
\frac{t_n(s_{N_0}+k\D)-t_n(s_{N_0}+(k-1)\D)}
{(s_{N_0}+k\D)^n}
=\frac{t_m(s_{N_0}+k\D)-t_m(s_{N_0}+(k-1)\D)}{(s_{N_0}+k\D)^m}
\ .
\eeq
This provides a set of non-trivial constraints on the theory
representations $t_n$ and $t_m$. If the theory representations $t_n$
and $t_m$ were exact, they would necessarily satisfy these constraints.
This will not, however, be true in general if the theory representation
is not exact, whether due to approximations or the use of a physically
unreliable form. The constraints in Eq.~(\ref{cond}) are operational
only in analyses involving both at least two weights and multiple $s_0$,
such as those provided by the multi-weight, multi-$s_0$,
block-diagonal-$Q^2$ fit analyses of Refs.~\cite{Boito:2014sta,Boito:2020xli}.

In Ref.~\cite{Pich:2022tca} a similar two-weight basis transformation
was carried out, but {\it omitting} the terms involving the theory
representations, $t(s_k)$, and focusing solely on the spectral integral sides
of the associated FESRs. From this, the authors demonstrated that such a
combined two-weight, multi-$s_0$ spectral integral set is linearly dependent,
with only one of the second-weight spectral integrals linearly independent of
the first-weight spectral integral set. At this point, however, they, albeit
implicitly, chose to consider, not the block-diagonal-$Q^2$ fits to the full
spectral integral set employed in Refs.~\cite{Boito:2014sta,Boito:2020xli},
but rather $\chi^2$ fits to the reduced, now linearly independent, set
obtained by throwing away all but one of the second set of spectral integrals.
With this restriction, and hence the second-weight theory representation
at only a single $s_0$, the second-weight differences on the RHS of
Eq.~(\ref{cond}) can no longer be formed, and access to the non-trivial
theory constraints present when the second-weight FESR is considered at
multiple $s_0$ is lost. The redundancy claim made in Ref.~\cite{Pich:2022tca}
would be correct for a $\chi^2$ fit to the restricted spectral integral set,
but this is irrelevant since the multi-weight, multi-$s_0$ block-diagonal
$Q^2$ fits employed in Refs.~\cite{Boito:2014sta,Boito:2020xli}, are not of
this type. The basic mathematical error made in Ref.~\cite{Pich:2022tca} is
to assume (without proof, and incorrectly, as it turns out) that the
redundancy result which would be valid for a $\chi^2$ fit to the reduced
spectral integral set also applies to the fundamentally different
block-diagonal $Q^2$ fits to the much larger set of spectral
integrals employed in Refs.~\cite{Boito:2014sta,Boito:2020xli}.

We emphasize that the block-diagonal-$Q^2$ fits to the full
(multi-weight, multi-$s_0$) set of spectral integrals employed in
Refs.~\cite{Boito:2014sta,Boito:2020xli} are chosen for a reason,
namely that they provide access precisely to non-trivial additional
$s_0$-dependent constraints on the theory representation, analogous
to those of Eq.~(\ref{cond}) above. These constraints would not be
accessible in a fit to the reduced subset of those spectral integrals
for which a standard $\chi^2$ fit is possible. The different $s_0$ scalings
of the different terms in the theory representations play an important
role in such tests.

As a specific example of the points made in the discussion above,
consider evaluating the theory side of the $w_2$ FESR using as input the
output for $\alpha_s$ and the DV parameters from the $w_0$ FESR fit. If
the DV model employed provides a perfect representation of DV effects
in QCD in the fit region, the nominally $D=6$ OPE $w_2$ FESR residual
obtained by subtracting from the $w_2$-weighted spectral integrals the sum
of $w_2$ $D=0$ perturbative and DV contributions evaluated using $w_0$ fit
results as input would be proportional to the $D=6$ OPE condensate $C_6$
and scale with $s_0$ as $1/s_0^3$. If, in contrast, the DV model does not
provide a good representation, the residual will not scale properly with
$s_0$ and, if one nonetheless tries to make it do so by using the QCD form
for the representation of the $D=6$ OPE contribution on the theory side of
the $w_2$ FESR, the $w_2$-only $\chi^2$ part of the block-diagonal $Q^2$
will be non-negligible. That this is so is because, in this case,
no single choice of $C_6$ will allow the theory and spectral integral
sides of the $w_2$ FESR to match for all $s_0$ in the fit window. The
combined two-weight, block-diagonal $Q^2$ fit will deal with this by
adjusting the $\alpha_s$ and DV parameter values obtained from the
$w_0$ fit, somewhat increasing the $w_0$ $\chi^2$ contribution to the
combined $Q^2$, while at the same time decreasing the corresponding
$w_2$ $\chi^2$ contribution. The minimized $Q^2$ of the block-diagonal
fit will have a $w_0$ $\chi^2$ component somewhat larger than that
obtained from the $\chi^2$ fit to the $w_0$ FESR alone. In contrast,
were the claim made in Ref.~\cite{Pich:2022tca} to be true, regardless
of whether the DV model provides a good or poor representation of
physical DV effects, the two-weight block-diagonal $Q^2$ fit would
return values for $\alpha_s$, the DV parameters, and hence of the
$\chi^2$ of the $w_0$ $\chi^2$ part of $Q^2$, {\it identical}
to those of the single-weight $w_0$ fit. The results of such
block-diagonal $Q^2$ fits explicitly bear out that this is not the
case.

To explore further the extent to which the multi-weight, multi-$s_0$ fits
of Refs.~\cite{Boito:2014sta,Boito:2020xli} provide a practically useful
implementation of the constraints analogous to those of Eq.~(\ref{cond})
above, it is useful to look at a specific example. We consider, for this
purpose, the combined two-weight, $w_0$ and $w_2$ $V$-channel analysis of
Ref.~\cite{Boito:2020xli}. Beginning with the $w_0$ FESR, $\alpha_s$
and the $V$-channel DV parameters can be obtained from a standard
$\chi^2$, multi-$s_0$ fit to the collection of $w_0$ spectral integrals.
Expanding the fit to include the $w_2$ spectral integrals, over the same
$s_0$ range, introduces one new theory fit parameter. In QCD, we expect
this to be the effective $D=6$ condensate, $C_{6,V}$, and the new NP
contributions to the theory sides of the $w_2$ FESRs to have the form
$-C_{6,V}/s_0^3$. According to the claim of Ref.~\cite{Pich:2022tca}, the
analysis including this second set of spectral integrals should be redundant:
the results for $\alpha_s$ and the DV parameters should be {\it identical} to
those obtained from the $w_0$ FESR fit and the $w_2$ spectral integral input
should serve only to fix the single new parameter $C_{6,V}$, in a redundant
manner, regardless of the $s_0$-dependence assumed for the NP condensate
contribution. As noted above, the two-weight fit results for $\alpha_s$ and
the DV parameters in Ref.~\cite{Boito:2020xli}, though very close to those
obtained from the $w_0$ FESR, are not identical. These (non-identical)
results, extracted from Ref.~\cite{Boito:2020xli}, are shown in
Table \ref{tab:noredundant}.\footnote{Table \ref{tab:prsoptwtredundancy}
below shows the results obtained from a multi-weight, {\it single}-$s_0$
tOPE fit, where the redundancy produced by the single-$s_0$ nature of the
analysis is, in contrast, immediately evident.} This already
establishes that the assumption of Ref.~\cite{Pich:2022tca} that the
redundancy result which would be true for a $\chi^2$ fit to the reduced
set of spectral integrals implicitly considered in Ref.~\cite{Pich:2022tca}
is also valid for the multi-weight, multi-$s_0$, block-diagonal-$Q^2$ fits
employed in Refs.~\cite{Boito:2014sta,Boito:2020xli} cannot be correct.

\begin{table}[h]
{\footnotesize
\begin{center}
\begin{tabular}{|l|c|l|l|c|c|c|c|}
\hline
weight &fit quality & $~~\a_s(m_\t^2)$ & $~~~~\,\d_V$ & $\g_V$ & $\a_V$
& $\b_V$ & $10^2\, C_{6V}$\\
\hline
$w_0$ & $\chi^2=$9.97 & 0.3056(64) & 3.61(30) & 0.52(18)
& $-$1.62(51) & 3.95(26) & --- \\
$w_0 \& w_2$ & $Q^2=$22.7 & 0.3073(69) & 3.50(31)
& 0.58(19) & $-$1.57(55) & 3.92(29) & $-$0.62(13) \\
\hline
\end{tabular}
\end{center}
\caption{{Comparison of the results obtained from the $s_0^{\rm min}=1.5863$
GeV$^2$ single-weight $w_0$ $\chi^2$ fit and combined two-weight $w_0$ and
$w_2$ block-diagonal $Q^2$ fit using the $V$ spectral function of
Ref.~\cite{Boito:2020xli}. $C_{6,V}$ is in units of GeV$^6$ and $\beta_V$
and $\gamma_V$ in units of GeV$^{-2}$. The results are clearly not
identical, demonstrating the non-redundancy of the expanded
two-weight, block-diagonal $Q^2$ fit.}}
\label{tab:noredundant}}
\end{table}

To drive home this point, and illustate the practical utility
of the additional theory constraints produced when multi-weight, multi-$s_0$
block-diagonal $Q^2$ fits are used, it helps to see what happens
if we consider a modified version of this two-weight analysis in which
the theory representation of at least one of the two FESRs is deliberately
chosen to be unphysical, i.e., to differ from QCD expectations.
We have thus carried out an exercise in which we compare the results
obtained for $\alpha_s$ in Ref.~\cite{Boito:2020xli} from
fits to the $V$-channel $w_0$ spectral integrals in windows
$s_0^{\rm min} \leq s_0 \leq 3.06\ \mathrm{GeV}^2$, with
$s_0^{\rm min}\geq 1.55$ GeV$^2$, to those obtained
from two simultaneous fits to the $w_0$ and $w_2$ moments in the same $s_0$
window, using, in one case, a form for the integrated NP contribution to the
$w_2$ FESR, $-C_{6,V}/s_0^{3}$, having the expected QCD scaling with
respect to $s_0$ and, in the other, a \emph{non-QCD} form,
$C^\prime /s_0^{5}$, having a different scaling.\footnote{This modified,
non-QCD form would, for example, result if one replaced the $C_{6,V}/z^3$
term expected in QCD in the vector, isovector polarization function
representation consisting of the sum of perturbative, DV and OPE condensate
contributions with the modified (and manifestly non-QCD) term
\begin{equation}
\Delta\Pi_{\rm non\mbox{-}QCD}(z)=
\frac{C^\prime}{z^5}\left(\log\left(-\frac{z}{\mu^2}\right)
-2\log^2\left(-\frac{z}{\mu^2}\right)\right)\ .
\end{equation}
It is easily verified that the FOPT ($\mu^2=s_0$) version of the $w_0$- and
$w_2$-weighted FESR integrals of $\Delta\Pi_{\rm non\mbox{-}QCD}(z)$ are,
respectively, zero and $C^\prime/s_0^5$. The resulting non-QCD
polarization function representation thus produces the same contribution
to the theory side of the $w_0$ FESR as does the expected QCD form, while
producing the altered NP contribution to the theory side of the $w_2$
FESR considered in this example.}
According to the argument of Ref.~\cite{Pich:2022tca}, since the new weight
$w_2$ introduces the single new free parameter ($C_{6,V}$ or $C^\prime$,
depending on which theory representation is chosen for the $w_2$ FESR),
the combined fits should simply determine the new parameter ($C_{6,V}$
or $C^\prime$), but otherwise be completely redundant,
leaving the result for $\alpha_s$ obtained from the $w_0$ fit
unchanged, regardless of whether the expected QCD or non-QCD scaling
with respect to $s_0$ is used.

\begin{figure}[!t]
\centering
\includegraphics[width=0.6\textwidth]{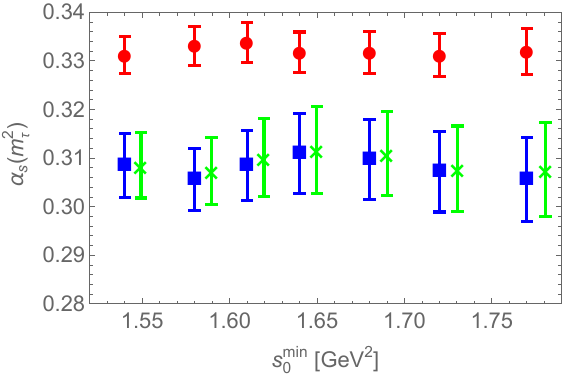}
\caption{Comparison of results for $\alpha_s(m_\tau^2)$ obtained from a fit
to the $V$-channel $w_0=1$ FESR in a window in $s_0$ (blue points)
to the results of simultaneous fits to the $w_0$ and $w_2(x)=1-x^2$
FESRs with either the expected QCD $s_0^{-3}$ scaling of the non-DV, NP
contributions to the $w_2$ FESR (green points) or a non-QCD $s_0^{-5}$
scaling of those contributions (red points). For presentational clarity,
a small horizontal offset has been applied to the green points to separate
them visually from the blue points.}
\label{fig:C6VTautology}
\end{figure}

Figure~\ref{fig:C6VTautology} shows the results of this exercise, as a
function of $s_0^{\rm min}$. The blue squares show the results for
$\alpha_s$ obtained from the single-weight, $w_0$, fit, the green
crosses those obtained from the combined $w_0$ and $w_2$ fit with
the expected QCD NP scaling, and the red circles those
obtained from the combined fit with the non-QCD NP scaling.
One sees that the results for $\alpha_s$ from the single-weight
($w_0$) fit and the combined ($w_0$ and $w_2$) weight fit with the
non-QCD NP scaling are far from equal, with a systematic
upward shift immediately evident in all of the red points.
When instead the expected QCD scaling is used, the results of
the single-weight and combined-weight fits (shown by the blue and green
points) are in excellent agreement. The first of these observations
establishes the fact that, contrary to the claim of Ref.~\cite{Pich:2022tca},
the implementation of the multi-$s_0$ constraints produced by adding
the $w_2$ FESR to the fit, using the block-diagonal-$Q^2$ fit form,
are highly non-trivial, allowing, for example, in this case, the non-QCD
NP scaling to be ruled out on self-consistency grounds alone.
In contrast, the agreement between the results of the single-weight
($w_0$) fit and combined two-weight ($w_0$ and $w_2$),
block-diagonal-$Q^2$ fit when the expected QCD NP scaling is used
in the new $w_2$ FESR represents a non-trivial self-consistency
test.\footnote{We also comment that the optimized $Q^2$ (fit quality)
values for the fits employing the non-QCD form, $C^\prime/s_0^{5}$,
are, as expected, much larger than those for the fits employing the
expected QCD form, $-C_{6,V}/s_0^{3}$, further confirming that the
data prefer the expected QCD behavior of the NP contribution.}

A second claim made in Ref.~\cite{Pich:2022tca}, based on a
misinterpretation of the expression~(\ref{basistransf}) for the weight
$w_0(x)=1$ ($n=0$), is that a fit to the data points given by
$\{I^{w_0}(s_{N_0}), \rho(s_k)\}$ results in a value of $\alpha_s$
that is determined by the spectral integral at $s_0^{\rm min}$, {\it i.e.},
$I^{w_0}(s_{N_0})$, with the DV parameters determined by fitting to the
spectral function $\rho(s_k)$ at all of the higher $s_0$ values, perhaps
casting doubt on the validity of perturbation theory since typical
values of $s_{N_0}=s_0^{\rm min}$, ranging from $1.55$~GeV$^2$ to
$1.9$~GeV$^2$~\cite{Boito:2020xli}, are relatively low. Note, however,
that, instead of starting at the lowest value of $s_0$, $s_0=s_{N_0}$ in
the analogue of Eq. (\ref{dataminusth}) and working up to
$s_0^{\rm max}=s_{N_0+N}$, one could instead consider starting at the
highest $s_0$ and working down, ending up with a different alternate
basis consisting of the $w_0$ moment at the highest $s_0$,
$I^{w_0}(s_{N_0+N})$, together with the spectral function values
$\rho (s_k)$ at all lower $s$. Following the reasoning of
Ref.~\cite{Pich:2022tca}, one would then have
to conclude that $\alpha_s$ was fixed by the spectral integral at the
highest scale $s_{N_0+N}\simeq m_\tau^2$ and the DV parameters by
fitting to the spectral function at the lower scales, which is precisely
the opposite conclusion to that of Ref.~\cite{Pich:2022tca}. It is thus
clear that both this alternate argument and the analogous original
argument in Ref.~\cite{Pich:2022tca} are incorrect. The correct
statement is that all of $\alpha_s$, the DV parameters and the OPE
condensates, when they contribute, are fixed collectively in the window
of $s_0$ values employed in the fit and that the consistency of results
across the different fits constitutes a valid, and non-trivial,
self-consistency check. Furthermore, the latest $\alpha_s$ determination
from the DV strategy~\cite{Boito:2020xli} is, in fact, the result of a
weighted average between fits with different $s_0^{\rm min}$, taking
into account the correlations between $\alpha_s$ values from different fits.

\section{\boldmath Redundancy issues for single-\texorpdfstring{$s_0$}{s0}
truncated OPE (tOPE) fit analyses}\label{tOPEredundancy}
In the previous section we showed, by means of the explicit $V$-channel
example involving the $w_0$ and $w_2$ FESRs, that adding a second (in
this case $w_2$) multi-$s_0$ FESR to an original (in this case $w_0$)
multi-$s_0$ FESR fit, provides additional non-trivial
constraints on the results obtained in the original single-weight fit.
We now turn to the question of what happens in the case of multi-weight,
single-$s_0$ FESR fits, of the type common in tOPE analyses.

A key observation in this regard is the following.{\footnote{A brief discussion
of this point also appears in the Appendix of Ref.~\cite{Pich:2022tca}.}}
Consider starting with a set of data $\{ d_k\}$, $k=1,\cdots ,N$, with
invertible covariance matrix $C$, and a set of corresponding
theoretical representations $\{ t_k\}$, depending on a set of $M<N$
theory parameters $\{ \eta_k\}$ which one fits with a standard $\chi^2$
minimization. Now expand the data set to include one more
data point, $d_{N+1}$, for which the corresponding theoretical
representation, $t_{N+1}$, depends on a single new
parameter $\eta_{M+1}$ and the resulting expanded covariance
matrix is still invertible{\footnote{As for example happens
if the new data point is not a linear combination of the previous
data points.}} so that a standard $\chi^2$ fit can be performed to the
expanded data set. The resulting extended fit then {\it is}
``redundant'' in the sense of Ref.~\cite{Pich:2022tca}: one simply has
one new data point with one new parameter to fix, and the results of
the extended fit for both the $\chi^2$ and the previously fitted
parameters $\eta_1,\cdots ,\, \eta_M$ are {\it identical} to those
obtained from the previous fit, with the single new
parameter $\eta_{M+1}$ serving only to ensure that the new theory
representation $t_{M+1}$ exactly matches the single new data point
$d_{N+1}$. Adding the new data point has produced no new information on the
previously fitted parameters and the single new parameter, $\eta_{M+1}$, is
tautologically determined and hence physically unconstrained. This
result, which is trivially obvious in the absence of correlations
between $d_{N+1}$ and the other data points, is also true when these
correlations are non-zero. For completeness, a proof of this fact is
provided in Appendix~\ref{app:redundancyproof}.

With this observation in hand, it is illuminating to revisit some of the
multi-weight, single-$s_0$ tOPE analyses reported in the literature. We
will show that the redundancy discussed above is a generic feature of
such analyses and necessitates a re-interpretation of the results they
produce. We focus, for definiteness, on the three different
$s_0=2.8\ {\rm GeV}^2$, five-weight, isovector $V+A$ channel analyses
discussed in Refs.~\cite{Pich:2016bdg,Pich:2022tca} involving weights all of
which are either doubly or triply pinched. These employ either (i) the
$km=00,\, 10,\, 11,\, 12$ and $13$ versions of the classic $km$ spectral
weights \cite{LeDiberder:1992zhd},
\begin{equation}
\label{classic}
w_{km}(x)=(1-x)^2(1+2x)(1-x)^k x^m\ ,\quad km=00,\ 10,\ 11,\ 12,\ 13\ ,
\end{equation}
(ii) the
$km=00,\, 10,\, 11,\, 12$ and $13$ versions of the modified
$km$ spectral weights,
\begin{equation}
\label{modified}
\hat{w}_{km}(x)=(1-x)^{2+k} x^m\ ,\quad km=00,\ 10,\ 11,\ 12,\ 13\ ,
\end{equation}
or (iii) the $m=1,\cdots ,5$ versions of the ``$(2m)$ optimal weights'',
\begin{equation}
\label{optimal}
w^{(2m)}(x)=1-(m+2)x^{m+1}+(m+1)x^{m+2}\ ,\quad m=1,\dots,5\ ,
\end{equation}
introduced in Ref.~\cite{Pich:2016bdg}. The classic $km$ spectral weight
choice is the one employed, with $s_0=m_\tau^2$, in the early
ALEPH~\cite{ALEPH:1998rgl} and OPAL~\cite{OPAL:1998rrm} analyses.
The lowest degree member of this set, $w_{00}(x)$, referred to as
the ``$\tau$ kinematic weight'' above, is the generalization to
$s_0\ne m_\tau^2$ of the kinematic weight $w_T(s;m_\tau^2)$ appearing in
Eq.~(\ref{taukinspectral}). The additional $km$ weights,
$\{ w_{1m}(s/m_\tau^2),\, m=0,\cdots ,\, 3\}$, were originally
introduced with the goal of allowing the $D=6$ and $8$ condensates
entering the $w_{00}$ FESR to be fitted, and hence $\alpha_s$
to be determined.

A statement sometimes found in the literature is that the
resulting five-weight, single-$s_0=m_\tau^2$ fit obtains $\alpha_s$
largely from the $w_{00}$ FESR and the condensates largely from the
remaining higher-degree-weight FESRs.\footnote{See, {\it e.g.},
Section 4 of Ref.~\cite{Pich:2016bdg} where the authors state that
`The strong coupling value turns out to be very stable in all fits
because it is basically determined by the lowest moment (...).'}
As we will show, this is not, in fact, the case. The redundancy
observation above allows us to see that the result for $\alpha_s$
is instead obtained from a combined fit to the FESRs of two
high-degree linear combinations of the $w_{km}$ in which all
NP effects are neglected and the theory representation consists
\emph{solely} of the perturbative contributions. Exactly analogous
statements are true for the analyses based on the modified-weight
$\hat{w}_{km}$ and optimal-weight $w^{(2m)}$ sets. To see this,
instead of jumping directly to the full five-weight fit for a given
class, it will be illuminating to arrive at that combined fit by
adding the single-weight FESRs one step at a time. This allows one to
see, as detailed below, that the combinations of high-degree weights
whose FESRs actually determine $\alpha_s$ in these analyses are, in all
three cases, linear combinations of the $w^{(23)}$, $w^{(24)}$ and
$w^{(25)}$ optimal weights. The associated condensate determinations
are, moreover, entirely redundant, and play no role in the corresponding
$\alpha_s$ determinations. This observation is relevant to assessing
the argument presented in Refs.~\cite{Pich:2016bdg,Pich:2022tca} that the
agreement between the results for $\alpha_s$ obtained from these
different analyses are non-trivial because of the ``very
different sensitivities to the vacuum condensates''.

We turn now to explicit proofs of the claims just made.
The discussion which follows is broken down into four subsections.
Sec.~\ref{kmspecwtsub} provides the relevant details for
the classic $km$ spectral weight moments. The redundancy in this case is
exposed through a change of basis for the moments, and is shown
to strongly limit the internal self-consistency tests available to this
analysis. The linear combinations of the $(23)$, $(24)$ and $(25)$ optimal
weights whose FESRs determine $\alpha_s$ in this analysis are also
explicitly identified. Secs.~\ref{whatkmwtsub} and \ref{optwtsub} provide
analogous, more streamlined versions of the discussion for the modified
$\hat{w}_{km}$ and $w^{(2m)}$ optimal weight analysis cases, where the
FESRs responsible for determining $\alpha_s$ are again found to involve
linear combinations of the $(23)$, $(24)$ and $(25)$ optimal weights with
only perturbative contributions included on the theory sides.
An exploration of the potential theoretical systematics induced by the
truncation in dimension of the tOPE approach can thus be carried
out using the, what turns out to be the somewhat larger, set of
multi-weight, single-$s_0$ self-consistency tests available in the
optimal weight case. The results of these explorations in the isovector
$V+A$ channel, carried out using ALEPH spectral data, show some clear
tensions. These are reported in Sec.~\ref{optwtsub}, which also contains
comments on the even stronger tensions found in the analogous isovector
$V$-channel analysis. Finally, in Sec.~\ref{condredundancysub},
we point out the existence of a sizable theoretical systematic uncertainty,
generic to such single-$s_0$ tOPE analyses, affecting the (redundant)
determinations of the NP OPE condensates.

\subsection[``\texorpdfstring{$km$}{km}'' spectral weights]{\boldmath
The classic ``\texorpdfstring{$km$}{km}'' spectral weight case}
\label{kmspecwtsub}

In this subsection, we consider the single-$s_0$ tOPE analysis using the
``$km$ spectral weights,'' defined in Eq.~(\ref{classic}).
As noted above, the four weights, $w_{1m}(s/m_\tau^2)$, were originally
introduced with the goal of allowing the $D=6$ and $8$ condensates
entering the $w_{00}$ FESR to be fitted, and hence $\alpha_s$
to be determined. However, even assuming (with $w_{00}$ doubly
pinched and the remaining $w_{km}$ triply pinched) that DV contributions
are negligible, this goal is not achievable without further assumptions
since the theory sides of the five $w_{km}$ FESRs involve, in addition
to $\alpha_s$, unsuppressed non-perturbative (NP) contributions
\begin{eqnarray}
\left[ I_{\rm th}^{w_{00}}(s_0)\right]_{\rm NP}&=&-3\, {\frac{C_6}{s_0^3}}
-2\, {\frac{C_8}{s_0^4}}\ ,\nonumber\\
\left[ I_{\rm th}^{w_{10}}(s_0)\right]_{\rm NP}&=&{\frac{C_4}{s_0^2}}
-3\, {\frac{C_6}{s_0^3}}-5\, {\frac{C_8}{s_0^4}}
-2\, {\frac{C_{10}}{s_0^5}}\ ,\nonumber\\
\left[ I_{\rm th}^{w_{11}}(s_0)\right]_{\rm NP}&=&-\, {\frac{C_4}{s_0^2}}
-\, {\frac{C_6}{s_0^3}}+3\, {\frac{C_8}{s_0^4}}+5\, {\frac{C_{10}}{s_0^5}}
+2\, {\frac{C_{12}}{s_0^6}}\ ,\nonumber\\
\left[ I_{\rm th}^{w_{12}}(s_0)\right]_{\rm NP}&=&{\frac{C_6}{s_0^3}}
+\, {\frac{C_8}{s_0^4}}
-3\, {\frac{C_{10}}{s_0^5}}-5\, {\frac{C_{12}}{s_0^6}}
-2\, {\frac{C_{14}}{s_0^7}}\ ,\nonumber\\
\left[ I_{\rm th}^{w_{13}}(s_0)\right]_{\rm NP}&=&
-\, {\frac{C_8}{s_0^4}}
-\, {\frac{C_{10}}{s_0^5}}+3\, {\frac{C_{12}}{s_0^6}}
+5\, {\frac{C_{14}}{s_0^7}}+2\, {\frac{C_{16}}{s_0^8}}
\ ,
\label{wkmspecwtnonpertcontribs}\end{eqnarray}
depending on seven additional parameters, the unknown condensates
$C_D$, with $D=4,\, 6,\, 8,\, 10,\, 12,\, 14,\, 16$. With only five
(single-$s_0$) spectral integrals as input, only four of the eight OPE
parameters in principle present can be fitted. The tOPE approach
deals with this problem by assuming contributions proportional to the four
highest-dimension condensates, $C_{D>8}$, can be safely neglected.
We now show, first, that the NP condensates retained in the analysis,
$C_{D\leq 8}$, are determined redundantly and play no role whatsoever
in the determination of $\alpha_s$, and, second, that $\alpha_s$ is
actually obtained from a combined fit to the FESRs of two high-degree
linear combinations of the $w_{km}$ having only perturbative contributions
on their theory sides. To see this, consider shifting from the initial
$\{ w_{km}\}$ basis to the alternate basis,
$\{ \hat{w}_k(s),\, k=1,\cdots ,\, 5\}$, with
\begin{eqnarray}
\hat{w}_1(x)&&=1-{\frac{15}{2}}x^4+12x^5-{\frac{17}{2}}x^6+3x^7
={\frac{3}{2}}w^{(23)}(x) - w^{(24)}(x) + {\frac{1}{2}}w^{(25)}(x)\, ,
\nonumber\\
\hat{w}_2(s)&&=1-9x^4+12x^5-4x^6
={\frac{9}{5}}w^{(23)}(x) - {\frac{4}{5}}w^{(24)}(x)\, ,\nonumber\\
\hat{w}_3(x)&&=1+2x^3-9x^4+6x^5
=\, -{\frac{1}{2}}w^{(22)}(x) + {\frac{3}{2}}w^{(23)}(x)\, ,\nonumber\\
\hat{w}_4(s)&&=1-3x^2+2x^3
=w^{(21)}(x)\ ,\nonumber\\
\hat{w}_5(x)&&=1+{\frac{2}{3}}x-{\frac{23}{3}}x^4+6x^5\nonumber\\
&&=\, -{\frac{1}{3}}w^{(20)}(x) - {\frac{1}{9}}w^{(21)}(x)
- {\frac{1}{18}}w^{(22)}(x) + {\frac{3}{2}}w^{(23)}(x)\, .
\label{altspecwtbasis}\end{eqnarray}
The $w^{(2m)}(x)$ appearing in the second line of each of the
expressions above are the $(2m)$ optimal weights.
The $\{ \hat{w}_n(x)\}$ basis is related to the $\{ w_{km}(x)\}$ basis by
\begin{eqnarray}
w_{00}(x)=&&\hat{w}_4(x)\ ,\nonumber\\
w_{10}(x)=&&{\frac{3}{2}}\hat{w}_3(x)+\hat{w}_4(x)
-{\frac{3}{2}}\hat{w}_5(x)\ ,\nonumber\\
w_{11}(x)=&&\, -{\frac{11}{6}}\hat{w}_3(x)+{\frac{1}{3}}\hat{w}_4(x)
+{\frac{3}{2}}\hat{w}_5(x)\ ,\nonumber\\
w_{12}(x)=&&{\frac{1}{2}}\hat{w}_2(x)-{\frac{1}{6}}\hat{w}_3(x)
-{\frac{1}{3}}\hat{w}_4(x)\ ,\nonumber\\
w_{13}(x)=&&\, -{\frac{2}{3}}\hat{w}_1(x)+{\frac{1}{6}}\hat{w}_2(x)
+{\frac{1}{2}}\hat{w}_3(x)\ .
\label{altkmspecwtbasis}\end{eqnarray}

Switching to the alternate basis, one sees that, with the tOPE
assumptions, (i) the theory sides of the $\hat{w}_1$ and $\hat{w}_2$
FESRs contain only $D=0$ OPE contributions and hence depend only on
$\alpha_s$; (ii) adding the $\hat{w}_3$ FESR introduces the single
new parameter $C_8$; (iii) further adding the $\hat{w}_4$ FESR
introduces the single new parameter $C_6$ and, finally, (iv) further
adding the $\hat{w}_5$ FESR introduces the last new parameter, $C_4$.

From the redundancy argument above, it follows that $\alpha_s$ is, in
fact, obtained from a combined fit to the FESRs of the two highest degree
weights, $\hat{w}_1$ and $\hat{w}_2$,{\footnote{Or any two linearly
independent combinations thereof.}} while the nominally $D=4$, $6$ and
$8$ condensates are obtained redundantly, their values serving simply
to make the theory and spectral integral sides of the $\hat{w}_3$,
$\hat{w}_4$ and $\hat{w}_5$ FESRs match exactly, regardless of the
reliability or unreliability of the assumed truncated OPE form. In
other words, even if the $s_0$ scalings used for the OPE terms in
the $\hat{w}_3$, $\hat{w}_4$ and $\hat{w}_5$ FESRs were totally
wrong, the value of $\alpha_s$ would still be the same as determined
from the tOPE versions of the $\hat{w}_1$ and $\hat{w}_2$ FESRs,
which contain no NP OPE contributions whatsoever. The tautological
nature of the nominal $C_4$, $C_6$ and $C_8$ determinations, moreover,
ensures that the analysis provides no evidence either for or against
the assumption that the corresponding nominally $D=4$, $6$ and $8$
FESR contributions are actually those of QCD, uncontaminated by
higher dimension effects or logarithmic corrections, and hence
no evidence either for or against the reliability of the assumed tOPE
truncation in dimension.

To expand on this latter point, suppose that residual $D>8$
OPE contributions are not completely negligible. The
$\alpha_s$ required by the tOPE version of the $\hat{w}_1$
FESR will then be contaminated by a combination of omitted
$D=10$, $12$, $14$ and $16$ NP contributions and that
required by the tOPE version of the $\hat{w}_2$ FESR by
a differently weighted combination of omitted $D=10$,
$12$ and $14$ contributions. The result for $\alpha_s$
from the combined $\hat{w}_1$ and $\hat{w}_2$ FESR fit
will be similarly contaminated. When one adds the $\hat{w}_3$
FESR to the fit, the new fit leaves this contaminated $\alpha_s$
determination untouched and fixes the nominally $D=8$ parameter
$C_{8}$ by adjusting it to make the theory side of
the $\hat{w}_3$ FESR, with $D=0$ contribution evaluated
using the contaminated $\alpha_s$ result as input, exactly
match the spectral integral side. The nominally $D=8$
quantity $C_{8}$ thus obtained is therefore also contaminated
by omitted $D>8$ NP contributions. The nominally $D=6$
quantity $C_6$ obtained by the exact matching of theory
and spectral integral sides of the $\hat{w}_4$ FESR and
nominally $D=4$ quantity $C_4$ obtained by the exact matching
of theory and spectral integral sides of the $\hat{w}_5$
FESR are similarly contaminated. The $D$-dependent ($s_0^{-D/2}$)
scaling of integrated OPE contributions of different dimensions, $D$,
for weights polynomial in $x=s/s_0$ aids in avoiding such higher
dimension contamination for multi-$s_0$ fits. This protection is,
of course, not available for single-$s_0$ fits.

The only self-consistency constraint at play in the conventional $km$
spectral weight analysis is that provided by the underlying
two-weight determination of $\alpha_s$, where the $\hat{w}_1$
and $\hat{w}_2$ FESRs involve differently weighted combinations
of any omitted, but in principle present, higher-dimension
condensate contributions. One thus has access to limited
checks of the self-consistency of the tOPE assumption by
considering, for example, the compatibility of the results
obtained from the $\hat{w}_1$ and $\hat{w}_2$ FESRs separately,
or between those obtained from the $\hat{w}_1$ FESR and the
combined fit to the $\hat{w}_1$ and $\hat{w}_2$ FESRs. If the
difference between such a pair of determinations is not
compatible with zero within its properly calculated correlated
error, this would signal the presence of an additional systematic
uncertainty in the analysis, resulting from a breakdown of the tOPE
assumption.

\subsection[Modified ``\texorpdfstring{$km$}{km}'' spectral weights]
{\boldmath The modified $\hat{w}_{km}$ ``\texorpdfstring{$km$}{km}''
spectral weight case}
\label{whatkmwtsub}

For the next of the single-$s_0$ analysis discussed in
Refs.~\cite{Pich:2016bdg,Pich:2022tca}, based on the five modified $km$
spectral weights, $\hat{w}_{km}$ defined in
Eq.~(\ref{modified}),{\footnote{Note that the $\tau$
kinematic weight, which is a linear
combination of $\hat{w}_{00}(x)$ and $\hat{w}_{10}(x)$,
is included in the analysis.}} the theory sides of the associated FESRs
depend, in addition to $\alpha_s$, on the six NP condensates $C_4$,
to $C_{14}$. The tOPE reduction from seven to four fit parameters is
accomplished by removing contributions proportional to $C_{10,12,14}$,
from the theory sides of the $\hat{w}_{11}$, $\hat{w}_{12}$ and
$\hat{w}_{13}$ FESRs.

An alternate choice of basis in this case is the optimal weight set
$\{ w^{(2m)}(x)=1-(m+2)x^{m+1}+(m+1)x^{m+2},\, m=0,\cdots ,\, 4\}$.{\footnote{For completeness, the alternate and original bases in this case are
related by
$\hat{w}_{00}(x)=w^{(20)}(x)$,
$\hat{w}_{10}(x)={\frac{3}{2}}w^{(20)}(x)-{\frac{1}{2}}w^{(21)}(x)$,
$\hat{w}_{11}(x)= -{\frac{1}{2}}w^{(20)}(x)+{\frac{5}{6}}w^{(21)}(x)
-{\frac{1}{3}}w^{(22)}(x)$,
$\hat{w}_{12}(x)= -{\frac{1}{3}}w^{(21)}(x)+{\frac{7}{12}}w^{(22)}(x)
-{\frac{1}{4}}w^{(23)}(x)$ and
$\hat{w}_{13}(x)= -{\frac{1}{4}}w^{(22)}(x)+{\frac{9}{20}}w^{(23)}(x)
-{\frac{1}{5}}w^{(24)}(x)$.}}
Switching to this new basis and removing NP contributions proportional
to $C_{10,12,14}$, as per the tOPE assumption, one sees that (i) the
theory sides of the $w^{(24)}(x)$ and $w^{(23)}(x)$ FESRs contain no
NP condensate contributions and hence depend only on $\alpha_s$;
(ii) adding the $w^{(22)}(x)$ FESR introduces the single new parameter
$C_8$; (iii) further adding the $w^{(21)}(x)$ FESR introduces the single
new parameter $C_6$ and, finally, (iv) further adding the $w^{(20)}(x)$
FESR introduces the last new parameter, $C_4$. The redundancy argument
above then implies that, as for the classic $km$ spectral weight
analysis, $\alpha_s$ is determined, not by the lowest-weight FESR(s),
but rather from a combined fit to the FESRs of, in this case,
the two highest-degree weights, $w^{(23)}$ and $w^{(24)}$ (or any
two linearly independent combinations thereof), with only
perturbative contributions included on the theory sides. The nominally
$D=4$, $6$ and $8$ condensates are also again obtained in a
redundant manner.
As was the case for the classic $km$ spectral weight analysis, the NP
OPE condensate contributions retained in the analysis play no
role in the determination of~$\alpha_s$.

\subsection[The \texorpdfstring{$2m$}{2m} optimal weight case]{The \boldmath $(2m)$ optimal weight case}
\label{optwtsub}

We now turn to the single-$s_0$ analysis involving the five
``optimal weights'' of Eq.~(\ref{optimal}),
introduced in Ref.~\cite{Pich:2016bdg}. The lowest
degree member of this set, $w^{(21)}$, is again the $\tau$
kinematic weight. The $w^{(2m)}$ are all doubly pinched{\footnote{Doubly
pinched polynomial weights necessarily produce a minimum of two NP
condensate contributions on the theory sides of the associated FESRs.
The $w^{(2m)}(x)$, like the analogous weights employed previously in
Refs.~\cite{Maltman:2008nf,Hudspith:2017vew}, provide examples where
this two-condensate minimum is actually achieved.}} and contain no
terms linear in $x$. Unsuppressed contributions proportional to the
condensate $C_4$ are thus absent from the theory sides of the
associated FESRs, leaving unsuppressed NP contributions, before
making any tOPE assumptions, given by
\begin{eqnarray}
\left[ I_{\rm th}^{w^{(21)}}(s_0)\right]_{\rm NP}&&= -3\, {\frac{C_6}{s_0^3}}
-2\, {\frac{C_8}{s_0^4}}\ ,\nonumber\\
\left[ I_{\rm th}^{w^{(22)}}(s_0)\right]_{\rm NP}&&=\ 4\, {\frac{C_8}{s_0^4}}
+3\, {\frac{C_{10}}{s_0^5}}\ ,\nonumber\\
\left[ I_{\rm th}^{w^{(23)}}(s_0)\right]_{\rm NP}&&=
-5\, {\frac{C_{10}}{s_0^5}} -4\, {\frac{C_{12}}{s_0^6}}\ ,\nonumber\\
\left[ I_{\rm th}^{w^{(24)}}(s_0)\right]_{\rm NP}&&=\ 6\,
{\frac{C_{12}}{s_0^6}}
+5\, {\frac{C_{14}}{s_0^7}}\ ,\nonumber\\
\left[ I_{\rm th}^{w^{(25)}}(s_0)\right]_{\rm NP}&&=
-7\, {\frac{C_{14}}{s_0^7}} -6\, {\frac{C_{16}}{s_0^8}}\ .
\label{optwtnpcontribs}\end{eqnarray}
The optimal-weight analyses reported in Refs.~\cite{Pich:2016bdg,Pich:2022tca}
take advantage of the absence of terms proportional to $C_4$ to include
also $C_{10}$ in the fits, throwing out only contributions proportional
to $C_{12}$, $C_{14}$ and $C_{16}$. Recall, however, that $D>8$ NP
contributions were already assumed negligible in the classic $km$
spectral weight and modified $km$ spectral weight analyses. The weight
$w^{(23)}$ is, moreover, a member of the alternate basis in the latter
case, and one of the two members of that basis for which NP contributions
to the theory sides of the associated FESRs are assumed negligible. If
one wishes to consider simultaneously all of the classic $km$ spectral
weight, modified $km$ spectral weight and optimal weight tOPE analyses,
one should strictly, for consistency, neglect NP contributions to the
$w^{(23)}$ FESR in the optimal weight analysis as well. The other option
is to retain $C_{10}$ in the optimal weight analysis but, for consistency,
jettison the classic $km$ spectral weight and modified $km$ spectral
weight analyses.

Let us consider first the former option and, as in the
classic $km$ spectral weight and modified $km$ spectral weight cases,
neglect $D>8$ contributions on the theory sides of the optimal-weight
analysis as well. The theory sides of the $w^{(2m)}$ optimal-weight
FESRs should then depend on only three theory parameters, $\alpha_s$,
$C_6$ and $C_8$, with the $w^{(23)}$, $w^{(24)}$ and $w^{(25)}$ FESRs
depending only on $\alpha_s$. A single-$s_0$ optimal-weight analysis
employing this version of the tOPE truncation will thus obtain
$\alpha_s$ solely from the combined fit to the $w^{(23)}$, $w^{(24)}$
and $w^{(25)}$ spectral integrals, using only the $D=0$ contributions
on the theory sides of the corresponding FESRs. The redundancy argument
above shows that both this result for $\alpha_s$ and the $\chi^2$ of
the fit will be left unchanged when the additional $w^{(22)}$ and
$w^{(21)}$ FESRs are added to the fit, regardless of whether the
$s_0$ scaling used for the OPE contributions to these weighted integrals
is right or not. As noted already above, because of the tOPE assumption,
the value of $\alpha_s$ so obtained is totally blind to the OPE in QCD
and, in particular, to whether $m_\tau$ is large enough or not to justify
the truncation assumed in this approach. The nominally $D=8$ and $D=6$ NP
condensate results are obtained, redundantly, from the $w^{(22)}$ and
$w^{(21)}$ FESRs, with the two values serving simply to ensure that
the nominally $D=6$ and $8$ contributions are such that, added to
the $D=0$ contributions evaluated using the $\alpha_s$ obtained from
the combined $w^{(23)}$, $w^{(24)}$ and $w^{(25)}$ fit, they produce
exact matches between the spectral integral and theory sides of the
$w^{(22)}$ and $w^{(21)}$ FESRs, regardless of whether the forms of
the theory representations used are correct or not.

If, instead, we choose to jettison the classic $km$ spectral weight
and modified $km$ spectral weight analyses, and consider the optimal-weight
case on its own, now with only $D>10$ contributions neglected, NP
contributions are absent from the theory sides of only the $w^{(24)}$
and $w^{(25)}$ FESRs, and $\alpha_s$ is obtained from a combined
fit to these two FESRs only. Adding the $w^{(23)}$ FESR now produces
a redundantly determined result for $C_{10}$, leaving the result
for $\alpha_s$ and the $\chi^2$ of the fit unchanged. Further
adding the $w^{(22)}$ FESR then produces the (redundantly determined)
result for $C_8$, with the values of $\alpha_s$, $\chi^2$ and
$C_{10}$ obtained in the previous steps unchanged. Finally, adding
the $w^{(21)}$ FESR produces the (redundantly determined) result for
the final retained condensate, $C_6$, leaving the results for
$\alpha_s$, $\chi^2$, $C_{10}$ and $C_8$ obtained in the previous
steps unchanged. These facts are illustrated, for the case of the
$s_0=2.80$ GeV$^2$ $V+A$ channel optimal-weight fit based on
ALEPH data, in Table~\ref{tab:prsoptwtredundancy} below.

\begin{table}[h]
{\footnotesize
\begin{center}
\begin{tabular}{|l|c|c|c|c|c|}
\hline
weights &$\chi^2$ & $\a_s(m_\t^2)$ & $10^3C_{10}$ & $10^3C_8$ & $10^3C_6$\\
\hline
$w^{(24)}, w^{(25)}$ & 3.068384 & 0.31685(0.00253)& --- & --- & --- \\
$w^{(23)}, w^{(24)}, w^{(25)}$ & 3.068384 & 0.31685(0.00253)
& 0.3464(0.1187) &--- &--- \\
$w^{(22)}, w^{(23)}, w^{(24)}, w^{(25)}$ & 3.068384 & 0.31685(0.00253)
& 0.3464(0.1187) & $-$0.8720(0.2107) &---\\
$w^{(21)}, w^{(22)}, w^{(23)}, w^{(24)}, w^{(25)}$ & 3.068384
& 0.31685(0.00253) & 0.3464(0.1187) & $-$0.8720(0.2107)
& 1.3771(0.2371)\\
\hline
\end{tabular}
\end{center}
\caption{{Illustration of the redundancy of the ALEPH-based, $s_0=2.8$
GeV$^2$, tOPE $V+A$-channel optimal-weight analysis. Results are quoted
for $\chi^2$, $\alpha_s(m_\tau^2)$ and the relevant condensates at each stage in the build-up of the final full five-weight analysis, with the $C_D$ in
units of GeV$^D$. The number of digits retained, which clearly
exceeds that required by the size of the accompanying errors, has been
chosen in order to emphasize the exact nature of the redundancy result
detailed in the text, and proven in Appendix~\ref{app:redundancyproof}.}}
\label{tab:prsoptwtredundancy}}
\end{table}

Returning to the case that all of the classic $km$ spectral weight,
modified $km$ spectral weight and optimal weight analyses are retained
and the self-consistent truncation at $D=8$ is employed, the optimal-weight
set provides a tOPE analysis option in which $\alpha_s$ can be obtained
from a combined fit to three, rather than just two, spectral integrals.
This is in contrast to the classic $km$ spectral weight and modified $km$
spectral weight cases, where $\alpha_s$ is determined from a fit to
only two such spectral integrals. With this tOPE truncation, the
optimal-weight analysis thus allows an extended version of the more
limited self-consistency checks available to the classic $km$ spectral
weight and modified $km$ spectral weight analyses to be performed, one
involving comparisons between $\alpha_s$ values obtained, for example,
from any of (i) the three individual single-weight $w^{(23)}$, $w^{(24)}$
and $w^{(25)}$ FESR determinations, (ii) the combined two-weight $w^{(24)}$
and $w^{(25)}$ FESR fit and (iii) the full combined three-weight $w^{(23)}$,
$w^{(24)}$ and $w^{(25)}$ FESR fit.

We explore such single-$s_0$ optimal-weight tests for the isovector
$V+A$ channel, employing as experimental input (i) the pion pole
contribution implied by $\pi_{\mu2}$ and the Standard Model, and (ii)
the 2013 ALEPH~\cite{Davier:2013sfa} continuum spectral distribution,
rescaled to account for the small subsequent change in the continuum
$V+A$ branching fraction. Appendix~\ref{app:aleph2013details} provides
a brief outline of the input to this rescaling. We follow
Refs.~\cite{Pich:2016bdg,Pich:2022tca} in avoiding the impact of the
large errors in the two highest ALEPH bins by working at
$s_0=2.8\ {\rm GeV}^2$, thus omitting contributions from those two bins.

The goal in carrying out this exploration is not to simply repeat
optimal-weight tOPE determinations of $\alpha_s$ already present in the
literature, but rather to investigate the internal consistency of
results obtained in those determinations in the light of the
improved understanding provided by the redundancy arguments above. We do
so by investigating whether the differences of pairs of $\alpha_s$
results obtained using different subsets of the five optimal-weight
FESRs, all employing the same underlying tOPE assumptions, are compatible
with zero within properly correlated errors. In evaluating such errors it
is critical to take into account the very strong correlations present in
such single-$s_0$ tOPE analyses. Not only are differently weighted spectral
integrals at the same $s_0$ highly correlated,{\footnote{In the
$s_0=2.8\ {\rm GeV}^2$, isovector $V+A$ case under consideration, for
example, the correlations are 99.0\% between the $w^{(23)}$- and
$w^{(24)}$-weighted integrals, 97.0\% between the $w^{(23)}$- and
$w^{(25)}$-weighted integrals, and 99.4\% between the $w^{(24)}$- and
$w^{(25)}$-weighted integrals.}} but, with the tOPE assumption, the
three FESRs responsible for determining $\alpha_s$ all have only $D=0$
OPE contributions on their theory sides, and thus theory errors which
are 100\% correlated. Such strong positive correlations produce
errors on the differences between two such determinations much
smaller than the quadrature sum of the individual errors. Not
infrequently in the literature pairs of $\alpha_s$ determinations are
assessed as displaying good agreement when the the quadrature
sums of their experimental and theory errors overlap (or nearly overlap).
Given the strong correlations generic to single-$s_0$ tOPE analyses,
however, such individual combined experiment-plus-theory errors, while
relevant to assessing the precision of the individual determinations,
are of no use, on their own, in assessing the compatibility of two
such determinations. Many quick eye-balled compatibility assessments
in the literature founder on this point. Experimental errors quoted
below are evaluated using the ALEPH $V+A$ covariance matrix.
Theory errors, stemming from the truncation of perturbation
theory and estimated using $c_{51}=283\pm 142$, are identified by
the subscript `$c_{51}$', and quoted only for the differences
which are the focus of the current investigation. The corresponding
experimental errors are identified by the subscript `ex'.

Turning finally to the results of the $s_0=2.8\ {\rm GeV}^2$,
optimal-weight tOPE analysis, we find, from the combined fit to
the $w^{(23)}$, $w^{(24)}$ and $w^{(25)}$ FESRs,
\begin{equation}
\alpha_s(m_\tau^2)=0.3125(23)_{\rm ex}\, ,
\label{3wttopeoptwtfitalphas}\end{equation}
with $\chi^2/{\rm dof}=11.6/2$ ($p$-value $0.3\%$),
{\it i.e.}, clearly not a good fit. Even though one could stop the
analysis at this point, one may still consider the analogous two-weight
fit to the $w^{(24)}$ and $w^{(25)}$ FESRs{\footnote{Because of the
redundancy argument, this is precisely the optimal-weight fit of
Ref.~\cite{Pich:2022tca}, with $C_{10}$ retained as a fit parameter.}}
which produces a $\chi^2/{\rm dof}$ of $3.1$ ($p$-value $7.8\%$) and
\begin{equation}
\alpha_s(m_\tau^2)=0.3168(27)_{\rm ex}\, .
\label{2wttopeoptwtfitalphas}\end{equation}
As an example of a test of the self-consistency of the underlying
tOPE assumptions, we compare the result of Eq.~(\ref{3wttopeoptwtfitalphas})
to that obtained using the $w^{(25)}$ FESR alone,
\begin{equation}
\alpha_s(m_\tau^2)=0.3228(43)_{\rm ex}\, .
\label{w25topeoptwtonlyalphas}\end{equation}
With correlations taken into account, the results of
Eqs.~(\ref{3wttopeoptwtfitalphas}) and (\ref{w25topeoptwtonlyalphas})
differ by $0.0103(37)_{\rm ex} (10)_{c_{51}}\, =\, 0.0103(38)$, where
experimental and theory errors have been combined in quadrature in the second
expression. The sum of the errors on the individual determinations, $0.0066$,
is significantly larger than the properly correlated $0.0038$ combined
error on their difference. The combined uncertainty is a factor
of $2.7$ times smaller than the central value of that difference, which
is, in turn, a factor of more than $4$ times larger than the error
produced by the combined, single-$s_0$, three-weight fit. It would
obviously be extremely incautious to treat that $0.0023$ experimental
error as a realistic representation of the experimental uncertainty
in the analysis, ignoring the potentially much larger systematic error
represented by the difference between the two results above.

As a further investigation of this point, consider the effect of
removing the contributions of either one or two additional
ALEPH bins from the optimal-weight $V+A$ spectral integrals, reducing
the $s_0$ employed from $2.8\ {\rm GeV}^2$ to either $2.6\ {\rm GeV}^2$
or $2.4\ {\rm GeV^2}$. These are the two minimum reductions possible
for the ALEPH binning, and represent decreases of only $\sim 60$ and
$120$ MeV in the maximum CM energy entering those integrals. The
motivation for this exploration is as follows. One of the reasons
given by the proponents of the tOPE approach for preferring the $V+A$
combination is that ``the flattening of the spectral function is
remarkably fast for the most inclusive $V+A$ channel, where perturbative
QCD seems to work even at quite low values of $s \sim 1.2$~GeV$^2$''
\cite{Pich:2022tca}. While the particular low value, $s \sim 1.2$~GeV$^2$,
suggested here looks somewhat over-optimistic for the ALEPH data
case, where the ratio of the DV to the $\alpha_s$-dependent,
dynamical QCD perturbative contribution to $\rho_{V+A}(s)$ is found to be
$\sim 1.27$ in the bin from $1.450$ to $1.475\ {\rm GeV}^2$ and
$\sim -0.75$ in the bin from $1.95$ to $2.00\ {\rm GeV}^2$, the
assumption that $\rho_{V+A}(s)$ is essentially purely perturbative
does become compatible with the ALEPH data, at least within
experimental errors, for $s\geq 2.4\,{\rm GeV}^2$. One could thus consider
the possibility, advocated by the proponents of the tOPE approach,
that DV contributions might be
safely negligible in single-$s_0$ $V+A$-channel analyses, so long as
$s_0\geq 2.4\ {\rm GeV}^2$. Assuming this to be the case,
$s_0=2.4\ {\rm GeV}^2$ and $2.6\ {\rm GeV}^2$ versions of the
self-consistency tests discussed above for $s_0=2.8\ {\rm GeV}^2$
will serve to further test the reliability of the assumed tOPE
truncation. Since any $D>8$ NP contributions omitted in the
above analysis grow with decreasing $s_0$, a sign that the difference
found above results from the omission of in-fact-non-negligible NP
condensate contributions would be an increase in the significance
of the discrepancy at these slightly lower $s_0$.

Carrying out the $s_0=2.6\ {\rm GeV}^2$ version of the single-$s_0$
optimal-weight tOPE analysis, we find, from the combined three-weight
fit, the result
\begin{equation}
\alpha_s(m_\tau^2)=0.3100(22)_{{\rm ex}}\, ,
\label{3wttopeoptwtfitalphas2pt6}\end{equation}
with a $\chi^2/{\rm dof}$, $18.7/2$ ($p$-value $9\times 10^{-5}$),
even worse than that of the corresponding $s_0=2.8\ {\rm GeV}^2$ fit,
again a rather disastrous fit.
The $\chi^2/{\rm dof}=4.5$ ($p$-value $3.4\%$) of the combined
two-weight $w^{(24)}$ and $w^{(25)}$ fit is, similarly, significantly
worse than that of the corresponding two-weight $s_0=2.8\ {\rm GeV^2}$
fit.{\footnote{For completeness, ignoring for the moment the very poor
three-weight fit quality and comparing the result of
Eq.~(\ref{3wttopeoptwtfitalphas2pt6}) to that obtained from the
$w^{(25)}$ FESR alone, $\alpha_s(m_\tau^2)=0.3202(34)_{{\rm ex}}$,
one finds, after accounting for correlations, a difference
$0.0102(27)_{\rm ex}(10)_{c_{51}}=0.0102(29)$. The central value
of the discrepancy is now $3.6$ times its combined uncertainty and $4.6$
times the nominal fit error in Eq.~(\ref{3wttopeoptwtfitalphas2pt6}).}}

For $s_0=2.4\ {\rm GeV}^2$, we find a combined three-weight fit
with an even more disastrous $\chi^2/{\rm dof}$ of $31.9/2$ ($p-$value
$\sim 10^{-7}$), and a combined two-weight, $w^{(24)}$ and $w^{(25)}$,
fit with a $\chi^2/{\rm dof}$ of $6.3$ ($p$-value
$\sim 1.2\%$).\footnote{Ignoring the (even more problematic)
three-weight fit quality and completing the comparison with the
$s_0=2.8\ {\rm GeV}^2$ and $2.6\ {\rm GeV}^2$ cases, we find a difference,
$0.0114(22)_{\rm ex}(11)_{c_{51}}=0.0114(24)$, between the three-weight
fit result, $\alpha_s(m_\tau^2)=0.3064(22)_{{\rm ex}}$, and single-weight
$w^{(25)}$ fit result, $\alpha_s(m_\tau^2)=0.3178(30)_{{\rm ex}}$,
with a central value $4.7$ times its combined uncertainty and $5.3$
times the nominal three-weight fit error.}

Clearly, the results just discussed point to non-trivial problems
for the tOPE assumptions, even in the purportedly safer $V+A$ channel.
The conclusions reached above for the tOPE based on the classic $km$
spectral-weight, modified spectral-weight, and optimal-weight
analyses of the $V+A$ channel are by no means unique. In
Appendix~\ref{app:improvedVchannel}, we revisit the determination
of $\alpha_s$ from an optimal-weight tOPE analysis of the improved
$V$ spectral function recently obtained in Ref.~\cite{Boito:2020xli}.
This analysis exposes much more severe problems with the tOPE-strategy
assumptions, with the combined two-weight $w^{(24)}$ and $w^{(25)}$ fit,
for example, producing a disastrous $\chi^2/{\rm dof}$ of $43.1$
($p$-value $5\times 10^{-11}$). Further details of this analysis
are provided in Appendix~\ref{app:improvedVchannel}.

To summarize, the results of the current section and
Appendix~\ref{app:improvedVchannel} clearly point to problems with
conventional tOPE assumptions and argue for the necessity of revisiting
that analysis framework with the goal of attempting to quantify, and
deal with, the uncertainties the above explorations have exposed.

\subsection{Redundancy and OPE condensates}\label{condredundancysub}

The redundancy observation detailed above, which, as we have seen, is
operative for multi-weight, single-$s_0$ tOPE fits (though not for
multi-weight, multi-$s_0$ block-diagonal $Q^2$ fits, as explained
in Sec.~\ref{sec:supposedtautology}) also serves to expose a serious
problem for the reliability of the associated redundantly determined
nominal NP condensates created by the absence of the $D$-dependent
$s_0$-scaling constraints implicitly present in multi-weight, multi-$s_0$
analyses. The general point is most easily understood by means of an example.

Let us consider the isovector $V+A$ channel optimal-weight analysis in
the form discussed in Refs.~\cite{Pich:2016bdg,Pich:2022tca},
in which only $D>10$ OPE contributions are assumed negligible.
The redundancy argument above then makes clear that, in this case,
(i) the result for $\alpha_s$, denoted $\alpha_s=\bar{\alpha}_s$,
is that given in Eq.~(\ref{2wttopeoptwtfitalphas}) above, resulting
from a combined fit to the $w^{(24)}$ and $w^{(25)}$ FESRs only;
(ii) the result for $C_{10}$, denoted $\bar{C}_{10}$, is that obtained
(redundantly) by matching the theory and spectral integral sides
of the $w^{(23)}$ FESR, using as input for the $D=0$ OPE contribution
the previously determined result $\alpha_s=\bar{\alpha}_s$;
(iii) the result for $C_8$, $\bar{C}_8$, is that obtained (redundantly)
by matching the theory and spectral integral sides of the $w^{(22)}$
FESR, using as input for the $D=0$ and $10$ contributions the previously
determined results $\alpha_s=\bar{\alpha}_s$ and $C_{10}=\bar{C}_{10}$;
and finally, (iv) the result for $C_6$, $\bar{C}_6$, is that obtained
(again redundantly) by matching the theory and spectral integral
sides of the $w^{(21)}$ FESR, using as input for the $D=0$ and $8$
contributions the previously determined results $\alpha_s =\bar{\alpha}_s$
and $C_8=\bar{C}_8$. Introducing the simplified notation
$I_{V+A}^{w^{(2m)}}(s_0)\equiv I_{V+A}^{(2m)}(s_0)$ for
the $w^{(2m)}$-weighted spectral integrals, and denoting by
$I_{V+A;D=0}^{({2m})}(s_0,\bar{\alpha}_s)$ the $\alpha_s=\bar{\alpha}_s$
versions of the corresponding $D=0$ OPE theory integrals, $\bar{C}_{10}$,
$\bar{C}_8$ and $\bar{C}_6$ are given by the expressions
\begin{eqnarray}
\bar{C}_{10}&&=\, -{\frac{s_0^5}{5}}\left[ I_{V+A}^{(23)}(s_0) -
I_{V+A;D=0}^{(23)}(s_0;\bar{\alpha}_s)\right],
\label{C10optwtope}\\
\bar{C}_8 &&=\, {\frac{s_0^4}{4}}\left[ I_{V+A}^{(22)}(s_0) -
I_{V+A;D=0}^{(22)}(s_0;\bar{\alpha}_s)
-{\frac{3\bar{C}_{10}}{s_0^5}}\right]\nonumber\\
&&= {\frac{s_0^4}{4}}\left[ I_{V+A}^{(22)}(s_0)
+{\frac{3}{5}}I_{V+A}^{(23)}(s_0)
-I_{V+A;D=0}^{(22)}(s_0;\bar{\alpha}_s)
-{\frac{3}{5}}I_{V+A;D=0}^{(23)}(s_0;\bar{\alpha}_s)\right] ,
\label{C8optwtope}\\
\bar{C}_6 &&=\, -{\frac{s_0^3}{3}}\left[ I_{V+A}^{(21)}(s_0) -
I_{V+A;D=0}^{(21)}(s_0;\bar{\alpha}_s)
+{\frac{2\bar{C}_{8}}{s_0^4}}\right]\, \nonumber\\
&&=\,-{\frac{s_0^3}{3}}\left[ I_{V+A}^{(21)}(s_0)+{\frac{1}{2}}
I_{V+A}^{(22)}(s_0)+{\frac{3}{10}}
I_{V+A}^{(23)}(s_0)\right.\nonumber\\
&&\qquad\left. -I_{V+A;D=0}^{(21)}(s_0;\bar{\alpha}_s)
-{\frac{1}{2}} I_{V+A;D=0}^{(22)}(s_0;\bar{\alpha}_s)
-{\frac{3}{10}} I_{V+A;D=0}^{(23)}(s_0;\bar{\alpha}_s)\right].
\label{C6optwtope}\end{eqnarray}
The right-hand sides of these equations involve differences between
the $w^{(23)}$-, $w^{(22)}$- and $w^{(21)}$-weighted spectral
integrals and their correspondingly weighted $D=0$ OPE counterparts.
Since the $D=0$ contributions dominate the theory representations,
very strong cancellations necessarily exist in these differences.
It follows that, if the result $\bar{\alpha}_s$ is contaminated
by omitted NP contributions, the $D=0$ contributions in
Eqs.~(\ref{C10optwtope})--(\ref{C6optwtope}) will be unphysical,
with related unphysical impact on the nominal $D=10$, $8$ and $6$
condensate results. Because of these strong cancellations, the
impact on the $C_D$ can be large, even if the contamination present
in $\bar{\alpha}_s$ is small.

To illustrate this point, imagine the true $\alpha_s(m_\tau^2)$ is not
the central optimal-weight fit result $\bar{\alpha}_s(m_\tau^2)=0.3168$
but, rather, either the single-weight $w^{(25)}$ $V+A$ channel result
$\alpha_s(m_\tau^2)=0.3228$ of Eq.~(\ref{w25topeoptwtonlyalphas}), or
the improved $V$-channel result, $0.3077$, obtained in the DV-strategy
analysis of Ref.~\cite{Boito:2020xli}. From the $s_0=2.8\ {\rm GeV}^2$
versions of Eqs.~(\ref{C10optwtope})--(\ref{C6optwtope}), we find,
using as input the central optimal-weight fit value,
$\bar{\alpha}_s(m_\tau^2)=0.3168$, the results
\begin{eqnarray}
&&C_{10}=\bar{C}_{10}=\, -0.0041(41)\ {\rm GeV}^{10}\nonumber\\
&&C_8=\bar{C}_8=\, 0.0016(26)\ {\rm GeV}^8\nonumber\\
&&C_6=\bar{C}_6=\, 0.0005(12)\ {\rm GeV}^6\, ,
\label{CDalpha3168}\end{eqnarray}
where the errors are those of the combinations of spectral integrals
appearing in the final forms of the right-hand sides of
Eqs.~(\ref{C10optwtope})--(\ref{C6optwtope}).
If, in contrast, the true value of $\alpha_s(m_\tau^2)$ were $0.3228$,
one would find, using this alternate input, the results
\begin{eqnarray}
&&C_{10}=\, 0.0033(41)\ {\rm GeV}^{10}\nonumber\\
&&C_8=\, -0.0037(26)\ {\rm GeV}^8\nonumber\\
&&C_6=\, 0.0033(12)\ {\rm GeV}^6\, ,
\label{CDalpha3228}\end{eqnarray}
while a true value $\alpha_s(m_\tau^2)=0.3077$ would produce the results
\begin{eqnarray}
&&C_{10}=\, -0.0151(41)\ {\rm GeV}^{10}\nonumber\\
&&C_8=\, -0.0093(26)\ {\rm GeV}^8\nonumber\\
&&C_6=\, -0.0036(12)\ {\rm GeV}^6\, .
\label{CDalpha3077}\end{eqnarray}
One observes very large shifts from the central tOPE values, including
large shifts in magnitude (by factors of $\sim 4$, $6$ and $7$ for
$C_{10}$, $C_8$ and $C_6$) in the latter case.

This exploration confirms the very strong sensitivity of single-$s_0$
tOPE condensate results to potential, even rather small, NP contamination
in the tOPE result for $\alpha_s$. Additional potential contaminations
associated with NP contributions omitted from the tOPE versions of the
theory sides of the $w^{(23)}$ and $w^{(22)}$ FESRs, and not included in
the above discussion, also, of course, exist at some level. The impact of
such additional contamination on the results for $C_{10}$, $C_8$ and
$C_6$ will also be amplified by the strong cancellation between the
$D=0$ OPE and spectral integrals. We conclude from the above exercise
that, because of the limitations resulting from the redundant nature
of the determination of the nominal dimension-$D$ NP condensates, $C_D$,
in multi-weight, single-$s_0$ tOPE analyses, tOPE results for these
condensates will generically suffer from large theoretical systematic
uncertainties not taken into account in existing tOPE analyses, and
where, at present, it is unclear how such uncertainties might be sensibly
estimated in the tOPE framework. Multi-$s_0$ analyses, in which
additional $s_0$-dependence constraints come into play, provide
an obvious strategy for dealing with this issue.

An important point that follows from the above exercise is
the observation that the small values of the
condensates produced by the strict tOPE $\alpha_s$ fit input, often
pointed to as physical evidence in support of the neglect of yet-higher
$D$ NP contributions, in fact, provide nothing of the sort. Rather, due
to the entirely tautological, and physically unconstrained, nature
of their determinations, they are subject to very large uncertainties and
cannot be realistically taken as establishing the smallness represented by
the central tOPE fit values. This is, in any case, irrelevant
as far as the determination of $\alpha_s$ is concerned since, as we
have seen above, the determination of the nominal NP condensates
plays no role whatsoever in the determination of $\alpha_s$ in such
tOPE analyses of $\tau$ decay data, which become entirely
perturbative in nature.

\subsection{Summary}

We close this section by reiterating the lessons learned from the
discussions above. First, multi-weight, multi-$s_0$ fits
are {\it not} redundant in the sense of Ref~\cite{Pich:2022tca}. Rather
the multi-weight, multi-$s_0$ nature provides highly non-trivial
self-consistency tests of the underlying theoretical representations.
This advantage is a result of the use of multiple $s_0$ values and in
no way restricted to the use of the particular DV model form employed
in previous DV-strategy analyses. Second, multi-weight, but single-$s_0$,
analyses, with their accompanying tOPE truncations, provide access to a
much more restricted set of self-consistency tests. They, in addition,
obtain results for $\alpha_s$ not from the FESRs with the lowest degree
weights considered, but rather those with the highest degree
considered, neglecting all non-perturbative contributions and without
any sensitivity to any of the OPE condensates of QCD. Furthermore,
different values of $\a_s$ obtained from these highest-degree weights
show non-trivial tensions, once all the correlations are properly
taken into account. FESRs involving high-degree weights are more
dependent on the reliability of the tOPE truncation assumption,
as well as being potentially more sensitive to issues associated
with high-$D$ OPE contributions and the asymptotic nature of the
OPE series. Finally, because of the strong enhancement of the
impact of NP-contamination-induced uncertainties in the associated
determination of $\alpha_s$, results for the NP OPE condensates $C_D$
obtained from single-$s_0$ tOPE fits are subject to potentially very
large theoretical systematic uncertainties. The enhancement in question
is a generic feature of the way in which the redundant determinations
of the $C_D$ work in such single-$s_0$ analyses, and (as far
as we can see) avoidable only by moving to multi-$s_0$ analyses.

\section{Conclusions}
\label{conclusions}
In this paper, we have investigated, in detail, several criticisms
of the DV strategy for the determination of the strong coupling from
hadronic $\tau$ decay data raised by the authors of
Ref.~\cite{Pich:2022tca}. This investigation led to an exploration
of some previously incompletely understood aspects of the alternate
tOPE strategy advocated by those authors, and a deeper understanding
of key shortcomings in that alternate approach.

The potential issues with the DV strategy raised in
Ref.~\cite{Pich:2022tca} varied from a claimed instability with
respect to variations of the form used to model DV contributions
to a supposed redundancy of the many consistency tests employed
in past DV-model analyses. Here, we proved these criticisms
to be, not just unfounded, but, in fact, incorrect. We briefly
mention the two most important such refuted claims.

In Sec.~\ref{sec:supposedtautology}, we addressed the claim that
DV-model-strategy analyses are equivalent to fits to the spectral
function plus just one spectral integral data point, with all
additional $s_0$-dependent spectral integrals, corresponding to all
additional weights considered in existing DV-strategy analyses in the
literature, being ``redundant'', in the sense of adding no new
information on the strong coupling itself. We showed this claim to
be incorrect, explicitly identifying the source of the error in the
argument of Ref.~\cite{Pich:2022tca}, which turns out to be the
assumption, stated without proof, that the result, which would
have been valid for standard $\chi^2$ fits to sets of spectral
integrals consisting of those involving a first weight at multiple
$s_0$, but a second weight at a single $s_0$ only, is also valid for
the necessarily non-$\chi^2$, block-diagonal $Q^2$ fits to spectral
integral sets with the same multiple $s_0$ employed for both weight
cases. That this assumption is incorrect is explicitly demonstrated
by the numerical results of two-weight fits using either expected QCD
or non-QCD forms for the theory representations, which show clearly
that results for the strong coupling do, in general, depend on the
choice of weights, the range of $s_0$ values used, and the forms
chosen for the corresponding theory representations. The fact that,
in the DV strategy approach, these different choices lead to
inconsistent results when the non-QCD form is employed on the
theory side, but consistent results when the expected QCD form is
used, constitutes a non-trivial success for the DV strategy approach.

In contrast, as also observed, but not fully investigated, in
Ref.~\cite{Pich:2022tca}, single-$s_0$ tOPE analyses do possess a
built-in redundancy, originating in the truncation of the OPE central
to that approach. We discussed this in detail in Sec.~\ref{tOPEredundancy},
where we showed that, once a set of weights is chosen and a truncation of
the OPE implemented, the value of $\alpha_s(m_\tau^2)$ is determined from
FESRs involving the highest, not the lowest, degree weights in the
analysis. These FESRs, moreover, involve theory sides which, in the tOPE
approximation, contain no NP contributions whatsoever, leading to $\alpha_s$
determinations which are based on perturbation theory alone. We also showed
that specific features of the redundant determinations of the NP OPE
condensates generic to single-$s_0$ tOPE analyses produce potentially
very large theoretical systematic uncertainties associated with strong
enhancements of the impact of residual NP contaminations of the
results obtained for $\alpha_s$, enhancements which appear, at
present, unavoidable in the single-$s_0$ tOPE framework. This adds to
previous questions raised regarding the reliability of the tOPE
approach~\cite{Boito:2016oam,Boito:2019iwh} which to date have yet
to be adequately addressed by the proponents of that approach.

Returning to the DV approach: this approach recognizes the need to make
use of a model for quark-hadron duality violations. This is unavoidable:
resonance contributions clearly visible in the spectral functions
used in the fits to obtain $\alpha_s(m_\tau^2)$ are not accounted for by
the OPE. It is thus important to test this approach, and one option is to
vary the model. Following Ref.~\cite{Boito:2017cnp}, one can add
subleading corrections of relative order $1/s$ to the leading asymptotic
form of the {\it ansatz} employed in the DV strategy, and this was tried
in Ref.~\cite{Pich:2022tca} using ALEPH $V$-channel data. We reproduced
this test in Appendix~\ref{sec:DV-parm}, but also demonstrated there
that using the much more precise data of Ref.~\cite{Boito:2020xli},
obtained employing CVC to combine ALEPH and OPAL $\tau$ data with
electroproduction results, leads unambiguously to the conclusion that
the result for $\alpha_s$ obtained using the DV model is very stable
against allowed versions of such $1/s$ corrections.

As a patient reader will note, when dealing with the issues discussed in
this paper, details are crucially important. We thus refrain from further
summarizing our findings in this concluding section, and instead refer
the reader to the detailed explanations of the various points contained
in the main body and appendices of this paper. We close by acknowledging
the utility of Ref.~\cite{Pich:2022tca} in providing details of previous,
often incompletely characterized, objections to the DV-model approach made
by the authors in recent years. This has allowed us to investigate these
one by one and show that, after taking into account the additional
information detailed above, none turn out to be substantiated. Our findings,
in fact, provide additional support for the use of multi-$s_0$ approaches
like the DV-model strategy, at least within the precision of currently
available data, while, at the same time, raising serious questions about the
reliability of the tOPE approach.

Both the DV and the tOPE approaches employ model assumptions
to deal with the non-perturbative effects affecting sum-rule
analyses of hadronic $\tau$ decay data. We have demonstrated that the
tOPE approach shows clear evidence for the presence of non-negligible
systematic uncertainties not taken into account in the literature
employing this approach, and it is unclear to us how to resolve
this issue.

While the DV-model approach has so far passed all the tests
we have applied to it, new ideas on how to further test this approach
remain of interest. We believe that further progress can be made both
experimentally and theoretically. Experimentally, more precise
spectral data may become available (for instance, for the two $4\pi$
modes, whose differential distributions could be measured at Belle II).
This would allow for the application of more stringent versions of the
tests we have applied to date. More important, perhaps, is theoretical
progress. Here lattice QCD may prove useful. Recently, methods
have been developed for directly computing weighted integrals of
inclusive spectral functions using lattice data (for an example related
to hadronic $\tau$ decays, see \rcite{Evangelista:2023fmt}).
A major advantage of this approach is that it would, in principle,
allow determinations of such integrals for $s_0>m_\tau^2$, a region
inaccessible, for kinematic reasons, to determinations based on
experimental hadronic $\tau$ decay data. This may allow for more
detailed studies of duality violations as a function of $s$ for $s$
in the region from below to well above $m_\tau^2$.

\vspace{1cm}
\noindent {\bf Acknowledgments}\\

This material is based upon work supported by the U.S. Department
of Energy, Office of Science, Office of Basic Energy Sciences Energy
Frontier Research Centers program under Award Number DE-SC-0013682 (MG).
DB's work was supported by the S\~ao Paulo Research Foundation (FAPESP)
Grant No. 2021/06756-6 and by CNPq Grant No. 308979/2021-4.
The work of KM is supported by a grant from the
Natural Sciences and Engineering Research Council of Canada. SP is
supported by the Spanish Ministry of Science, Innovation and Universities
(project PID2020-112965GB-I00/AEI/10.13039/501100011033) and by Departament
de Recerca i Universitats de la Generalitat de Catalunya, Grant No 2021
SGR 00649. IFAE is partially funded by the CERCA program of the Generalitat
de Catalunya.

\appendix
\section{Logarithms in the condensates and asymptoticity of the OPE}
\label{sec:OPE-logs}

In Ref.~\cite{Boito:2020xli} the set of polynomials $w(x)$, $x=s/s_0$,
chosen for use in the FESRs of Eq.~(\ref{cauchy}) was $w_0(x)=1$,
$w_2(x)=1-x^2$, $w_3(x)=(1-x)^2 (1+2x)$, and $w_4(x)=(1-x^2)^2$. This
choice was made in order to restrict the unsuppressed OPE terms
entering the analysis to dimensions $D\leq 10$. The set also constitutes
a complete linearly independent basis for polynomials up to degree 4
without a linear term. The restriction to relatively low dimensions is in
keeping with the expectation that the OPE is an asymptotic expansion,
while the absence of a linear term is motivated by the results of
Ref.~\cite{Beneke:2012vb}, which showed evidence for bad
perturbative behavior in the case of weights with a linear
term, which are strongly sensitive to the gluon-condensate
renormalon~\cite{Boito:2020hvu,Benitez-Rathgeb:2022yqb}. Moreover, using
the unpinched weight $w(x)=1$ in different combinations with the
other three weights, which are either singly and doubly pinched, allows
for an estimate of the systematic error due to the DV parametrization,
since the differently weighted spectral integrals have different
sensitivities to DVs.\footnote{The contrary claim made in
Ref.~\cite{Pich:2022tca} (that the addition of FESRs involving the
other three weights would add no new constraints on the DV form)
has been shown to be incorrect, and results from the mathematical
error identified above. See Sec.~\ref{sec:supposedtautology} for more
details on this point.} In addition, different weight combinations
receive contributions from different terms in the OPE. For instance, a
combined fit to the $w_0(x)=1$ and $w_2(x)=1-x^2$ FESRs determines,
in addition to $\alpha_s$, $C_6$ in Eq. ~(\ref{OPE}). A combined
fit to the $w_0$ and $w_3(x)=(1-x)^2(1+2x)$ FESRs, similarly, determines
$\alpha_s$, $C_6$ and $C_8$, while a combined fit to the $w_0$ and
$w_4(x) =(1-x^2)^2$ FESRs determines $\alpha_s$, $C_6$ and $C_{10}$.
The values obtained for $\alpha_s$ and $C_6$ from the three different
fits can be tested for consistency, thus further testing the method.

In spite of the successful consistency checks carried out in
Ref.~\cite{Boito:2020xli}, in Sec. 4.1 of Ref.~\cite{Pich:2022tca},
the validity of the DV-strategy for the extraction of $\alpha_s$ is
questioned with the argument that the fitted OPE condensates are
too large, which could invalidate the assumption that the logarithmic
$s$-dependence of Wilson coefficients in the OPE can be neglected. In order
to investigate this question, we write the $C_{2k}$ coefficients of the
OPE, including leading logarithmic corrections, as
\beq
C_{2k}(s)\equiv C_{D}(s) = C^{(0)}_D(\mu^2)\left( 1 +
L^{(1)}_D \log\left(\frac{-s}{\mu^2}\right)\right),
\eeq
where $L^{(1)}_D$ is suppressed by at least one power of $\alpha_s$.
The presence of the logarithm alters the form of the monomial-weighted
contributions of the dimension $D$ term, $ C_{D}(z)/(-z)^{D/2}$, to the
polynomial-weighted OPE FESR integrals. In the resulting altered form,
\beq
\lbl{newcauchy}
-\frac{1}{2 \pi i} \oint_{|z|=s_0}\frac{dz}{s_0}\left( \frac{z}{s_0}\right)^n
\frac{C_{2p}(z)}{(-z)^p} = -\ \frac{ C^{(0)}_{2p}(s_0)}{(-s_0)^p}
\left( \delta_{p,n+1}+ \frac{ L^{(1)}_{2p} }{1+n-p}\,
\overline{\delta}_{p,n+1}\right)\, ,
\eeq
where $\overline{\delta}_{p,n+1}=1- \delta_{p,n+1}$, the tight connection
that existed, in the absence of the logarithm, between the power, $n$, of
the monomial and the dimension $p$ of the OPE condensate for which the
monomial-weighted integral survives, is lost. Naively, these logarithmic
corrections contribute to all the FESRs when $p\gtrless n+1$, an
observation which led the authors of Ref.~\cite{Pich:2022tca} to conclude
that even though individual logarithmic contributions might be quite
small on account of the aforementioned $\alpha_s$ suppression, their
sum might be large, given the expected divergent nature of the series,
thus invalidating the usual implementation of the FESR. But, is this true?

The answer is no. Note that the FESR in Eq. (\ref{cauchy}) always
exists and is well defined for any power of $n$ in the
polynomial $w(s)=(s/s_0)^n$ so long as one uses the exact function
for $\Pi(z)$ on the right-hand side. This is nothing but Cauchy's
theorem. On the other hand, as is well known, an asymptotic expansion,
such as the OPE, diverges at large orders and approximating the
exact $\Pi(z)$ by an expansion including terms $C_{2p}(z)/(-z)^p$
up to some large $p=N$ becomes incorrect when $N$ is too large, with
or without the logarithmic corrections. Of course, how large a value
of $p$ is too large is connected to how large $|z|=s_0$ is in the
problem at hand. Since an asymptotic expansion is valid only
when truncated at sufficiently low orders, this renders the
OPE useless if the weight in question has a degree, $n$, such
that Eq.~(\ref{newcauchy}) requires $p$ to lie in the region where
the OPE is no longer valid. This problem is not solved by
setting the right-hand side of Eq.~(\ref{newcauchy}) to zero when
$p$ is large, as is done in the tOPE approach, since, were the exact
form of $\Pi (z)$ to have been used on the theory side of the FESR
in question, non-perturbative contributions to the $(z/s_0)^n$-weighted
part of that contour integral would not be zero.
When the power $n$ is small, however, one may legitimately
truncate the logarithmic corrections $L^{(1)}_D$ in the Wilson coefficients
at low dimension, as corresponds to an asymptotic expansion, after
including the non-logarithmic terms $C^{(0)}_D$ to match the
left-hand side of the FESR. The bottom line is that one should
restrict oneself to FESRs involving weights with low degree $n$ and
include, \emph{not} the infinite set of logarithmic corrections
in Eq.~(\ref{newcauchy}), but only those up to the dimension
of the retained leading contributions, $D=2(n+1)$.
The question that remains to be answered is whether in a realistic
situation not including any logarithmic correction at all is a sufficiently
good approximation. This, of course, depends on the numerical
size of these corrections, to which we now turn.

In Sec.~4.1 of their article, the authors of Ref.~\cite{Pich:2022tca}
decided to employ the following estimate
\beq\label{eq:CD-assumption-PRS}
|L^{(1)}_D| \approx 0.2 \qquad \mbox{(Ref.~\cite{Pich:2022tca})},
\eeq
claiming to show that this value would invalidate any
analysis based on the DV strategy. Let us consider how reliable
this estimate is.

Quantitative knowledge of the $L^{(1)}_D$ coefficients is limited to
$D=2,\, 4$ and $6$. Since the (perturbative) $D=2$ contributions
proportional to $m_{u,d}^2$ are already known to be tiny, the first case
to consider is that of the $D=4$ condensates. Here it turns out that both
the numerically dominant gluon condensate and subleading quark condensate
terms have logarithmic corrections suppressed by at least two powers of
$\alpha_s$. In the case of the gluon condensate contribution, one
finds\cite{Braaten:1991qm}
\beq
\lbl{gluon condensate}
L^{(1)}_{D=4}=\frac{11}{8}\left(\frac{\alpha_s(s_0)}{\pi}\right)^2
\simeq 0.012
\eeq
where we have taken $\alpha_s(s_0)\simeq 0.3$. Because of the
smallness of the logarithmic term --- and given the accepted
range of values for the gluon condensate~\cite{Beneke:2008ad} --- it
is safe to neglect the gluon condensate contribution as long as
the weight function used in the FESR does not contain the monomial
$x$. This was verified explicitly in Ref.~\cite{Beneke:2008ad} in
the context of the kinematic moment. A quantitative investigation
of the gluon-condensate logarithmic contribution in the case of
FESRs with $w=1$ shows that it is always at least one order of
magnitude smaller than the last included term in perturbation
theory. It is, therefore, also completely safe to ignore it in
the analyses of Ref.~\cite{Boito:2020xli}.

We now turn to an investigation of the $D=6$ OPE contribution,
where the leading contribution to $\mathcal{O}(\alpha_s)$ is
given by that of four-quark condensates. Since these
four-quark condensates are not very precisely known, we will
estimate them using factorization, based on the large $N_c$
expansion in the $SU(3)$ limit.
In this approximation, $u,d$ and $s$ quarks share a common
quark condensate value $\langle\bar{q}q\rangle$ and the
$D=6$ contribution to the OPE of $\Pi_V^{(1+0)}$ becomes ({\it cf}.
Appendix \ref{app:dim-6 condensates})
\beq\label{eq:log-d-6}
(-s)^3\Pi_V^{(1+0)}(s)\Big|_{D=6} \simeq -\frac{28}{9}\pi
\,\rho\, \alpha_s(s_0) \langle \bar qq \rangle^2\left( 1-\frac{19}{63}
\left(\frac{\alpha_s(s_0)}{\pi}\right)\log\left(\frac{-s}{s_0}\right) \right).
\eeq
The factor $\rho$ here parametrizes possible deviations from the
factorization approximation. From the result (\ref{eq:log-d-6}), we
read off that
\beq
L^{(1)}_{D=6}=-\frac{19}{63}\left(\frac{\alpha_s(s_0)}{\pi}\right)\simeq 0.03,
\eeq
again an order of magnitude smaller than the estimate
(\ref{eq:CD-assumption-PRS}). This value of $L^{(1)}_{D=6}$ is small
enough not to affect the results found with the DV strategy in
Ref.~\cite{Boito:2020xli}.

To obtain a numerical estimate of the impact of $D=6$ logarithmic
correction we recall that $C_{6,V}= -\frac{28}{9}\pi\, \rho\, \alpha_s(s_0)
\langle \bar qq \rangle^2 \simeq -0.7\times 10^{-2}
~\mathrm{GeV}^6$, as extracted from the fits in Ref.~\cite{Boito:2020xli}.
An illustration of the impact of this correction is provided by
the resulting $D=6$ contribution to the theory side of the $w_0$ FESR
which, in the absence of this correction, had no $D=6$ term. With the
input above, we obtain, taking $n=0$ and $p=3$ in Eq.~(\ref{newcauchy}),
\beq
\lbl{dim6}
\delta^{(6)}_{w_0}(s_0)\simeq \frac{38}{63} \pi
\alpha_s(s_0) \frac{C_{6,V}}{s_0^3} = 1\times 10^{-3},
\eeq
where $\delta^{(6)}_{w_0}(s_0)$ is the correction appearing in
\beq\lbl{Ith-parametrization}
I_{V;{\rm th}}^{w}(s_0)=\frac{1}{4\pi^2}\left[ \delta^{(0)}_w(s_0)
+\sum_{D=4,6,8...}\delta^{(D)}_{w}(s_0)
+\delta^{\rm (DV)}_{w}(s_0)\right]\ ,
\eeq
with $\delta^{(0)}_w(s_0)$ and $\delta^{\rm (DV)}_w(s_0)$ the
corresponding perturbative and DV corrections, and
the numerical result in Eq.~(\ref{dim6}) obtained using
$\alpha_s\simeq 0.3$ and $s_0=s_0^{\rm min}=1.55\, \mathrm{GeV}^2$,
the lowest $s_0$ in the analysis. This result is about five times
smaller than the estimated $\alpha_s^5$ term in the perturbative contribution
of Eqs.~(\ref{pertth}) and we have checked that it does not affect the
results for $\alpha_s$ within the errors quoted in Ref.~\cite{Boito:2020xli}.

In principle OPE terms with $D>6$ will also have logarithmic
corrections, but little is known about the size of the associated
$L^{(1)}_{D}$ coefficients. This is why in Ref.~\cite{Boito:2020xli}
the results for OPE contributions obtained from FESR analyses
involving different weight combinations were cross-checked for
self-consistency: while in the absence of logarithmic corrections, the
$w_0$ FESR receives no condensate contributions, the $w_2$
FESR receives a $D=6$ contribution, the $w_3$ FESR $D=6$ and $8$
contributions and the $w_4$ FESR $D=6$ and $10$ contributions. Three
of the four FESRs considered in Ref.~\cite{Boito:2020xli} are thus
sensitive to $C_{6,V}$, while all four are sensitive to $\alpha_s$.
Excellent consistency was found for the values of $\alpha_s$ and
$C_{6,V}$ obtained from fits to different sets of these FESRs.

Based on the available knowledge about the logarithmic corrections in OPE
condensates of with $D=4$ and $6$, and given the consistency checks already
performed in our previous analyses, we conclude that the neglect of those
small, $\alpha_s$-suppressed, logarithmic contributions is justified and
should not affect the $\alpha_s$ results obtained with the DV strategy.
The exercise we performed also suggests that the estimate employed in
Ref.~\cite{Pich:2022tca} for $L_D^{(1)}$ is likely to significantly
overestimate the size of logarithmic contributions. If correct, logarithmic
corrections of this size would potentially affect {\it any} of the existing
analyses of $\alpha_s$ from $\tau\rightarrow ({\rm hadrons})+\nu_\tau$.

\section{\label{app:dim-6 condensates} Contribution from the
dimension-six quark condensates}

Following Ref.~\cite{Braaten:1991qm}, the list of dimension-six four-quark
condensates contributing to $L^{(1)}_{D=6}$ to $\mathcal{O}(\alpha_s)$ is
\begin{align}
O_1&=\bar{u}\gamma_\m \gamma_5 T^a d\ \bar{d}\gamma^\m \gamma_5 T^a u\, ,
\nn \\
O_2& =\bar{u}\gamma_\m T^a d \ \bar{d}\gamma^\m T^a u\, , \nn \\
O_3&=\bar{u}\gamma_\m d\ \bar{d}\gamma^\m u\, , \nn \\
O_4&=(\bar{u}\gamma_\m T^a u+ \bar{d}\gamma_\m T^a d) (\bar{u}\gamma^\m
T^a u+ \bar{d}\gamma^\m T^a u + \bar{s}\gamma^\m T^a s)\, , \nn \\
O_5&=(\bar{u}\gamma_\m \gamma_5 T^a u+ \bar{d}\gamma_\m \gamma_5 T^a d)
(\bar{u}\gamma^\m \gamma_5 T^a u+ \bar{d}\gamma^\m \gamma_5 T^a u +
\bar{s}\gamma^\m \gamma_5 T^a s)\, , \nn \\
O_6&=(\bar{u}\gamma_\m \gamma_5 u+ \bar{d}\gamma_\m \gamma_5 d)
(\bar{u}\gamma^\m \gamma_5 u+ \bar{d}\gamma^\m \gamma_5 u +
\bar{s}\gamma^\m \gamma_5 s)\, , \nn \\
O_7&=(\bar{u}\gamma_\m T^a u+ \bar{d}\gamma_\m T^a d +\bar{s}\gamma_\m
T^a s) (\bar{u}\gamma^\m T^a u+ \bar{d}\gamma^\m T^a u +
\bar{s}\gamma^\m T^a s)\, ,
\end{align}
where $T^a $ are color $SU(3)$ generators with the normalization
$\sum_a T^a_{\a \b} T^a_{\g \d}=\delta_{\a \d} \d_{\b \g}/2
+ \mathcal{O}(1/N_c)$. In the large-$N_c$ approximation, these
operators can be expressed, after Fierzing, in terms of
$\langle\bar{q}q\rangle$ condensates with the result:
\begin{eqnarray}
\langle O_1\rangle &=&\frac{1}{2} \langle\bar{u}u\rangle
\langle\bar{d}d\rangle\ ,\ \langle O_2\rangle=-\frac{1}{2}
\langle\bar{u}u\rangle \langle\bar{d}d\rangle \ , \ \langle
O_4\rangle=-\frac{1}{2} \langle\bar{u}u\rangle^2-\frac{1}{2}
\langle\bar{d}d\rangle^2 \nonumber\\
\langle O_5\rangle &=& \frac{1}{2} \langle\bar{u}u\rangle^2
+ \frac{1}{2} \langle\bar{d}d\rangle^2\ ,\ \langle O_7\rangle
=-\frac{1}{2} \langle\bar{u}u\rangle^2 -\frac{1}{2}\langle\bar{d}d\rangle^2
- \frac{1}{2}\langle\bar{s}s\rangle^2 ,
\end{eqnarray}
with $ \langle O_3\rangle= \langle O_6\rangle=0$. These condensates give
the following contribution to the vector two-point correlator
\begin{eqnarray}
(-s)^3\Pi_{V; ud}^{(1+0)} (s)&=& -8 \p^2\left[1-
\frac{9}{8}\left(\frac{\a_s}{\p}\right) L\right]
\left(\frac{\a_s}{\p}\right)\langle O_1\rangle+
5 \pi^2 L \left(\frac{\a_s}{\p}\right)^2 \langle O_2\rangle \nonumber\\
&&-\frac{8\p^2}{9}\left[1- \frac{95}{72} L
\left(\frac{\a_s}{\p}\right)\right] \left(\frac{\a_s}{\p}\right)
\langle O_4\rangle + \frac{5 \p^2}{9} L \left(\frac{\a_s}{\p}\right)^2
\langle O_5\rangle \nonumber\\
&& +\frac{24\p^2}{81} L \left(\frac{\a_s}{\p}\right)^2 \langle O_7\rangle,
\end{eqnarray}
with $L=\log\left( -s/s_0\right), \a_s\equiv \a_s(s_0)$ and
$ \langle O_i\rangle\equiv \langle O_i\rangle(s_0)$. A further
simplification may be obtained in the $SU(3)$ limit where
$\langle\bar{u}u\rangle =\langle\bar{d}d\rangle
=\langle\bar{s}s\rangle\equiv \langle\bar{q}q\rangle$,
\begin{equation}
(-s)^3 \Pi_{V; ud}^{(1+0)} (s)=- \frac{28}{9}\p \,
\rho\, \a_s \langle\bar{q}q\rangle^2
\left( 1- \frac{19}{63}\a_s L \right),
\end{equation}
where $\rho$ is a factor parametrizing the possible deviations from
the large-$N_c$ and $SU(3)$ limits. This is the result quoted in
Eq.~(\ref{eq:log-d-6}).

\section{Sensitivity of \texorpdfstring{\boldmath $\alpha_s$}{alpha_s}
to corrections in the DV parametrization}
\label{sec:DV-parm}

One of the core arguments of Ref.~\cite{Pich:2022tca} against the DV
strategy for the extraction of $\alpha_s$ from
$\tau\rightarrow {\rm hadrons}+\nu_\tau$ is an
alleged strong dependence of the final results on details of the
DV parametrization, Eq.~(\ref{DV-parametrization}). An investigation
of this argument must address two critical issues. First, any
modifications of the parametrization used must be physically
reasonable, in light of what is known about the QCD spectrum.
Second, it is important to confront those modified parametrizations with
the best available data set, taking into account all uncertainties
and correlations. To simplify the analysis, we focus
on the $V$ channel, where fewer DV parameters are required than
for the $V+A$ combination, using the new $V$ spectral function
of \rcite{Boito:2020xli}, since it is more precise than the earlier
ALEPH~\cite{Davier:2013sfa} version.

In Ref.~\cite{Pich:2022tca}, as in Ref.~\cite{Pich:2016bdg}, it was
assumed that, in testing the sensitivity of the extracted $\alpha_s$
to the assumed DV form, one was free to multiply
Eq.~(\ref{DV-parametrization}) by a factor $s^n$ (with $n>0$),
although general observations about transseries and the concrete
result in Eq.~(\ref{DV-parametrization-gen1}) do not support such
modifications of the DV parametrization. Using the ALEPH data, the
authors found that the $p$-values of their fits increased continuously
with increasing $n$. Choosing to stop at $n=8$, they obtained a
value $\alpha_s(m_\tau^2)=0.314$ from a fit with minimum $s_0$ value
$s_0^{\rm min}=1.55$ GeV$^2$. This result, when compared to that
obtained with the ALEPH $V+A$ input and unmodified DV form used in
Ref.~\cite{Boito:2014sta}, $\alpha_s(m_\tau^2)=0.298(10)$~\cite{Pich:2022tca},
was considered by these authors as a proof of the existence of
an unaccounted-for systematic in the DV strategy approach.\footnote{This
criterion is somewhat surprising, given that the same authors
characterize the considerably larger $0.348-0.314=0.034$ difference in
CIPT-based $\alpha_s$ results reported in Table 2 of Ref.~\cite{Pich:2022tca}
as showing ``amazing stability.''} In fact, the $\alpha_s$ statistical
error, not quoted in Ref.~\cite{Pich:2022tca}, for their fit result
with the $s^8$ factor in front of Eq.~(\ref{DV-parametrization}) is
$\pm 0.015$ (see Ref.~\cite{Pich:2016bdg}), which is too large to allow
a definite conclusion to be drawn. Furthermore, the pattern of
$p$-value behavior in these fits is reversed when the new, more precise,
$V$ spectral function of Ref.~\cite{Boito:2020xli} is used in similar fits.
The more precise data, therefore, also disfavors the multiplication of the
DV parametrization by positive powers of $s$, independent of (and confirming)
the fact that the form used for the modification is theoretically
unmotivated.

A different case is the inclusion of the $1/s$ correction present
in (\ref{DV-parametrization-gen1}). This correction, which is in
principle allowed, requires the introduction a new parameter, $c$,
and produces the generalized DV form
\beq
\lbl{DV-v2}
\r_{V}^{\rm DV}(s;c)=\left(1+\frac{c}{s}
\right)e^{-\d_{V}-\g_{V} s}\sin\left(\a_{V}+\b_{V} s \right)\ .
\eeq
We use this form to study how $\alpha_s$ changes in the presence of a non-zero
value for $c$.\footnote{The full, finite-$s$ form of the additional
sub-leading $s\rightarrow\infty$ correction written schematically as
$\mathcal{O}(1/\log(s))$ in Eq.~(\ref{DV-parametrization-gen1}) is too
slowly varying with $s$ to be visible in the data.}

Ref.~\cite{Pich:2022tca} reports on results using the DV strategy to
fit the ALEPH vector~\cite{Davier:2013sfa} data with two very specific
values of $c$: namely $-1.35$~GeV$^2$ and $-2.00$~GeV$^2$, using the
weight function $w(s)=1$. Note that, for such values of $c$,
the ``correction'' is, in fact, comparable in magnitude (and
of opposite sign) to the quantity it is nominally ``correcting'' and
hence dramatically alters the $s$-dependence of
$\r_{V}^{\rm DV}$ in the region of $s$ used in the fits.
We have repeated this exercise, with these large negative $c$ values, and
found numerical agreement for the central $\alpha_s$ values obtained in
these. The results, including now the uncertainties not quoted in
Ref.~\cite{Pich:2022tca}, are: $\alpha_s(m_\tau^2)=0.319\pm 0.016$
for $c=-1.35$~GeV$^2$ and $\alpha_s(m_\tau^2)=0.260\pm 0.089$ for
$c=-2.00$~GeV$^2$. Note that the statistical uncertainty in the value
of $\alpha_s(m_\tau^2)$ with $c=-2.00$~GeV$^2$ is extremely large.
Taking these errors into account, it becomes clear that the
ALEPH data is, in fact, not sufficiently precise to test the
inclusion of subleading terms of the type of Eq.~(\ref{DV-v2}).

\begin{figure}[!t]
\centering
\includegraphics[width=0.49\textwidth]{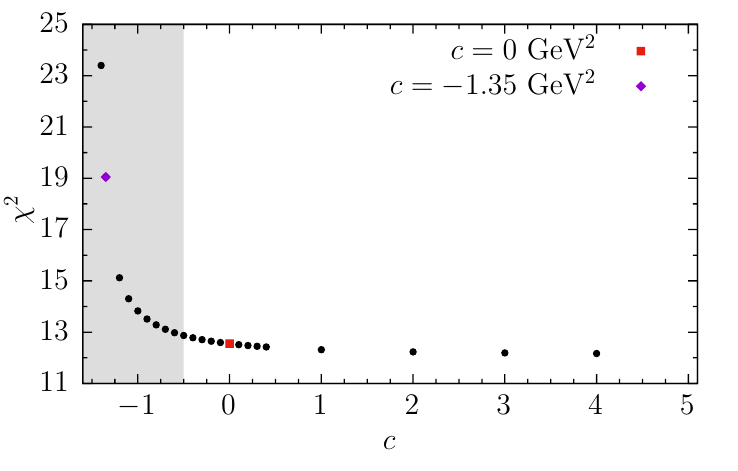}
\includegraphics[width=0.49\textwidth]{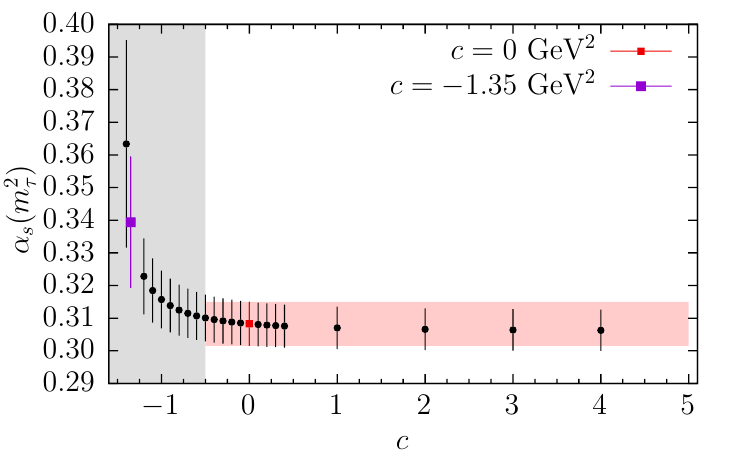}
\caption{Results for fits to the improved $V$ spectral
function of Ref.~\cite{Boito:2020xli}, with $w(x)=1$, using 20 $s_0$
values (15 degrees of freedom) and with the DV parametrization of
Eq.~(\ref{DV-v2}), as a function of $c$. Results using the unmodified
model ($c=0$), obtained in Ref.~\cite{Boito:2020xli}, are shown in red.
On the left-hand side the $\chi^2$ values are shown while on the
right-hand side we show the $\alpha_s(m_\tau^2)$ values with the
statistical uncertainty. In gray, we highlight the region where the
$\chi^2$ has a steep increase.}
\label{fig:c-parm-res}
\end{figure}

The improved $V$ spectral function of Ref.~\cite{Boito:2020xli} once more
proves useful here, making possible an updated version of this exploration
with significantly reduced errors. Fitting for $c$ is not practical because,
even with the improved data, there are flat directions in the $\chi^2$
landscape, preventing the fit from disentangling all of the DV parameters.

It is, however, easy to scan over $c$, including many more values than
the two considered in Ref.~\cite{Pich:2022tca}. We have performed fits
with the weight function $w(x)=1$, using $20$ $s_0$ values ranging from
$1.55$~GeV$^2$ to $3.06$~GeV$^2$, fixing $c$ in each of the fits. In the
left-hand panel of Fig.~\ref{fig:c-parm-res}, we show the $\chi^2$ values
for these fits as a function of the parameter $c$. A sharp increase in
$\chi^2$ is evident for large negative values of $c$, such as those
considered in Ref.~\cite{Pich:2022tca}. Combined with the observation
that such $c$ produce, rather than corrections, strong modifications of
the leading asymptotic form, this rapid $\chi^2$ increase clearly disfavors
such large negative $c$.{\footnote{In fact, in contrast to what was
seen in Ref.~\cite{Pich:2022tca}, where the larger errors of the ALEPH
$V$-channel data employed were found to permit successful fits at larger
negative $c$, our study, using the significantly reduced-error $V$-channel
input of Ref.~\cite{Boito:2020xli}, was unable to find meaningful
fits for $c< -1.45\ {\rm GeV}^2$.}} The associated larger values of
$\alpha_s$, shown in the right-hand panel of Fig.~\ref{fig:c-parm-res},
are similarly disfavored and do not constitute an instability (notice
also their larger uncertainties). For $c>0$ one obtains a slightly
better fit than with $c=0$, but the values of $\alpha_s$, as shown
in the right-hand panel of Fig.~\ref{fig:c-parm-res}, show excellent
stability for $c>0$ and are completely compatible with the result
obtained for $c=0$, which is the central result of the analysis of
Ref.~\cite{Boito:2020xli} (shown in red in the figure, and with a red
band in the right-hand panel). The results are stable for such
$c$ even when $c$ is large and the ``correction'' it produces to
the leading behavior dictated by Eq.~(\ref{DV-v2}) is no longer small.
Such a large $c>0$ could affect, {\it e.g.}, the extracted values for
the DV parameters but, as Fig.~\ref{fig:c-parm-res} shows, the impact
on $\alpha_s$ is very small.
The shift in the central value of $\alpha_s$ when comparing results for
$c=0$ and $c=2$, for example, is about 25\% of the statistical uncertainty
and would not be noticeable in the final results of our original
analysis~\cite{Boito:2020xli}.

The argument disfavoring large negative $c$ to this point is based
on a comparison of single-weight $w_0$-FESR-fit $\chi^2$ results only.
Considering also the $w_2$ FESR over the same range of $s_0$ is expected
to yield additional constraints, as per the discussion in
Sec.~\ref{sec:supposedtautology}. Consistency of the $w_0$ and $w_2$
FESRs can be tested by looking, for example, at (i) the ``restricted $w_2$
$\chi^2$'' produced by minimizing the $w_2$ $\chi^2$ with respect to
$C_{6V}$ only, holding $\alpha_s$ and the DV parameters fixed at their
single-weight $w_0$ fit values, or (ii) the $s_0$ dependence of the
rescaled $w_2$ residual,
\begin{equation}
R(s_0)\equiv s_0^3\left[I_{\rm th;(D=0+DV)}^{w_2}(s_0)\vert_{w_0}
-I^{w_2}(s_0)\right]\, ,
\end{equation}
where $I_{\rm th;(D=0+DV)}^{w_2}(s_0)\vert_{w_0}$ denotes the sum of $D=0$
and DV contributions to the $w_2$ FESR evaluated using $w_0$-fit results for
$\alpha_s$ and the DV parameters. $R(s_0)$ should be equal to $C_{6V}$,
independent of $s_0$, if the DV model employed is physically reasonable,
producing differences $\Delta R=R(s_0^{(1)})-R(s_0^{(2)})$ compatible with
zero within errors for pairs of $s_0$ values in the $s_0$ fit window. The
``restricted $w_2$ $\chi^2$'' is found to (i) start out at $14.7$ for $c=0$,
(ii) increase slowly for increasingly negative, but still small, $c$,
(iii) show a rapid acceleration in that increase beginning around $c=-1.0$
GeV$^2$, and (iv) reach $45.8$ at $c=-1.4$ GeV$^2$. The shortcomings of
the modified DV model for values of $c$ as large and negative as those
considered in Ref.~\cite{Pich:2022tca}, not surprisingly, also show up
in the associated $\Delta R$, which start out $\lesssim 1\sigma$ from
zero on average for $c=0$, grow monotonically as $c$ becomes more negative,
and reach values differing from zero by as much as $\sim 5\sigma$ for
$c=-1.4$ GeV$^2$.

Given (i) that large negative $c$ are disfavored by data, (ii) that
the results of our original analysis for $\alpha_s$, Ref.~\cite{Boito:2020xli},
are stable against the inclusion of sub-leading corrections proportional to
$c$ for other $c$ and (iii) that an (arbitrary) multiplication of the DV
parametrization by positive powers of $s$ is unsupported by more precise
experimental data, the conclusions drawn in Sec. 4.3 of
Ref.~\cite{Pich:2022tca} are seen to now no longer hold.

\section{\label{app:redundancyproof} Redundancy of the tOPE}

Let $d_i$, $i=1,\dots,n+1$ be a set of data points, $t_i(\h_\a)$ and $t_{n+1}(\eta_\alpha,\eta)$
theory representations depending on parameters $\h_\a$, $\a=1,\dots,A$,
and an additional parameter $\h$, with the dependence on $\h$ appearing only in $t_{n+1}$, and $C$ the data covariance matrix.
We denote the upper left-hand $n\times n$ block of $C$ by $D$, and
we define
\beq
\lbl{chi2n}
\c^2_n=
\sum_{i,j=1}^n(d_i-t_i(\h_\a))D^{-1}_{i,j} (d_j-t_j(\h_\a))\ ,
\eeq
and
\bqry
\lbl{chi2np1}
\c^2_{n+1}&=&\sum_{i,j=1}^n(d_i-t_i(\h_\a))C^{-1}_{i,j}
(d_j-t_j(\h_\a))\\
&&+2\sum_{i=1}^n(d_i-t_i(\h_\a))C^{-1}_{i,n+1}
(d_{n+1}-t_{n+1}(\h_\a,\h))\nonumber\\
&&+
(d_{n+1}-t_{n+1}(\h_\a,\h))C^{-1}_{n+1,n+1}(d_{n+1}-
t_{n+1}(\h_\a,\h))\ .\nonumber
\eqry
Note that only $t_{n+1}$ depends on the extra parameter $\h$, the other $t_i$
do not. We then minimize $\c^2_{n+1}$ with respect to $\h$, making use of this:
\beq
\lbl{minchi2}
-\frac{\partial\c^2_{n+1}}{\partial\h}=\left(\sum_{i=1}^n(d_i-t_i)
C^{-1}_{i,n+1}
+(d_{n+1}-t_{n+1})C^{-1}_{n+1,n+1}\right)\frac{\partial
t_{n+1}}{\partial\h}=0\ ,
\eeq
which implies that
\beq
\lbl{minchi2cons}
\sum_{i=1}^n(d_i-t_i(\h_\a))C^{-1}_{i,n+1}
+(d_{n+1}-t_{n+1}(\h_\a,\h))C^{-1}_{n+1,n+1}=0\ .
\eeq
Substituting this into \seEq{chi2np1} yields
\bqry
\lbl{chi2np1sub}
\c^2_{n+1}&=&\sum_{i,j=1}^n(d_i-t_i(\h_\a))C^{-1}_{i,j}
(d_j-t_j(\h_\a))\\
&&-\sum_{i,j=1}^n(d_i-t_i(\h_\a))C^{-1}_{i,n+1}\,\frac{1}
{C^{-1}_{n+1,n+1}}\,C^{-1}_{n+1,j}(d_j-t_j(\h_\a))\ .\nonumber
\eqry
Note that $C^{-1}_{i,j}$ is always the $(i,j)$ element of the inverse,
while $\frac{1}{C^{-1}_{n+1,n+1}}$ is (obviously) the inverse of
the $(n+1,n+1)$ element of the inverse of $C$. Equation\seneq{chi2np1sub}
is equal to $\c^2_n$ defined in \seEq{chi2n} if
\beq
\lbl{D}
D^{-1}_{i,j}=C^{-1}_{i,j}-C^{-1}_{i,n+1}\,\frac{1}{C^{-1}_{n+1,n+1}}\,
C^{-1}_{n+1,j}\ .
\eeq
This is, indeed, the case. We first diagonalize
$D$. Then $C$ can be written as
\beq
\lbl{Cpdiag}
C=
\left(\begin{array}{ccccc}
\l_1 & 0 & 0 & \dots & a_1 \\
0 &\l_2 & 0 & \dots & a_2\\
0 &0 &\l_3 & \dots & a_3\\
. & & & & \\
. & & & & \\
a_1 &a_2 & a_3& \dots & a_{n+1}\\
\end{array}\right)\ ,
\eeq
with $\lambda_i$ the eigenvalues of $D$.
Denoting the components of the inverse $C^{-1}_{i,j}=b_{i,j}$, we have the
following set of equations determining the $b_{i,j}$ (we use throughout
that $b_{i,j}=b_{j,i}$):
\bqry
\lbl{bs}
&&\l_i b_{i,j}+a_i b_{n+1,j}=\d_{ij}\ ,\qquad i=1,\dots,n\ ,\\
&&\l_i b_{i,n+1}+a_i b_{n+1,n+1}=0\ ,\qquad i=1,\dots,n\ ,\nonumber\\
&&\sum_{i=1}^n a_i b_{i,j}+a_{n+1}b_{n+1,j}=0\ ,\nonumber\\
&&\sum_{i=1}^n a_i b_{i,n+1}+a_{n+1}b_{n+1,n+1}=1\ .\nonumber
\eqry
The second of these equations implies that
\beq
\lbl{2nd}
b_{i,n+1}=-\frac{a_i}{\l_i}\,b_{n+1,n+1}\ ,
\eeq
and thus that
\beq
\lbl{2ndterm}
\frac{b_{i,n+1}b_{n+1,j}}{b_{n+1,n+1}}=\frac{a_i a_j}{\l_i\l_j}\,b_{n+1,n+1} .
\eeq
The first of Eqs.~(\ref{bs}) gives
\beq
\lbl{1st}
b_{i,j}=\frac{1}{\l_i}\left(\d_{ij}-a_i b_{n+1,j}\right)
=\frac{1}{\l_i}\,\d_{ij}+\frac{a_i}{\l_i}\frac{a_j}{\l_j}b_{n+1,n+1}\ ,
\eeq
where in the second step we used \seEq{2nd}. Subtracting \seEq{2ndterm} from
\seEq{1st} we find that
\beq
\lbl{res}
C^{-1}_{i,j}-C^{-1}_{i,n+1}\,\frac{1}{C^{-1}_{n+1,n+1}}\,C^{-1}_{n+1,j}
=b_{i,j}-\frac{b_{i,n+1}b_{n+1,j}}{b_{n+1,n+1}}=\frac{1}{\l_i}\,\d_{ij}\ ,
\eeq
which is precisely the inverse of $D$.

From \seEq{minchi2cons} we find, using \seEq{2nd}, that
\beq
\lbl{tnp1}
t_{n+1}(\h_\a,\h) =d_{n+1}- \sum_{i=1}^n\frac{a_i}{\l_i}(d_i-t_i(\h_\a))\ .
\eeq
This gives a solution for the parameter $\h$ once $\c_n^2$ has been minimized
to find the other parameters $\h_\a$.

\section{\label{app:aleph2013details} Details of the 2013 ALEPH
isovector \boldmath \texorpdfstring{$V+A$}{V+A} input}
With all spin-$0$, non-pion-pole contributions to the experimental
hadronic $\tau$ decay spectral distribution numerically negligible,
Eqs.~(\ref{R}) and (\ref{taukinspectral}) imply the following relation
between the continuum (non-pion-pole) isovector $V+A$ isovector
spectral function, $\rho_{V+A;ud}^{1+0;\rm cont}(s)$, and the corresponding
differential branching fraction distribution, $dB_{V+A;ud}^{\rm cont}/ds$
\begin{equation}
\rho_{V+A;ud}^{\rm 1+0;cont}(s)={\frac{m_\tau^2}{12\pi^2\vert V_{ud}\vert^2
S_{\rm EW} B_e w_T(s;m_\tau^2)}}\, {\frac{dB_{V+A;ud}^{\rm cont}}{ds}}\ ,
\label{rhofromdBds}\end{equation}
where $B_e$ is the $\tau^- \rightarrow e^-\nu_\tau\bar{\nu}_e$
branching fraction. The publicly accessible 2013 ALEPH experimental
distribution comes in the form of the values and covariances of
$\Delta B^{V+A;ud}_k$, the contributions to the differential
non-strange hadronic branching fraction, where $k$ labels the
(variable width) ALEPH bins and the normalization is such that
the $\Delta B^{V+A;ud}_k$ sum to the central value of the
then-current result for the continuum non-strange branching fraction,
$B_{V+A;ud}^{\rm cont}=0.51116$. That branching
fraction is related to the single-prong electron, muon and $\pi$ decay
branching fractions, $B_e$, $B_\mu$ and $B_\pi$, and the inclusive
strange decay branching fraction, $B_{V+A;us}$, by
\begin{equation}
B_{V+A;ud}^{\rm cont}=1-B_e-B_\mu-B_{V+A;us}-B_\pi\, .
\label{contudbfdefn}\end{equation}
We use this relation, with current input, to update the normalization
of the 2013 ALEPH distribution. Since the analyses we consider
are purely Standard Model (SM) in nature, we take $B_\mu/B_e$ to have its
SM value and $B_\pi$ to be given by SM expectations based on the result for
$\vert V_{ud}\vert f_\pi$ obtained from the $\pi_{\mu 2}$ decay
width. $B_e$ is, for similar reasons, taken to be the most recent
SM lepton-universality-improved 2022 HFLAV result~\cite{HFLAV:2022esi},
$B_e=0.17812(22)$. Finally, for $B_{V+A;us}$ we take the 2022 HFLAV
value, modified by replacing the less precise experimental
$\tau^-\rightarrow K^-\nu_\tau$ branching fraction with its
SM expectation based on the result for $\vert V_{us}\vert f_K$ implied
by the $K_{\mu 2}$ decay width. This input produces an updated central
value $B_{V+A;ud}^{\rm cont}=0.51097$, and hence a rescaling of $0.99962$
for the ALEPH $\Delta B^{V+A;ud}_k$ distribution and the square of this
factor for the associated covariances. This rescaled distribution is
converted to the
corresponding $dB_{V+A;ud}^{\rm cont}/ds$ distribution using the 2013
ALEPH bin widths, and thence to the continuum isovector $V+A$ spectral
function using the values $B_e=0.17812(22)$~\cite{HFLAV:2022esi},
$m_\tau = 1.77686(12)$ GeV~\cite{ParticleDataGroup:2022pth},
$S_{\rm EW}=1.0201(3)$\cite{Erler:2002mv} and
$\vert V_{ud}\vert = 0.97373(31)$~\cite{ParticleDataGroup:2022pth}
for the external inputs required in Eq.~(\ref{rhofromdBds}). The
value of $f_\pi$ corresponding to this choice of $\vert V_{ud}\vert$,
used in determining the pion-pole contributions to the various
$V+A$ spectral integrals, is $0.09216$ GeV.

\section{\label{app:improvedVchannel} Further evidence against
tOPE assumptions: Optimal moments and the improved V channel}

As a final investigation of the implications of the redundancy
observation for the single-$s_0$ tOPE approach, we consider an
optimal-weight analysis of the isovector $V$ channel, using as
experimental input the recently improved $V$ spectral function
obtained in Ref.~\cite{Boito:2020xli}. The optimal-weight analysis
is chosen because of the somewhat expanded set of self-consistency
checks it provides. We remind the reader that $\chi^2/{\rm dof}$ values
for $V$ channel, ALEPH-data-based, single-$s_0$, classic $km$
spectral weight tOPE fits reported in the literature are at least
as good as those of the corresponding $V+A$ channel fits, with
Ref.~\cite{Pich:2016bdg}, for example, quoting $\chi^2/{\rm dof}\sim 0.4$
and $\sim 2.4$ for the $V$ and $V+A$ channels, respectively.
One should, however, bear in mind that in tOPE analyses with the classic
$km$ spectral weights, which are sensitive to the $D=4$ gluon condensate,
some tension is typically observed between results for the gluon condensate
obtained from separate $V$ and $V+A$ fits. A tension between $V$- and
$V+A$-channel gluon condensate results would not be unexpected were the
analyses to suffer from channel-dependent single-$s_0$ tOPE contamination
of the (redundantly determined) nominal NP condensates. It is thus of
interest to revisit previous single-$s_0$ $V$-channel analyses, taking
advantage of (i) the significant reduction in spectral function errors
in the upper part of the spectrum produced by the analysis of
Ref.~\cite{Boito:2020xli}, and (ii) the improved understanding, provided
by the redundancy discussion above, of the self-consistency tests available
to single-$s_0$ tOPE fits. Because the binning of the $V$ spectral data
differs from that of the ALEPH $V+A$ distribution, it is not possible
to exactly match the choice, $s_0=2.8\ {\rm GeV}^2$, considered in the
$V+A$ discussion above. We thus choose, for our $V$ channel analysis,
$s_0=2.8821\ {\rm GeV}^2$, the first $s$ above $2.8\ {\rm GeV}^2$
in the $V$-channel data compilation of Ref.\cite{Boito:2020xli}.
The results of the single-weight, $w^{(23)}$, $w^{(24)}$ and
$w^{(25)}$ $\alpha_s$ determinations,
\begin{eqnarray}
\left[\alpha_s(m_\tau^2)\right]_{w^{(23)}}&&=0.3188(43)_{{\rm ex}}\ ,
\nonumber\\
\left[\alpha_s(m_\tau^2)\right]_{w^{(24)}}&&=0.3094(49)_{{\rm ex}}\ ,
\nonumber\\
\left[\alpha_s(m_\tau^2)\right]_{w^{(25)}}&&=0.3046(54)_{{\rm ex}}\ ,
\label{udvecoptwtsinglewtalphas}\end{eqnarray}
have discrepancies,
\begin{eqnarray}
\left[\alpha_s(m_\tau^2)\right]_{w^{(23)}}
-\left[\alpha_s(m_\tau^2)\right]_{w^{(24)}}&&=0.00944(86)_{{\rm ex}}\ ,
\nonumber\\
\left[\alpha_s(m_\tau^2)\right]_{w^{(23)}}
-\left[\alpha_s(m_\tau^2)\right]_{w^{(25)}}&&=0.01422(162)_{{\rm ex}}\ ,
\nonumber\\
\left[\alpha_s(m_\tau^2)\right]_{w^{(24)}}
-\left[\alpha_s(m_\tau^2)\right]_{w^{(25)}}&&=0.00477(78)_{{\rm ex}}\ ,
\label{udvecoptwtsinglewtalphasdiffs}\end{eqnarray}
so large that a sensible combined three-weight fit is impossible,
while even the combined two-weight, $w^{(24)}$ and $w^{(25)}$, fit
is problematic, producing a $\chi^2/{\rm dof}$ of $43.1$. This example
makes clear that standard tOPE assumptions, unconstrained by the
use of an extended multi-$s_0$ analysis, are not generally valid
at scales at or just below the $\tau$ mass.

\bibliographystyle{jhep}
\bibliography{References}

\providecommand{\href}[2]{#2}\begingroup\raggedright\begin{thebibliography}{10}

\bibitem{ParticleDataGroup:2022pth}
{\scshape Particle Data Group} collaboration, \emph{{Review of Particle
  Physics}}, \href{https://doi.org/10.1093/ptep/ptac097}{\emph{PTEP} {\bfseries
  2022} (2022) 083C01}.

\bibitem{Tsai:1971vv}
Y.-S. Tsai, \emph{{Decay Correlations of Heavy Leptons in $e^+ e^- \to \ell^+
  \ell^-$}}, \href{https://doi.org/10.1103/PhysRevD.13.771}{\emph{Phys. Rev. D}
  {\bfseries 4} (1971) 2821}.

\bibitem{Shankar:1977ap}
R.~Shankar, \emph{{Determination of the Quark-Gluon Coupling Constant}},
  \href{https://doi.org/10.1103/PhysRevD.15.755}{\emph{Phys. Rev. D} {\bfseries
  15} (1977) 755}.

\bibitem{Floratos:1978jb}
E.~G. Floratos, S.~Narison and E.~de~Rafael, \emph{{Spectral Function Sum Rules
  in Quantum Chromodynamics. 1. Charged Currents Sector}},
  \href{https://doi.org/10.1016/0550-3213(79)90359-6}{\emph{Nucl. Phys. B}
  {\bfseries 155} (1979) 115}.

\bibitem{Nachtmann:1978zf}
O.~Nachtmann and W.~Wetzel, \emph{{Quantum Chromodynamics and the Decay of the
  $\tau$ Lepton}},
  \href{https://doi.org/10.1016/0370-2693(79)90530-6}{\emph{Phys. Lett. B}
  {\bfseries 81} (1979) 229}.

\bibitem{Narison:1979pf}
S.~Narison and E.~de~Rafael, \emph{{Spectral Function Sum Rules in Quantum
  Chromodynamics. 2. Flavor Components of the Electromagnetic Current}},
  \href{https://doi.org/10.1016/0550-3213(80)90032-2}{\emph{Nucl. Phys. B}
  {\bfseries 169} (1980) 253}.

\bibitem{Krasnikov:1982ea}
N.~V. Krasnikov, A.~A. Pivovarov and N.~N. Tavkhelidze, \emph{{The Use of
  Finite Energy Sum Rules for the Description of the Hadronic Properties of
  QCD}}, \href{https://doi.org/10.1007/BF01577186}{\emph{Z. Phys. C} {\bfseries
  19} (1983) 301}.

\bibitem{Schilcher:1983ae}
K.~Schilcher and M.~D. Tran, \emph{{Duality in Semileptonic $\tau$ Decay}},
  \href{https://doi.org/10.1103/PhysRevD.29.570}{\emph{Phys. Rev. D} {\bfseries
  29} (1984) 570}.

\bibitem{Bertlmann:1984ih}
R.~A. Bertlmann, G.~Launer and E.~de~Rafael, \emph{{Gaussian Sum Rules in
  Quantum Chromodynamics and Local Duality}},
  \href{https://doi.org/10.1016/0550-3213(85)90475-4}{\emph{Nucl. Phys. B}
  {\bfseries 250} (1985) 61}.

\bibitem{Braaten:1988hc}
E.~Braaten, \emph{{QCD Predictions for the Decay of the tau Lepton}},
  \href{https://doi.org/10.1103/PhysRevLett.60.1606}{\emph{Phys. Rev. Lett.}
  {\bfseries 60} (1988) 1606}.

\bibitem{Braaten:1988ea}
E.~Braaten, \emph{{The Perturbative QCD Corrections to the Ratio R for tau
  Decay}}, \href{https://doi.org/10.1103/PhysRevD.39.1458}{\emph{Phys. Rev. D}
  {\bfseries 39} (1989) 1458}.

\bibitem{Narison:1988ni}
S.~Narison and A.~Pich, \emph{{QCD Formulation of the tau Decay and
  Determination of Lambda (MS)}},
  \href{https://doi.org/10.1016/0370-2693(88)90830-1}{\emph{Phys. Lett. B}
  {\bfseries 211} (1988) 183}.

\bibitem{Braaten:1991qm}
E.~Braaten, S.~Narison and A.~Pich, \emph{{QCD analysis of the tau hadronic
  width}}, \href{https://doi.org/10.1016/0550-3213(92)90267-F}{\emph{Nuclear
  Physics B} {\bfseries 373} (1992) 581}.

\bibitem{Poggio:1975af}
E.~C. Poggio, H.~R. Quinn and S.~Weinberg, \emph{{Smearing the Quark Model}},
  \href{https://doi.org/10.1103/PhysRevD.13.1958}{\emph{Phys. Rev. D}
  {\bfseries 13} (1976) 1958}.

\bibitem{Bigi:1998kc}
I.~I.~Y. Bigi, M.~A. Shifman, N.~Uraltsev and A.~I. Vainshtein, \emph{{Heavy
  flavor decays, OPE and duality in two-dimensional 't Hooft model}},
  \href{https://doi.org/10.1103/PhysRevD.59.054011}{\emph{Phys. Rev. D}
  {\bfseries 59} (1999) 054011}
  [\href{https://arxiv.org/abs/hep-ph/9805241}{{\ttfamily hep-ph/9805241}}].

\bibitem{Cata:2005zj}
O.~Cat{\`a}, M.~Golterman and S.~Peris, \emph{{Duality violations and spectral
  sum rules}},
  \href{https://doi.org/10.1088/1126-6708/2005/08/076}{\emph{Journal of High
  Energy Physics} {\bfseries 08} (2005) 076}
  [\href{https://arxiv.org/abs/hep-ph/0506004}{{\ttfamily hep-ph/0506004}}].

\bibitem{Cata:2008ye}
O.~Cat\`a, M.~Golterman and S.~Peris, \emph{{Unraveling duality violations in
  hadronic tau decays}},
  \href{https://doi.org/10.1103/PhysRevD.77.093006}{\emph{Phys. Rev. D}
  {\bfseries 77} (2008) 093006}
  [\href{https://arxiv.org/abs/0803.0246}{{\ttfamily 0803.0246}}].

\bibitem{ALEPH:1997fek}
{\scshape ALEPH} collaboration, \emph{{Measurement of the spectral functions of
  vector current hadronic tau decays}},
  \href{https://doi.org/10.1007/s002880050523}{\emph{Z. Phys. C} {\bfseries 76}
  (1997) 15}.

\bibitem{ALEPH:1998rgl}
{\scshape ALEPH} collaboration, \emph{{Measurement of the spectral functions of
  axial-vector hadronic tau decays and determination of $\alpha_s(M_\tau^2)$}},
  \href{https://doi.org/10.1007/s100520050217}{\emph{Eur. Phys. J. C}
  {\bfseries 4} (1998) 409}.

\bibitem{OPAL:1998rrm}
{\scshape OPAL} collaboration, \emph{{Measurement of the strong coupling
  constant alpha(s) and the vector and axial vector spectral functions in
  hadronic tau decays}},
  \href{https://doi.org/10.1007/s100529901061}{\emph{Eur. Phys. J. C}
  {\bfseries 7} (1999) 571}
  [\href{https://arxiv.org/abs/hep-ex/9808019}{{\ttfamily hep-ex/9808019}}].

\bibitem{LeDiberder:1992zhd}
F.~Le~Diberder and A.~Pich, \emph{{Testing QCD with tau decays}},
  \href{https://doi.org/10.1016/0370-2693(92)91380-R}{\emph{Physics Letters B}
  {\bfseries 289} (1992) 165}.

\bibitem{Pich:2016bdg}
A.~Pich and A.~Rodr\'iguez-S\'anchez, \emph{{Determination of the QCD coupling
  from ALEPH $\tau$ decay data}},
  \href{https://doi.org/10.1103/PhysRevD.94.034027}{\emph{Physical Review D}
  {\bfseries 94} (2016) 034027}
  [\href{https://arxiv.org/abs/1605.06830}{{\ttfamily 1605.06830}}].

\bibitem{Ayala:2022cxo}
C.~Ayala, G.~Cvetic and D.~Teca, \emph{{Borel\textendash{}Laplace sum rules
  with \ensuremath{\tau} decay data, using OPE with improved anomalous
  dimensions}}, \href{https://doi.org/10.1088/1361-6471/acbd65}{\emph{J. Phys.
  G} {\bfseries 50} (2023) 045004}
  [\href{https://arxiv.org/abs/2206.05631}{{\ttfamily 2206.05631}}].

\bibitem{Boito:2011qt}
D.~Boito, O.~Cat{\`a}, M.~Golterman, M.~Jamin, K.~Maltman, J.~Osborne et~al.,
  \emph{{A new determination of $\alpha_s$ from hadronic $\tau$ decays}},
  \href{https://doi.org/10.1103/PhysRevD.84.113006}{\emph{Physical Review D}
  {\bfseries 84} (2011) 113006}
  [\href{https://arxiv.org/abs/1110.1127}{{\ttfamily 1110.1127}}].

\bibitem{Boito:2012cr}
D.~Boito, M.~Golterman, M.~Jamin, A.~Mahdavi, K.~Maltman, J.~Osborne et~al.,
  \emph{{An Updated determination of $\alpha_s$ from $\tau$ decays}},
  \href{https://doi.org/10.1103/PhysRevD.85.093015}{\emph{Phys. Rev. D}
  {\bfseries 85} (2012) 093015}
  [\href{https://arxiv.org/abs/1203.3146}{{\ttfamily 1203.3146}}].

\bibitem{Boito:2014sta}
D.~Boito, M.~Golterman, K.~Maltman, J.~Osborne and S.~Peris, \emph{{Strong
  coupling from the revised ALEPH data for hadronic $\tau$ decays}},
  \href{https://doi.org/10.1103/PhysRevD.91.034003}{\emph{Physical Review D}
  {\bfseries 91} (2015) 034003}
  [\href{https://arxiv.org/abs/1410.3528}{{\ttfamily 1410.3528}}].

\bibitem{Boito:2020xli}
D.~Boito, M.~Golterman, K.~Maltman, S.~Peris, M.~V. Rodrigues and W.~Schaaf,
  \emph{{Strong coupling from an improved $\tau$ vector isovector spectral
  function}}, \href{https://doi.org/10.1103/PhysRevD.103.034028}{\emph{Physical
  Review D} {\bfseries 103} (2021) 034028}
  [\href{https://arxiv.org/abs/2012.10440}{{\ttfamily 2012.10440}}].

\bibitem{Boito:2017cnp}
D.~Boito, I.~Caprini, M.~Golterman, K.~Maltman and S.~Peris,
  \emph{{Hyperasymptotics and quark-hadron duality violations in QCD}},
  \href{https://doi.org/10.1103/PhysRevD.97.054007}{\emph{Physical Review D}
  {\bfseries 97} (2018) 054007}
  [\href{https://arxiv.org/abs/1711.10316}{{\ttfamily 1711.10316}}].

\bibitem{Boito:2016oam}
D.~Boito, M.~Golterman, K.~Maltman and S.~Peris, \emph{{Strong coupling from
  hadronic $\tau$ decays: A critical appraisal}},
  \href{https://doi.org/10.1103/PhysRevD.95.034024}{\emph{Phys. Rev. D}
  {\bfseries 95} (2017) 034024}
  [\href{https://arxiv.org/abs/1611.03457}{{\ttfamily 1611.03457}}].

\bibitem{Pich:2016yfh}
A.~Pich, \emph{{Precision physics with QCD}},
  \href{https://doi.org/10.1051/epjconf/201713701016}{\emph{EPJ Web Conf.}
  {\bfseries 137} (2017) 01016}
  [\href{https://arxiv.org/abs/1612.05010}{{\ttfamily 1612.05010}}].

\bibitem{Pich:2018jiy}
A.~Pich, \emph{{Tau-decay determination of the strong coupling}},
  \href{https://doi.org/10.21468/SciPostPhysProc.1.036}{\emph{SciPost Phys.
  Proc.} {\bfseries 1} (2019) 036}
  [\href{https://arxiv.org/abs/1811.10067}{{\ttfamily 1811.10067}}].

\bibitem{Pich:2020gzz}
A.~Pich, \emph{{Precision physics with inclusive QCD processes}},
  \href{https://doi.org/10.1016/j.ppnp.2020.103846}{\emph{Prog. Part. Nucl.
  Phys.} {\bfseries 117} (2021) 103846}
  [\href{https://arxiv.org/abs/2012.04716}{{\ttfamily 2012.04716}}].

\bibitem{Pich:2022tca}
A.~Pich and A.~Rodriguez-Sanchez, \emph{{Violations of quark-hadron duality in
  low-energy determinations of $\alpha_{s}$}},
  \href{https://doi.org/10.1007/JHEP07(2022)145}{\emph{JHEP} {\bfseries 07}
  (2022) 145} [\href{https://arxiv.org/abs/2205.07587}{{\ttfamily
  2205.07587}}].

\bibitem{Pivovarov:1991rh}
A.~A. Pivovarov, \emph{{Renormalization group analysis of the tau lepton decay
  within QCD}}, \href{https://doi.org/10.1007/BF01625906}{\emph{Zeitschrift
  für Physik C} {\bfseries 53} (1992) 461}
  [\href{https://arxiv.org/abs/hep-ph/0302003}{{\ttfamily hep-ph/0302003}}].

\bibitem{LeDiberder:1992jjr}
F.~Le~Diberder and A.~Pich, \emph{{The perturbative QCD prediction to R(tau)
  revisited}}, \href{https://doi.org/10.1016/0370-2693(92)90172-Z}{\emph{Phys.
  Lett. B} {\bfseries 286} (1992) 147}.

\bibitem{Pich:2013lsa}
A.~Pich, \emph{{Precision Tau Physics}},
  \href{https://doi.org/10.1016/j.ppnp.2013.11.002}{\emph{Prog. Part. Nucl.
  Phys.} {\bfseries 75} (2014) 41}
  [\href{https://arxiv.org/abs/1310.7922}{{\ttfamily 1310.7922}}].

\bibitem{Beneke:2008ad}
M.~Beneke and M.~Jamin, \emph{{$\alpha_s$ and the tau hadronic width:
  fixed-order, contour-improved and higher-order perturbation theory}},
  \href{https://doi.org/10.1088/1126-6708/2008/09/044}{\emph{Journal of High
  Energy Physics} {\bfseries 09} (2008) 044}
  [\href{https://arxiv.org/abs/0806.3156}{{\ttfamily 0806.3156}}].

\bibitem{Beneke:2012vb}
M.~Beneke, D.~Boito and M.~Jamin, \emph{{Perturbative expansion of tau hadronic
  spectral function moments and $\alpha_s$ extractions}},
  \href{https://doi.org/10.1007/JHEP01(2013)125}{\emph{Journal of High Energy
  Physics} {\bfseries 01} (2013) 125}
  [\href{https://arxiv.org/abs/1210.8038}{{\ttfamily 1210.8038}}].

\bibitem{Hoang:2020mkw}
A.~H. Hoang and C.~Regner, \emph{{Borel representation of $\tau$ hadronic
  spectral function moments in contour-improved perturbation theory}},
  \href{https://doi.org/10.1103/PhysRevD.105.096023}{\emph{Physical Review D}
  {\bfseries 105} (2022) 096023}
  [\href{https://arxiv.org/abs/2008.00578}{{\ttfamily 2008.00578}}].

\bibitem{Hoang:2021nlz}
A.~H. Hoang and C.~Regner, \emph{{On the difference between FOPT and CIPT for
  hadronic tau decays}},
  \href{https://doi.org/10.1140/epjs/s11734-021-00257-z}{\emph{European
  Physical Journal Special Topics} {\bfseries 230} (2021) 2625}
  [\href{https://arxiv.org/abs/2105.11222}{{\ttfamily 2105.11222}}].

\bibitem{Benitez-Rathgeb:2022yqb}
M.~A. Benitez-Rathgeb, D.~Boito, A.~H. Hoang and M.~Jamin, \emph{{Reconciling
  the contour-improved and fixed-order approaches for \ensuremath{\tau}
  hadronic spectral moments. Part I. Renormalon-free gluon condensate scheme}},
  \href{https://doi.org/10.1007/JHEP07(2022)016}{\emph{Journal of High Energy
  Physics} {\bfseries 07} (2022) 016}
  [\href{https://arxiv.org/abs/2202.10957}{{\ttfamily 2202.10957}}].

\bibitem{Benitez-Rathgeb:2022hfj}
M.~A. Benitez-Rathgeb, D.~Boito, A.~H. Hoang and M.~Jamin, \emph{{Reconciling
  the contour-improved and fixed-order approaches for \ensuremath{\tau}
  hadronic spectral moments. Part II. Renormalon norm and application in
  \ensuremath{\alpha}$_{s}$ determinations}},
  \href{https://doi.org/10.1007/JHEP09(2022)223}{\emph{Journal of High Energy
  Physics} {\bfseries 09} (2022) 223}
  [\href{https://arxiv.org/abs/2207.01116}{{\ttfamily 2207.01116}}].

\bibitem{Gracia:2023qdy}
N.~G. Gracia, A.~H. Hoang and V.~Mateu, \emph{{Mathematical aspects of the
  asymptotic expansion in contour improved perturbation theory for hadronic tau
  decays}}, \href{https://doi.org/10.1103/PhysRevD.108.034013}{\emph{Phys. Rev.
  D} {\bfseries 108} (2023) 034013}
  [\href{https://arxiv.org/abs/2305.10288}{{\ttfamily 2305.10288}}].

\bibitem{Golterman:2023oml}
M.~Golterman, K.~Maltman and S.~Peris, \emph{{Difference between fixed-order
  and contour-improved perturbation theory}},
  \href{https://doi.org/10.1103/PhysRevD.108.014007}{\emph{Phys. Rev. D}
  {\bfseries 108} (2023) 014007}
  [\href{https://arxiv.org/abs/2305.10386}{{\ttfamily 2305.10386}}].

\bibitem{Beneke:2023wkq}
M.~Beneke and H.~Takaura, \emph{{Gradient-flow renormalon subtraction and the
  hadronic tau decay series}},  9, 2023,
  \href{https://arxiv.org/abs/2309.10853}{{\ttfamily 2309.10853}}.

\bibitem{Huston:2023ofk}
J.~Huston, K.~Rabbertz and G.~Zanderighi, \emph{{Quantum Chromodynamics}},
  \href{https://arxiv.org/abs/2312.14015}{{\ttfamily 2312.14015}}.

\bibitem{Davier:2013sfa}
M.~Davier, A.~H{\"o}cker, B.~Malaescu, C.-Z. Yuan and Z.~Zhang, \emph{{Update
  of the ALEPH non-strange spectral functions from hadronic $\tau$ decays}},
  \href{https://doi.org/10.1140/epjc/s10052-014-2803-9}{\emph{European Physical
  Journal C} {\bfseries 74} (2014) 2803}
  [\href{https://arxiv.org/abs/1312.1501}{{\ttfamily 1312.1501}}].

\bibitem{BaBar:2007ceh}
{\scshape BaBar} collaboration, \emph{{Measurements of $e^{+} e^{-} \to K^{+}
  K^{-} \eta$, $K^{+} K^{-} \pi^0$ and $K^0_{s} K^\pm \pi^\mp$ cross- sections
  using initial state radiation events}},
  \href{https://doi.org/10.1103/PhysRevD.77.092002}{\emph{Phys. Rev. D}
  {\bfseries 77} (2008) 092002}
  [\href{https://arxiv.org/abs/0710.4451}{{\ttfamily 0710.4451}}].

\bibitem{BaBar:2017zmc}
{\scshape BaBar} collaboration, \emph{{Measurement of the
  ${e}^{+}{e}^{{-}}{\rightarrow}{{\pi}}^{+}{{\pi}}^{{-}}{{\pi}}^{0}{{\pi}}^{0}$
  cross section using initial-state radiation at BABAR}},
  \href{https://doi.org/10.1103/PhysRevD.96.092009}{\emph{Phys. Rev. D}
  {\bfseries 96} (2017) 092009}
  [\href{https://arxiv.org/abs/1709.01171}{{\ttfamily 1709.01171}}].

\bibitem{Achasov:2016zvn}
M.~N. Achasov et~al., \emph{{Updated measurement of the $e^+e^- \to \omega
  \pi^0 \to \pi^0\pi^0\gamma$ cross section with the SND detector}},
  \href{https://doi.org/10.1103/PhysRevD.94.112001}{\emph{Phys. Rev. D}
  {\bfseries 94} (2016) 112001}
  [\href{https://arxiv.org/abs/1610.00235}{{\ttfamily 1610.00235}}].

\bibitem{SND:2014rfi}
{\scshape SND} collaboration, \emph{{Measurement of the $e^+e^- \to
  \eta\pi^+\pi^-$ cross section in the center-of-mass energy range 1.22-2.00
  GeV with the SND detector at the VEPP-2000 collider}},
  \href{https://doi.org/10.1103/PhysRevD.91.052013}{\emph{Phys. Rev. D}
  {\bfseries 91} (2015) 052013}
  [\href{https://arxiv.org/abs/1412.1971}{{\ttfamily 1412.1971}}].

\bibitem{Achasov:2017kqm}
M.~N. Achasov et~al., \emph{{Measurement of the $e^+e^- \to \eta\pi^+\pi^-$
  cross section with the SND detector at the VEPP-2000 collider}},
  \href{https://doi.org/10.1103/PhysRevD.97.012008}{\emph{Phys. Rev. D}
  {\bfseries 97} (2018) 012008}
  [\href{https://arxiv.org/abs/1711.08862}{{\ttfamily 1711.08862}}].

\bibitem{BaBar:2018erh}
{\scshape BaBar} collaboration, \emph{{Study of the process $e^+e^- \to
  \pi^+\pi^-\eta $ using initial state radiation}},
  \href{https://doi.org/10.1103/PhysRevD.97.052007}{\emph{Phys. Rev. D}
  {\bfseries 97} (2018) 052007}
  [\href{https://arxiv.org/abs/1801.02960}{{\ttfamily 1801.02960}}].

\bibitem{BaBar:2018rkc}
{\scshape BaBar} collaboration, \emph{{Study of the reactions
  $e^+e^-\to\pi^+\pi^-\pi^0\pi^0\pi^0\gamma$ and
  $\pi^+\pi^-\pi^0\pi^0\eta\gamma$ at center-of-mass energies from threshold to
  4.35 GeV using initial-state radiation}},
  \href{https://doi.org/10.1103/PhysRevD.98.112015}{\emph{Phys. Rev. D}
  {\bfseries 98} (2018) 112015}
  [\href{https://arxiv.org/abs/1810.11962}{{\ttfamily 1810.11962}}].

\bibitem{Gribanov:2019qgw}
S.~S. Gribanov et~al., \emph{{Measurement of the
  $e^{+}e^{-}\rightarrow\eta\pi^{+}\pi^{-}$ cross section with the CMD-3
  detector at the VEPP-2000 collider}},
  \href{https://doi.org/10.1007/JHEP01(2020)112}{\emph{JHEP} {\bfseries 01}
  (2020) 112} [\href{https://arxiv.org/abs/1907.08002}{{\ttfamily
  1907.08002}}].

\bibitem{BaBar:2006vzy}
{\scshape BaBar} collaboration, \emph{{The $e^+e^- \to 3(\pi^+ \pi^-), 2(\pi^+
  \pi^- \pi^0)$ and $K^+ K^- 2(\pi^+ \pi^-)$ cross sections at center-of-mass
  energies from production threshold to 4.5-GeV measured with initial-state
  radiation}}, \href{https://doi.org/10.1103/PhysRevD.73.052003}{\emph{Phys.
  Rev. D} {\bfseries 73} (2006) 052003}
  [\href{https://arxiv.org/abs/hep-ex/0602006}{{\ttfamily hep-ex/0602006}}].

\bibitem{CMD-3:2013nph}
{\scshape CMD-3} collaboration, \emph{{Study of the process $e^+e^-\to
  3(\pi^+\pi^-)$ in the c.m.energy range 1.5--2.0 gev with the cmd-3
  detector}}, \href{https://doi.org/10.1016/j.physletb.2013.04.065}{\emph{Phys.
  Lett. B} {\bfseries 723} (2013) 82}
  [\href{https://arxiv.org/abs/1302.0053}{{\ttfamily 1302.0053}}].

\bibitem{Achasov:2019nws}
M.~N. Achasov et~al., \emph{{Recent results from SND detector at VEPP-2000
  collider}}, \href{https://doi.org/10.1051/epjconf/201921204002}{\emph{EPJ Web
  Conf.} {\bfseries 212} (2019) 04002}.

\bibitem{CMD-3:2017tgb}
{\scshape CMD-3} collaboration, \emph{{Study of the process $e^+e^-\to
  \pi^+\pi^-\pi^0\eta$ in the c.m. energy range 1394-2005 MeV with the CMD-3
  detector}}, \href{https://doi.org/10.1016/j.physletb.2017.08.019}{\emph{Phys.
  Lett. B} {\bfseries 773} (2017) 150}
  [\href{https://arxiv.org/abs/1706.06267}{{\ttfamily 1706.06267}}].

\bibitem{Achasov:2019duv}
M.~N. Achasov et~al., \emph{{Measurement of the $e^+e^- \to
  \pi^+\pi^-\pi^0\eta$ cross section below $\sqrt{s}=2$ GeV}},
  \href{https://doi.org/10.1103/PhysRevD.99.112004}{\emph{Phys. Rev. D}
  {\bfseries 99} (2019) 112004}
  [\href{https://arxiv.org/abs/1903.09307}{{\ttfamily 1903.09307}}].

\bibitem{BaBar:2007qju}
{\scshape BaBar} collaboration, \emph{{The $e^+ e^- \to 2(\pi^+ \pi^-) \pi^0,$
  $2(\pi^+ \pi^-) \eta$, $K^+ K^- \pi^+ \pi^- \pi^0$ and $K^+ K^- \pi^+ \pi^-
  \eta$ Cross Sections Measured with Initial-State Radiation}},
  \href{https://doi.org/10.1103/PhysRevD.76.092005}{\emph{Phys. Rev. D}
  {\bfseries 76} (2007) 092005}
  [\href{https://arxiv.org/abs/0708.2461}{{\ttfamily 0708.2461}}].

\bibitem{Achasov:2016eyg}
M.~N. Achasov et~al., \emph{{Study of the process $e^+e^-\to\omega\eta\pi^0$ in
  the energy range $\sqrt{s} <2$ GeV with the SND detector}},
  \href{https://doi.org/10.1103/PhysRevD.94.032010}{\emph{Phys. Rev. D}
  {\bfseries 94} (2016) 032010}
  [\href{https://arxiv.org/abs/1606.06481}{{\ttfamily 1606.06481}}].

\bibitem{BaBar:2018qry}
{\scshape BaBar} collaboration, \emph{{Measurement of the spectral function for
  the $\tau^-\to K^-K_S\nu_{\tau}$ decay}},
  \href{https://doi.org/10.1103/PhysRevD.98.032010}{\emph{Phys. Rev. D}
  {\bfseries 98} (2018) 032010}
  [\href{https://arxiv.org/abs/1806.10280}{{\ttfamily 1806.10280}}].

\bibitem{Baikov:2008jh}
P.~A. Baikov, K.~G. Chetyrkin and J.~H. Kuhn, \emph{{Order $\alpha^4_s$ QCD
  Corrections to $Z$ and $\tau$ Decays}},
  \href{https://doi.org/10.1103/PhysRevLett.101.012002}{\emph{Physical Review
  Letters} {\bfseries 101} (2008) 012002}
  [\href{https://arxiv.org/abs/0801.1821}{{\ttfamily 0801.1821}}].

\bibitem{Herzog:2017dtz}
F.~Herzog, B.~Ruijl, T.~Ueda, J.~A.~M. Vermaseren and A.~Vogt, \emph{{On Higgs
  decays to hadrons and the R-ratio at N$^{4}$LO}},
  \href{https://doi.org/10.1007/JHEP08(2017)113}{\emph{Journal of High Energy
  Physics} {\bfseries 08} (2017) 113}
  [\href{https://arxiv.org/abs/1707.01044}{{\ttfamily 1707.01044}}].

\bibitem{Boito:2018rwt}
D.~Boito, P.~Masjuan and F.~Oliani, \emph{{Higher-order QCD corrections to
  hadronic $\tau$ decays from Pad{\'e} approximants}},
  \href{https://doi.org/10.1007/JHEP08(2018)075}{\emph{Journal of High Energy
  Physics} {\bfseries 08} (2018) 075}
  [\href{https://arxiv.org/abs/1807.01567}{{\ttfamily 1807.01567}}].

\bibitem{Caprini:2019kwp}
I.~Caprini, \emph{{Higher-order perturbative coefficients in QCD from series
  acceleration by conformal mappings}},
  \href{https://doi.org/10.1103/PhysRevD.100.056019}{\emph{Physical Review D}
  {\bfseries 100} (2019) 056019}
  [\href{https://arxiv.org/abs/1908.06632}{{\ttfamily 1908.06632}}].

\bibitem{Jamin:2021qxb}
M.~Jamin, \emph{{Higher-order behaviour of two-point current correlators}},
  \href{https://doi.org/10.1140/epjs/s11734-021-00266-y}{\emph{European Physics
  Journal Special Topics} {\bfseries 230} (2021) 2609}
  [\href{https://arxiv.org/abs/2106.01614}{{\ttfamily 2106.01614}}].

\bibitem{Chetyrkin:2017lif}
K.~G. Chetyrkin, J.~H. K{\"u}hn, A.~Maier, P.~Maierhofer, P.~Marquard,
  M.~Steinhauser et~al., \emph{{Addendum to ``Charm and bottom quark masses: An
  update''}}, \href{https://doi.org/10.1103/PhysRevD.96.116007}{\emph{Phys.
  Rev.} {\bfseries D96} (2017) 116007}
  [\href{https://arxiv.org/abs/1710.04249}{{\ttfamily 1710.04249}}].

\bibitem{Herzog:2017ohr}
F.~Herzog, B.~Ruijl, T.~Ueda, J.~A.~M. Vermaseren and A.~Vogt, \emph{{The
  five-loop beta function of Yang-Mills theory with fermions}},
  \href{https://doi.org/10.1007/JHEP02(2017)090}{\emph{Journal of High Energy
  Physics} {\bfseries 02} (2017) 090}
  [\href{https://arxiv.org/abs/1701.01404}{{\ttfamily 1701.01404}}].

\bibitem{Boito:2020hvu}
D.~Boito and F.~Oliani, \emph{{Renormalons in integrated spectral function
  moments and $\alpha_s$ extractions}},
  \href{https://doi.org/10.1103/PhysRevD.101.074003}{\emph{Physical Review D}
  {\bfseries 101} (2020) 074003}
  [\href{https://arxiv.org/abs/2002.12419}{{\ttfamily 2002.12419}}].

\bibitem{Shifman:2000jv}
M.~A. Shifman, \emph{{Quark hadron duality}},  in \emph{{8th International
  Symposium on Heavy Flavor Physics}}, vol.~3, (Singapore), pp.~1447--1494,
  World Scientific, 7, 2000,
  \href{https://doi.org/10.1142/9789812810458_0032}{DOI}
  [\href{https://arxiv.org/abs/hep-ph/0009131}{{\ttfamily hep-ph/0009131}}].

\bibitem{Peris:2016jah}
S.~Peris, D.~Boito, M.~Golterman and K.~Maltman, \emph{{The case for duality
  violations in the analysis of hadronic $\tau$ decays}},
  \href{https://doi.org/10.1142/S0217732316300317}{\emph{Mod. Phys. Lett. A}
  {\bfseries 31} (2016) 1630031}
  [\href{https://arxiv.org/abs/1606.08898}{{\ttfamily 1606.08898}}].

\bibitem{Cata:2008ru}
O.~Cat\`a, M.~Golterman and S.~Peris, \emph{{Possible duality violations in tau
  decay and their impact on the determination of alpha(s)}},
  \href{https://doi.org/10.1103/PhysRevD.79.053002}{\emph{Phys. Rev. D}
  {\bfseries 79} (2009) 053002}
  [\href{https://arxiv.org/abs/0812.2285}{{\ttfamily 0812.2285}}].

\bibitem{Boito:2012nt}
D.~Boito, M.~Golterman, M.~Jamin, K.~Maltman and S.~Peris, \emph{{Low-energy
  constants and condensates from the $\tau$ hadronic spectral functions}},
  \href{https://doi.org/10.1103/PhysRevD.87.094008}{\emph{Phys. Rev. D}
  {\bfseries 87} (2013) 094008}
  [\href{https://arxiv.org/abs/1212.4471}{{\ttfamily 1212.4471}}].

\bibitem{Boito:2015fra}
D.~Boito, A.~Francis, M.~Golterman, R.~Hudspith, R.~Lewis, K.~Maltman et~al.,
  \emph{{Low-energy constants and condensates from ALEPH hadronic
  \ensuremath{\tau} decay data}},
  \href{https://doi.org/10.1103/PhysRevD.92.114501}{\emph{Phys. Rev. D}
  {\bfseries 92} (2015) 114501}
  [\href{https://arxiv.org/abs/1503.03450}{{\ttfamily 1503.03450}}].

\bibitem{Gonzalez-Alonso:2010kpl}
M.~Gonz\'alez-Alonso, A.~Pich and J.~Prades, \emph{{Violation of Quark-Hadron
  Duality and Spectral Chiral Moments in QCD}},
  \href{https://doi.org/10.1103/PhysRevD.81.074007}{\emph{Phys. Rev. D}
  {\bfseries 81} (2010) 074007}
  [\href{https://arxiv.org/abs/1001.2269}{{\ttfamily 1001.2269}}].

\bibitem{Gonzalez-Alonso:2016ndl}
M.~Gonz\'alez-Alonso, A.~Pich and A.~Rodr\'\i{}guez-S\'anchez, \emph{{Updated
  determination of chiral couplings and vacuum condensates from hadronic $\tau$
  decay data}}, \href{https://doi.org/10.1103/PhysRevD.94.014017}{\emph{Phys.
  Rev. D} {\bfseries 94} (2016) 014017}
  [\href{https://arxiv.org/abs/1602.06112}{{\ttfamily 1602.06112}}].

\bibitem{Masjuan:2012gc}
P.~Masjuan, E.~Ruiz~Arriola and W.~Broniowski, \emph{{Systematics of radial and
  angular-momentum Regge trajectories of light non-strange
  q\textbackslash{}bar{q}-states}},
  \href{https://doi.org/10.1103/PhysRevD.85.094006}{\emph{Phys. Rev. D}
  {\bfseries 85} (2012) 094006}
  [\href{https://arxiv.org/abs/1203.4782}{{\ttfamily 1203.4782}}].

\bibitem{Peris:2021jap}
S.~Peris, \emph{{Violation of quark\textendash{}hadron duality: The missing
  oscillation in the OPE}},
  \href{https://doi.org/10.1140/epjs/s11734-021-00255-1}{\emph{Eur. Phys. J.
  ST} {\bfseries 230} (2021) 2691}.

\bibitem{Maltman:2008nf}
K.~Maltman and T.~Yavin, \emph{{$\alpha_s(M_Z^2)$ from hadronic tau decays}},
  \href{https://doi.org/10.1103/PhysRevD.78.094020}{\emph{Phys. Rev. D}
  {\bfseries 78} (2008) 094020}
  [\href{https://arxiv.org/abs/0807.0650}{{\ttfamily 0807.0650}}].

\bibitem{Hudspith:2017vew}
R.~J. Hudspith, R.~Lewis, K.~Maltman and J.~Zanotti, \emph{{A resolution of the
  inclusive flavor-breaking $\tau$ $|V_{us}|$ puzzle}},
  \href{https://doi.org/10.1016/j.physletb.2018.03.074}{\emph{Phys. Lett. B}
  {\bfseries 781} (2018) 206}
  [\href{https://arxiv.org/abs/1702.01767}{{\ttfamily 1702.01767}}].

\bibitem{Boito:2019iwh}
D.~Boito, M.~Golterman, K.~Maltman and S.~Peris, \emph{{Evidence against naive
  truncations of the OPE from $e^+e^- \to$ hadrons below charm}},
  \href{https://doi.org/10.1103/PhysRevD.100.074009}{\emph{Phys. Rev. D}
  {\bfseries 100} (2019) 074009}
  [\href{https://arxiv.org/abs/1907.03360}{{\ttfamily 1907.03360}}].

\bibitem{Evangelista:2023fmt}
{\scshape Extended Twisted Mass} collaboration, \emph{{Inclusive hadronic decay
  rate of the \ensuremath{\tau} lepton from lattice QCD}},
  \href{https://doi.org/10.1103/PhysRevD.108.074513}{\emph{Phys. Rev. D}
  {\bfseries 108} (2023) 074513}
  [\href{https://arxiv.org/abs/2308.03125}{{\ttfamily 2308.03125}}].

\bibitem{HFLAV:2022esi}
{\scshape HFLAV} collaboration, \emph{{Averages of b-hadron, c-hadron, and
  \ensuremath{\tau}-lepton properties as of 2021}},
  \href{https://doi.org/10.1103/PhysRevD.107.052008}{\emph{Phys. Rev. D}
  {\bfseries 107} (2023) 052008}
  [\href{https://arxiv.org/abs/2206.07501}{{\ttfamily 2206.07501}}].

\bibitem{Erler:2002mv}
J.~Erler, \emph{{Electroweak radiative corrections to semileptonic tau
  decays}}, {\emph{Rev. Mex. Fis.} {\bfseries 50} (2004) 200}
  [\href{https://arxiv.org/abs/hep-ph/0211345}{{\ttfamily hep-ph/0211345}}].

\end{thebibliography}\endgroup

\end{document}